\documentclass[journal,a4paper]{IEEEtran}
\IEEEoverridecommandlockouts
\usepackage{relsize}
\usepackage{diagbox}
\usepackage{multirow}
\usepackage[utf8]{inputenc} 
\usepackage[T1]{fontenc}
\usepackage{url}              
\usepackage{cite}             

\usepackage[cmex10]{amsmath}  
\usepackage{mleftright}       
\mleftright                   

\usepackage{graphicx}         
\usepackage{booktabs}         

\usepackage[caption=false,font=footnotesize]{subfig}

\usepackage{stfloats}

\usepackage{tikz}
\usepackage{pgfplotstable}
\usepackage{pgfplots}
\usetikzlibrary{plotmarks}
  \usetikzlibrary{arrows.meta}
  \usepgfplotslibrary{patchplots}
  \usepackage{grffile}
  \pgfplotsset{plot coordinates/math parser=false}
  \newlength\figureheight
  \newlength\figurewidth
   \pgfplotsset{compat=1.11,
    /pgfplots/ybar legend/.style={
    /pgfplots/legend image code/.code={%
       \draw[##1,/tikz/.cd,yshift=-0.25em]
        (0cm,0cm) rectangle (3em,8pt);},
   },
}

\usepackage[utf8]{inputenc}
\usepackage{relsize}
\usepackage{amsfonts, amssymb, cuted}
\usepackage{cleveref}
\usepackage{mathtools}
\usepackage{algorithm}
\usepackage[noend]{algpseudocode}
\usepackage{xcolor}
\usepackage{soul}
\usepackage{nicefrac}
\usepackage{enumitem}
\usepackage{acro}
\usepackage{pifont}

\title{Cache-Aided Communications in MISO Networks with Dynamic User Behavior}
\author{\IEEEauthorblockN{Milad Abolpour,~\IEEEmembership{Student Member,~IEEE}, MohammadJavad Salehi,~\IEEEmembership{ Member,~IEEE}, and\\Antti T\"olli,~\IEEEmembership{Senior Member,~IEEE}
\thanks{The authors are with the Centre for Wireless Communications, University of Oulu, Finland. \textrm{E-mail: \{firstname.lastname\}@oulu.fi}. This research has been supported by the Academy of Finland, 6G Flagship program under Grant  346208, 343586 (CAMAIDE), and by the Finnish-American Research and Innovation Accelerator (FARIA) program. Part of this work has been presented at the IEEE International Symposium on Information Theory (ISIT) 2023, in Taipei, Taiwan.}}
}

\begin{document}

\newtheorem{lemma}{Lemma}
\newtheorem{corollary}{Corollary}

\newcommand{\diag}{{\mbox{diag}}}
\newcommand{\herm}{^{\mbox{\scriptsize H}}}
\newcommand{\sherm}{^{\mbox{\scriptsize H}}}
\newcommand{\tran}{^{\mbox{\scriptsize T}}}
\newcommand{\stran}{^{\mbox{\tiny T}}}

\newcommand{\vbar}{\raisebox{.17ex}{\rule{.04em}{1.35ex}}}
\newcommand{\vbarind}{\raisebox{.01ex}{\rule{.04em}{1.1ex}}}
\newcommand{\R}{\ifmmode{\rm I}\hspace{-.2em}{\rm R} \else ${\rm I}\hspace{-.2em}{\rm R}$ \fi}
\newcommand{\T}{\ifmmode{\rm I}\hspace{-.2em}{\rm T} \else ${\rm I}\hspace{-.2em}{\rm T}$ \fi}
\newcommand{\N}{\ifmmode{\rm I}\hspace{-.2em}{\rm N} \else \mbox{${\rm I}\hspace{-.2em}{\rm N}$} \fi}
\newcommand{\B}{\ifmmode{\rm I}\hspace{-.2em}{\rm B} \else \mbox{${\rm I}\hspace{-.2em}{\rm B}$} \fi}
\newcommand{\Hil}{\ifmmode{\rm I}\hspace{-.2em}{\rm H} \else \mbox{${\rm I}\hspace{-.2em}{\rm H}$} \fi}
\newcommand{\C}{\ifmmode\hspace{.2em}\vbar\hspace{-.31em}{\rm C} \else \mbox{$\hspace{.2em}\vbar\hspace{-.31em}{\rm C}$} \fi}
\newcommand{\Cind}{\ifmmode\hspace{.2em}\vbarind\hspace{-.25em}{\rm C} \else \mbox{$\hspace{.2em}\vbarind\hspace{-.25em}{\rm C}$} \fi}
\newcommand{\Q}{\ifmmode\hspace{.2em}\vbar\hspace{-.31em}{\rm Q} \else \mbox{$\hspace{.2em}\vbar\hspace{-.31em}{\rm Q}$} \fi}
\newcommand{\Z}{\ifmmode{\rm Z}\hspace{-.28em}{\rm Z} \else ${\rm Z}\hspace{-.28em}{\rm Z}$ \fi}

\newcommand{\sgn}{\mbox{sgn}}
\newcommand{\var}{\mbox{var}}
\renewcommand{\Re}{\mbox{Re}}
\renewcommand{\Im}{\mbox{Im}}

\renewcommand{\vec}[1]{\bf{#1}}     
\newcommand{\vecsc}[1]{\mbox{\bf \scriptsize #1}}
\newcommand{\itvec}[1]{\mbox{\boldmath{$#1$}}}
\newcommand{\itvecsc}[1]{\mbox{\boldmath{$\scriptstyle #1$}}}
\newcommand{\gvec}[1]{\mbox{\boldmath{$#1$}}}

\newcommand{\balpha}{\mbox{\boldmath{$\alpha$}}}
\newcommand{\bbeta}{\mbox{\boldmath{$\beta$}}}
\newcommand{\bgamma}{\mbox{\boldmath{$\gamma$}}}
\newcommand{\bdelta}{\mbox{\boldmath{$\delta$}}}
\newcommand{\bepsilon}{\mbox{\boldmath{$\epsilon$}}}
\newcommand{\bvarepsilon}{\mbox{\boldmath{$\varepsilon$}}}
\newcommand{\bzeta}{\mbox{\boldmath{$\zeta$}}}
\newcommand{\boldeta}{\mbox{\boldmath{$\eta$}}}
\newcommand{\btheta}{\mbox{\boldmath{$\theta$}}}
\newcommand{\bvartheta}{\mbox{\boldmath{$\vartheta$}}}
\newcommand{\biota}{\mbox{\boldmath{$\iota$}}}
\newcommand{\blambda}{\mbox{\boldmath{$\lambda$}}}
\newcommand{\bmu}{\mbox{\boldmath{$\mu$}}}
\newcommand{\bnu}{\mbox{\boldmath{$\nu$}}}
\newcommand{\bxi}{\mbox{\boldmath{$\xi$}}}
\newcommand{\bpi}{\mbox{\boldmath{$\pi$}}}
\newcommand{\bvarpi}{\mbox{\boldmath{$\varpi$}}}
\newcommand{\brho}{\mbox{\boldmath{$\rho$}}}
\newcommand{\bvarrho}{\mbox{\boldmath{$\varrho$}}}
\newcommand{\bsigma}{\mbox{\boldmath{$\sigma$}}}
\newcommand{\bvarsigma}{\mbox{\boldmath{$\varsigma$}}}
\newcommand{\btau}{\mbox{\boldmath{$\tau$}}}
\newcommand{\bupsilon}{\mbox{\boldmath{$\upsilon$}}}
\newcommand{\bphi}{\mbox{\boldmath{$\phi$}}}
\newcommand{\bvarphi}{\mbox{\boldmath{$\varphi$}}}
\newcommand{\bchi}{\mbox{\boldmath{$\chi$}}}
\newcommand{\bpsi}{\mbox{\boldmath{$\psi$}}}
\newcommand{\bomega}{\mbox{\boldmath{$\omega$}}}

\newcommand{\bolda}{\mbox{\boldmath{$a$}}}
\newcommand{\bb}{\mbox{\boldmath{$b$}}}
\newcommand{\bc}{\mbox{\boldmath{$c$}}}
\newcommand{\bd}{\mbox{\boldmath{$d$}}}
\newcommand{\bolde}{\mbox{\boldmath{$e$}}}
\newcommand{\boldf}{\mbox{\boldmath{$f$}}}
\newcommand{\bg}{\mbox{\boldmath{$g$}}}
\newcommand{\bh}{\mbox{\boldmath{$h$}}}
\newcommand{\bp}{\mbox{\boldmath{$p$}}}
\newcommand{\bq}{\mbox{\boldmath{$q$}}}
\newcommand{\br}{\mbox{\boldmath{$r$}}}
\newcommand{\bs}{\mbox{\boldmath{$s$}}}
\newcommand{\bt}{\mbox{\boldmath{$t$}}}
\newcommand{\bu}{\mbox{\boldmath{$u$}}}
\newcommand{\bv}{\mbox{\boldmath{$v$}}}
\newcommand{\bw}{\mbox{\boldmath{$w$}}}
\newcommand{\bx}{\mbox{\boldmath{$x$}}}
\newcommand{\by}{\mbox{\boldmath{$y$}}}
\newcommand{\bz}{\mbox{\boldmath{$z$}}}

\newtheorem{thm}{Theorem}
\newtheorem{lem}{Lemma}
\newtheorem{prop}[thm]{Proposition}
\newtheorem{cor}{Corollary}
\newtheorem{conj}{Conjecture}[section]
\newtheorem{exmp}{Example}
\newtheorem{defn}{Definition}

\newtheorem{rem}{Remark}

\newcommand{\CA}[0]{{\mathcal{A}}}
\newcommand{\CB}[0]{{\mathcal{B}}}
\newcommand{\CC}[0]{{\mathcal{C}}}
\newcommand{\CD}[0]{{\mathcal{D}}}
\newcommand{\CE}[0]{{\mathcal{E}}}
\newcommand{\CF}[0]{{\mathcal{F}}}
\newcommand{\CG}[0]{{\mathcal{G}}}
\newcommand{\CH}[0]{{\mathcal{H}}}
\newcommand{\CI}[0]{{\mathcal{I}}}
\newcommand{\CJ}[0]{{\mathcal{J}}}
\newcommand{\CK}[0]{{\mathcal{K}}}
\newcommand{\CL}[0]{{\mathcal{L}}}
\newcommand{\CM}[0]{{\mathcal{M}}}
\newcommand{\CN}[0]{{\mathcal{N}}}
\newcommand{\CO}[0]{{\mathcal{O}}}
\newcommand{\CP}[0]{{\mathcal{P}}}
\newcommand{\CQ}[0]{{\mathcal{Q}}}
\newcommand{\CR}[0]{{\mathcal{R}}}
\newcommand{\CS}[0]{{\mathcal{S}}}
\newcommand{\CT}[0]{{\mathcal{T}}}
\newcommand{\CU}[0]{{\mathcal{U}}}
\newcommand{\CV}[0]{{\mathcal{V}}}
\newcommand{\CW}[0]{{\mathcal{W}}}
\newcommand{\CX}[0]{{\mathcal{X}}}
\newcommand{\CY}[0]{{\mathcal{Y}}}
\newcommand{\CZ}[0]{{\mathcal{Z}}}

\newcommand{\Ba}[0]{{\mathbf{a}}}
\newcommand{\Bb}[0]{{\mathbf{b}}}
\newcommand{\Bc}[0]{{\mathbf{c}}}
\newcommand{\Bd}[0]{{\mathbf{d}}}
\newcommand{\Be}[0]{{\mathbf{e}}}
\newcommand{\Bf}[0]{{\mathbf{f}}}
\newcommand{\Bg}[0]{{\mathbf{g}}}
\newcommand{\Bh}[0]{{\mathbf{h}}}
\newcommand{\Bi}[0]{{\mathbf{i}}}
\newcommand{\Bj}[0]{{\mathbf{j}}}
\newcommand{\Bk}[0]{{\mathbf{k}}}
\newcommand{\Bl}[0]{{\mathbf{l}}}
\newcommand{\Bm}[0]{{\mathbf{m}}}
\newcommand{\Bn}[0]{{\mathbf{n}}}
\newcommand{\Bo}[0]{{\mathbf{o}}}
\newcommand{\Bp}[0]{{\mathbf{p}}}
\newcommand{\Bq}[0]{{\mathbf{q}}}
\newcommand{\Br}[0]{{\mathbf{r}}}
\newcommand{\Bs}[0]{{\mathbf{s}}}
\newcommand{\Bt}[0]{{\mathbf{t}}}
\newcommand{\Bu}[0]{{\mathbf{u}}}
\newcommand{\Bv}[0]{{\mathbf{v}}}
\newcommand{\Bw}[0]{{\mathbf{w}}}
\newcommand{\Bx}[0]{{\mathbf{x}}}
\newcommand{\By}[0]{{\mathbf{y}}}
\newcommand{\Bz}[0]{{\mathbf{z}}}

\newcommand{\BA}[0]{{\mathbf{A}}}
\newcommand{\BB}[0]{{\mathbf{B}}}
\newcommand{\BC}[0]{{\mathbf{C}}}
\newcommand{\BD}[0]{{\mathbf{D}}}
\newcommand{\BE}[0]{{\mathbf{E}}}
\newcommand{\BF}[0]{{\mathbf{F}}}
\newcommand{\BG}[0]{{\mathbf{G}}}
\newcommand{\BH}[0]{{\mathbf{H}}}
\newcommand{\BI}[0]{{\mathbf{I}}}
\newcommand{\BJ}[0]{{\mathbf{J}}}
\newcommand{\BK}[0]{{\mathbf{K}}}
\newcommand{\BL}[0]{{\mathbf{L}}}
\newcommand{\BM}[0]{{\mathbf{M}}}
\newcommand{\BN}[0]{{\mathbf{N}}}
\newcommand{\BO}[0]{{\mathbf{O}}}
\newcommand{\BP}[0]{{\mathbf{P}}}
\newcommand{\BQ}[0]{{\mathbf{Q}}}
\newcommand{\BR}[0]{{\mathbf{R}}}
\newcommand{\BS}[0]{{\mathbf{S}}}
\newcommand{\BT}[0]{{\mathbf{T}}}
\newcommand{\BU}[0]{{\mathbf{U}}}
\newcommand{\BV}[0]{{\mathbf{V}}}
\newcommand{\BW}[0]{{\mathbf{W}}}
\newcommand{\BX}[0]{{\mathbf{X}}}
\newcommand{\BY}[0]{{\mathbf{Y}}}
\newcommand{\BZ}[0]{{\mathbf{Z}}}

\newcommand{\Bra}[0]{{\Bar{a}}}
\newcommand{\Brb}[0]{{\Bar{b}}}
\newcommand{\Brc}[0]{{\Bar{c}}}
\newcommand{\Brd}[0]{{\Bar{d}}}
\newcommand{\Bre}[0]{{\Bar{e}}}
\newcommand{\Brf}[0]{{\Bar{f}}}
\newcommand{\Brg}[0]{{\Bar{g}}}
\newcommand{\Brh}[0]{{\Bar{h}}}
\newcommand{\Bri}[0]{{\Bar{i}}}
\newcommand{\Brj}[0]{{\Bar{j}}}
\newcommand{\Brk}[0]{{\Bar{k}}}
\newcommand{\Brl}[0]{{\Bar{l}}}
\newcommand{\Brm}[0]{{\Bar{m}}}
\newcommand{\Brn}[0]{{\Bar{n}}}
\newcommand{\Bro}[0]{{\Bar{o}}}
\newcommand{\Brp}[0]{{\Bar{p}}}
\newcommand{\Brq}[0]{{\Bar{q}}}
\newcommand{\Brr}[0]{{\Bar{r}}}
\newcommand{\Brs}[0]{{\Bar{s}}}
\newcommand{\Brt}[0]{{\Bar{t}}}
\newcommand{\Bru}[0]{{\Bar{u}}}
\newcommand{\Brv}[0]{{\Bar{v}}}
\newcommand{\Brw}[0]{{\Bar{w}}}
\newcommand{\Brx}[0]{{\Bar{x}}}
\newcommand{\Bry}[0]{{\Bar{y}}}
\newcommand{\Brz}[0]{{\Bar{z}}}

\newcommand{\BrA}[0]{{\Bar{A}}}
\newcommand{\BrB}[0]{{\Bar{B}}}
\newcommand{\BrC}[0]{{\Bar{C}}}
\newcommand{\BrD}[0]{{\Bar{D}}}
\newcommand{\BrE}[0]{{\Bar{E}}}
\newcommand{\BrF}[0]{{\Bar{F}}}
\newcommand{\BrG}[0]{{\Bar{G}}}
\newcommand{\BrH}[0]{{\Bar{H}}}
\newcommand{\BrI}[0]{{\Bar{I}}}
\newcommand{\BrJ}[0]{{\Bar{J}}}
\newcommand{\BrK}[0]{{\Bar{K}}}
\newcommand{\BrL}[0]{{\Bar{L}}}
\newcommand{\BrM}[0]{{\Bar{M}}}
\newcommand{\BrN}[0]{{\Bar{N}}}
\newcommand{\BrO}[0]{{\Bar{O}}}
\newcommand{\BrP}[0]{{\Bar{P}}}
\newcommand{\BrQ}[0]{{\Bar{Q}}}
\newcommand{\BrR}[0]{{\Bar{R}}}
\newcommand{\BrS}[0]{{\Bar{S}}}
\newcommand{\BrT}[0]{{\Bar{T}}}
\newcommand{\BrU}[0]{{\Bar{U}}}
\newcommand{\BrV}[0]{{\Bar{V}}}
\newcommand{\BrW}[0]{{\Bar{W}}}
\newcommand{\BrX}[0]{{\Bar{X}}}
\newcommand{\BrY}[0]{{\Bar{Y}}}
\newcommand{\BrZ}[0]{{\Bar{Z}}}

\newcommand{\Sfa}[0]{{\mathsf{a}}}
\newcommand{\Sfb}[0]{{\mathsf{b}}}
\newcommand{\Sfc}[0]{{\mathsf{c}}}
\newcommand{\Sfd}[0]{{\mathsf{d}}}
\newcommand{\Sfe}[0]{{\mathsf{e}}}
\newcommand{\Sff}[0]{{\mathsf{f}}}
\newcommand{\Sfg}[0]{{\mathsf{g}}}
\newcommand{\Sfh}[0]{{\mathsf{h}}}
\newcommand{\Sfi}[0]{{\mathsf{i}}}
\newcommand{\Sfj}[0]{{\mathsf{j}}}
\newcommand{\Sfk}[0]{{\mathsf{k}}}
\newcommand{\Sfl}[0]{{\mathsf{l}}}
\newcommand{\Sfm}[0]{{\mathsf{m}}}
\newcommand{\Sfn}[0]{{\mathsf{n}}}
\newcommand{\Sfo}[0]{{\mathsf{o}}}
\newcommand{\Sfp}[0]{{\mathsf{p}}}
\newcommand{\Sfq}[0]{{\mathsf{q}}}
\newcommand{\Sfr}[0]{{\mathsf{r}}}
\newcommand{\Sfs}[0]{{\mathsf{s}}}
\newcommand{\Sft}[0]{{\mathsf{t}}}
\newcommand{\Sfu}[0]{{\mathsf{u}}}
\newcommand{\Sfv}[0]{{\mathsf{v}}}
\newcommand{\Sfw}[0]{{\mathsf{w}}}
\newcommand{\Sfx}[0]{{\mathsf{x}}}
\newcommand{\Sfy}[0]{{\mathsf{y}}}
\newcommand{\Sfz}[0]{{\mathsf{z}}}

\newcommand{\SfA}[0]{{\mathsf{A}}}
\newcommand{\SfB}[0]{{\mathsf{B}}}
\newcommand{\SfC}[0]{{\mathsf{C}}}
\newcommand{\SfD}[0]{{\mathsf{D}}}
\newcommand{\SfE}[0]{{\mathsf{E}}}
\newcommand{\SfF}[0]{{\mathsf{F}}}
\newcommand{\SfG}[0]{{\mathsf{G}}}
\newcommand{\SfH}[0]{{\mathsf{H}}}
\newcommand{\SfI}[0]{{\mathsf{I}}}
\newcommand{\SfJ}[0]{{\mathsf{J}}}
\newcommand{\SfK}[0]{{\mathsf{K}}}
\newcommand{\SfL}[0]{{\mathsf{L}}}
\newcommand{\SfM}[0]{{\mathsf{M}}}
\newcommand{\SfN}[0]{{\mathsf{N}}}
\newcommand{\SfO}[0]{{\mathsf{O}}}
\newcommand{\SfP}[0]{{\mathsf{P}}}
\newcommand{\SfQ}[0]{{\mathsf{Q}}}
\newcommand{\SfR}[0]{{\mathsf{R}}}
\newcommand{\SfS}[0]{{\mathsf{S}}}
\newcommand{\SfT}[0]{{\mathsf{T}}}
\newcommand{\SfU}[0]{{\mathsf{U}}}
\newcommand{\SfV}[0]{{\mathsf{V}}}
\newcommand{\SfW}[0]{{\mathsf{W}}}
\newcommand{\SfX}[0]{{\mathsf{X}}}
\newcommand{\SfY}[0]{{\mathsf{Y}}}
\newcommand{\SfZ}[0]{{\mathsf{Z}}}


\renewcommand{\Re}{\mbox{Re}}
\renewcommand{\Im}{\mbox{Im}}

\renewcommand{\vec}[1]{\bf{#1}}     

\newcommand{\FillGray}[3]{\filldraw[gray!50](#3-1+0.1,#1-#2+0.1) rectangle (#3-0.1,#1-#2+1-0.1)}
\newcommand{\FillBlack}[3]{\filldraw[black!70](#3-1+0.1,#1-#2+0.1) rectangle (#3-0.1,#1-#2+1-0.1)}
\newcommand{\FillHatch}[3]{\fill[pattern=crosshatch, pattern color=black!65](#3-1,#1-#2)rectangle(#3,#1-#2+1)}
\ExplSyntaxOn

\NewDocumentCommand \vect { s o m }
 {
  \IfBooleanTF {#1}
   { \vectaux*{#3} }
   { \IfValueTF {#2} { \vectaux[#2]{#3} } { \vectaux{#3} } }
 }
\DeclarePairedDelimiterX \vectaux [1] {\lbrack} {\rbrack}
 { \, \dbacc_vect:n { #1 } \, }
\cs_new_protected:Npn \dbacc_vect:n #1
 {
  \seq_set_split:Nnn \l_tmpa_seq { , } { #1 }
  \seq_use:Nn \l_tmpa_seq { \enspace }
 }
\ExplSyntaxOff
\maketitle

\begin{abstract}
Coded caching (CC) can substantially enhance network performance by leveraging memory as an additional communication resource. However, the use of CC is challenging in various practical applications due to dynamic user behavior. The existing solutions, based on shared caching, cannot directly handle all scenarios where users freely enter and depart the network at any time as they are constrained by specific conditions on network parameters. This paper proposes a universally applicable shared-caching scheme for dynamic setups without any restriction on network parameters. The closed-form expressions for the achievable degrees of freedom (DoF) are computed for the resulting generalized scheme, and are shown to achieve the existing optimal bounds of the shared-cache model. Furthermore, a successive-interference-cancellation-free extension based on a fast iterative optimized beamformer design is devised to optimize the use of excess spatial dimensions freed by cache-aided interference cancellation. Extensive numerical experiments are carried out to assess the performance of the proposed scheme. In particular, the results demonstrate that while a dynamic setup may achieve a DoF substantially lower than the optimal DoF of shared caching, our proposed scheme significantly improves the performance at the finite signal-to-noise ratio compared to unicasting, which only benefits from the local caching gain.
\end{abstract}

\section{Introduction}
The increasing volume and diversity of multimedia content, resulting from emerging applications such as wireless immersive viewing, requires network providers to develop their infrastructure to support data communications at higher rates and with lower latency~\cite{cisco2018ciscob,salehi2022enhancing,mahmoodi2021non}. To facilitate the efficient delivery of such multimedia content, coded caching (CC) has been proposed to increase the data rates and reduce the communication load during peak traffic times~\cite{maddah2014fundamental}. A large amount of memory is available in modern network devices. The CC models provide the opportunity to employ this distributed memory as a supplementary communication resource, enabling them  
to improve the scalability, latency, and cost-effectiveness in various real-world applications such as large-scale video-on-demand (VoD)~\cite{9361706}, extended reality (XR)~\cite{salehi2022enhancing,mahmoodi2021non}, and coded distributed computing~\cite{8051074}. Incorporating CC into a single-stream downlink network boosts the achievable rate by a multiplicative factor proportional to the cumulative cache size in the entire network via multicasting carefully designed codewords to different user groups. In a network of $K$ users, each equipped with a cache memory large enough to store a portion $\gamma$ of the entire library, CC can enable the so-called \emph{global caching gain} of $K\gamma + 1$. Interestingly, this new gain is also additive with the spatial multiplexing gain, and hence, in a multi-input single-output (MISO) setup with $L$ antennas and with the spatial multiplexing gain of $\alpha\leq L$ at the transmitter, the delivery speed-up factor of CC is increased to $K\gamma + \alpha$~\cite{shariatpanahi2016multi,tolli2017multi,shariatpanahi2018physical,lampiris2019bridgingb,10086693,8374074,10206943}, appealing given the critical role of multi-antenna connectivity in next-generation networks~\cite{rajatheva2020white}. This so-called combined degree-of-freedom (DoF) of $K\gamma + \alpha$ in MISO-CC setups is even shown to be information-theoretically optimal for uncoded cache placement and single-shot data delivery. Of course, a larger DoF is achievable by using interference alignment techniques \cite{Maddah-Ali2019Cache-AidedChannels,cao2017fundamental,Hachem2018DegreesNetworks,8278043}. However, such schemes are complex and pose implementation challenges, contrasting our primary goal of practical applicability in this~paper.

Despite these promising gains, a practical implementation of CC is still challenging due to various critical impediments. In this paper, we consider one such bottleneck: the applicability of CC techniques in dynamic setups where the users may freely enter or depart the network~\cite{salehi2021low,Abolpour2022CodedNetworks,10206827}. 
We use the existing \emph{shared-cache} model to develop a generalized low-complexity solution to enable the CC gain for every dynamic setup, coupled with an iterative optimized beamformer design for enhanced performance over the entire signal-to-noise ratio (SNR) range.

\subsection{Prior Art}

\noindent\textbf{The shared-cache model:} One of the most critical and extensively investigated challenges for the practical CC deployment is the complexity of the subpacketization process \cite{yan2017placement,shangguan2018centralized,9477627,jin2019new,tang2018coded,michel2020placement,9457641}. That is, in a network with $K$ users, each file should be split into many smaller parts, the number of which grows exponentially with $K$. This process limits the real achievable gain under practical constraints, such that in a network with the cache ratio $\gamma=0.01$ and $K=1000$ users, enabling the caching gain of $K\gamma=10$ requires each file to be split into $\binom{1000}{10}>10^{23}$ mini-files, which is not possible in practice~\cite{lampiris2018adding}. While the subpacketization complexity issue is inherent in single-antenna scenarios~\cite{8080217}, in multi-antenna setups, it is possible to address this challenge through the incorporation of signal-level decoding strategies where the undesired terms are regenerated from the local memory and removed before the received signal is retrieved at the receiver~\cite{lampiris2018adding}.\footnote{The signal-level approach is not only beneficial in reducing the subpacketization; it also enables a straightforward extension of CC schemes to multi-input multi-output (MIMO) setups~\cite{10096035}.} A systematic extension of the signal-level scheme in~\cite{lampiris2018adding} was later provided as the \emph{shared-cache} model~\cite{parrinello2019fundamental}, where there exist $P \le K$ \emph{caching profiles}, and $\eta_{p}$ users are assigned to profile  $p \in \left\lbrace 1,\cdots,P \right\rbrace$. Even though multiple users with a cache ratio of $\gamma$ can be assigned to the same profile and cache exactly the same data, the scaling behavior of CC gains in MISO setups can be close to the optimal case of a dedicated cache profile for every user~~\cite{parrinello2020extending,parrinello2019fundamental}. More precisely, for small-antenna arrays with $\alpha \leq  \frac{K}{P}$~\cite{parrinello2019fundamental} and large-antenna arrays with $\alpha > \frac{K}{P}$~\cite{parrinello2020extending}, the shared-cache model enables the optimal DoF of $\alpha \left( 1+P\gamma \right)$ and $K\gamma+\alpha$, respectively.

\noindent\textbf{CC in dynamic networks:} In order to implement CC schemes in actual applications, these schemes must be capable of effectively handling a dynamic population of users departing and entering the network at any time. The conventional CC schemes require the placement phase to be designed based on the number of users known in advance, which is impossible in dynamic networks. 
The authors in \cite{9834866, 10206467} have applied the maximum distance separable (MDS) codes in the content placement phase to manage the networks with dynamic conditions and have demonstrated that using MDS coding not only handles the dynamicity of networks but also can guarantee the privacy of users. However, implementation of MDS coding in the placement phase requires having a very large finite field. Interestingly, the shared-cache model developed for reducing the subpacketization complexity can also be used to efficiently address this issue, as the cache placement phase is built upon knowledge of the number of profiles $P$, which can be chosen independently from the number of users $K$. To this end, the authors in \cite{salehi2021low,Abolpour2022CodedNetworks,10206827} applied the shared caching idea to address the dynamic user behavior by utilizing the users' cache ratio $\gamma$ to design the content placement phase. In this sense, for a given $\gamma$ value, they selected an appropriate number of caching profiles $P$ ensuring that $P\gamma$ is an integer.  However, unfortunately, the existing models are not dynamic in the true sense, as they are applicable only under specific constraints. For example, \cite{parrinello2019fundamental} is applicable only if the spatial multiplexing gain is very small, i.e., $\alpha \leq \min_{p} \eta_{p}$, while \cite{salehi2021low} and \cite{Abolpour2022CodedNetworks} can only support the large-antenna array networks with $\alpha > \hat{\eta}$, given that $\alpha \geq K\gamma$ and $1 \leq \hat{\eta} \leq \max_{p} \eta_{p}$. Accordingly, no existing scheme in the literature can handle network dynamicity properly in the following cases: {I) small-antenna array:} $\alpha \leq \hat{\eta}$, with $\alpha \geq \min_{p} \eta_{p}$, and
{II) large-antenna array:} $\alpha> \hat{\eta}$, either with non-integer $\nicefrac{\alpha}{\hat{\eta}}$ or with $\alpha < K\gamma$.  Therefore, a universal shared-cache setup that supports any user-to-profile association is not yet available. The primary objective of this paper is to design a transmission strategy that not only supports the regions already covered but also supports the regions so far omitted in the literature. Moreover, this strategy should minimize the DoF loss in the dynamic networks, which is caused by the non-uniform association of users to the caching profiles. 

\noindent\textbf{CC in finite-SNR:}
Initial multi-antenna CC schemes focused on improving the DoF, using zero-forcing (ZF) precoders at the transmitter to completely null out the inter-stream interference \cite{shariatpanahi2018physical,shariatpanahi2017multi}. Nevertheless, the DoF metric is an appropriate performance indicator only at the high-SNR regime~\cite{10052083}, and ZF beamformers should also be replaced with optimized beamformers in finite-SNR communications~\cite{tolli2018multicast}. Accordingly, the work~\cite{tolli2017multi} focused on improving both performance and reducing the complexity of the beamforming design for MISO-CC schemes by presenting optimized beamformers that effectively controlled both the spatial multiplexing gain and the size of the multiple-access channel that is encountered during decoding. Similarly, the authors in~\cite{zhao2020multi} presented a CC method that constrained the number of messages each user receives per time slot to reduce the complexity without significantly compromising the transmission rates.

Another possibility for reducing the beamforming complexity is to use signal-level schemes that enable replacing the intricate multi-group multicast beamforming design with a considerably simpler multi-user unicast solution~\cite{salehi2020lowcomplexity,9923619}. This facilitates the deployment of CC schemes for very large networks while maintaining an enhanced performance in the finite-SNR regime. In addition, the signal-level decoding strategies allow for simplification of the receiver as a successive interference cancellation (SIC) structure is not required. In this regard, a sub-optimal beamforming design (ZF precoding) via a SIC-free data delivery scheme was studied in~\cite{9834007}.

In cache-aided MISO networks with dynamic user behavior, the achievable DoF is adversely affected by non-uniform user distributions. However, this negative impact can be mitigated by optimized beamforming techniques, which allow for greater beamforming gain in the finite-SNR regime~\cite{10052083}. To this end, the early results in~\cite{salehi2021low} demonstrated how the joint application of the shared caching model and optimized linear beamforming design can handle the impact of non-uniform distribution of users and, at the same time, boost the finite-SNR performance. 

\subsection{Our Contributions}
In this paper, a universal cache-assisted MISO system is designed to handle dynamic setups with any network parameters by leveraging the shared caching model, which also keeps the subpacketization overhead low. Moreover, we present a novel SIC-free data delivery algorithm tailored for networks with dynamic user behavior and propose a fast iterative beamforming design to optimize its finite-SNR performance. In this regard, the contributions of  this paper are three-fold:

    \noindent
    $\bullet$ Assuming the shared-cache model for content placement phase, two transmission strategies are presented for the content delivery phase able to support any user-to-profile association, including the regions not covered in the existing literature, i.e.: \textit{i)} uneven user association with $\alpha> \hat{\eta}$ and non-integer $\frac{\alpha}{\hat{\eta}}$ (unlike~\cite{parrinello2020extending}), \textit{ii)} $\min_{p} \eta_{p} \leq \alpha \leq \hat{\eta}$ (unlike~\cite{parrinello2019fundamental}), and \textit{iii)}~$\alpha > \hat{\eta}$ with $\alpha < K\gamma$ (unlike~\cite{Abolpour2022CodedNetworks}). 
   We note that the proposed shared caching model can be applied to any baseline CC scheme, as long as the constraints of the baseline scheme are satisfied. For instance, for networks with the number of files $N$ exceeding the number of users $K$, applying the proposed shared caching model to the introduced YMA scheme in~\cite{8226776} can outperform the baseline scheme \cite{maddah2014fundamental}, as YMA removes the linearly dependent multicast messages that may occur when multiple users request the same file.  
    
    \noindent
     $\bullet$ Closed-form expressions are obtained for the DoF, revealing the DoF loss caused by non-uniformness in users' distribution. Particularly, for the uniform user associations, the proposed scheme achieves the optimal DoF not only in the regions covered in the literature but also when $\alpha \geq \hat{\eta}$ and $\nicefrac{\alpha}{\hat{\eta}}$ is non-integer.
     
    \noindent
     $\bullet$ In contrast to the approaches outlined in~\cite{parrinello2020extending,parrinello2019fundamental}, which assume the utilization of SIC structure for decoding requested files, a novel data delivery algorithm is introduced, completely eliminating the need for SIC. Consequently, optimized linear precoders are devised to maximize the symmetric rate performance at the finite-SNR regime. In particular, it is shown that for highly non-uniform user-to-profile associations, even though the DoF of our scheme may fall substantially and tends to $\alpha$, i.e., the DoF of the unicasting scheme, it still provides a noticeable performance improvement over unicasting at finite SNR. Indeed, by employing our proposed scheme, users benefit from the CC gain and can remove part of the interference to decode their desired files by using their cache content. This cache-aided interference removal frees up the spatial multiplexing gain and enables a larger beamforming gain, resulting in improving the rate in power-limited finite-SNR communications.  
    
 \textit{Notation:} 
 $\left[ a:b \right]$ shows the set $\lbrace a,\cdots,b \rbrace$, $[a]=\lbrace 1,\cdots,a \rbrace$, $\left\vert \CA \right\vert$ is the cardinality of the set $\CA$, and for $\Lambda \subseteq \CA$, $\CA_{\backslash \Lambda}$ represents $\lbrace x \in \CA : x \notin \Lambda \rbrace$.  $\CA(i)$ is the $i$-the element of the set $\CA$. $\left( \CA \Vert \CA \right)_{n}$ denotes $n$ concatenations of $\CA$ with itself,\footnote{We use generalized multiset definition, where we allow the same elements to be repeated in the sets, e.g., $\left\lbrace a, a \right\rbrace$ cannot be reduced to $\left\lbrace a \right\rbrace$.} and $\CA \Vert \CB$ is the concatenation of $\CA$ and $\CB$. Throughout the paper, bold lower-case letters show vectors. $\Bv^{\rm H}$ is the conjugate-transpose (Hermitian) of the vector $\Bv$. The letter $\SfA$ represents the triple $(r,c,l)$ for some $r,c,l \in \mathbb{N}$, and $\SfB$ expresses the quintuple $(r,c,l,m,s)$ for some $r,c,l,m,s \in \mathbb{N}$.
 $f^{*}$ represents the phantom (non-existent) users.  $\%$ sign demonstrates the mod operator, for which $c \% c= c$ and $\left(d+c \right) \% c =d\%c$. Furthermore, for integers $a$ and $b$, $\binom{a}{b} = \frac{a!}{b! (b-a)!}$.  $\left\lVert \cdot \right\rVert$ denotes the norm operator. Finally, $\lfloor y \rfloor$ and $\lceil y \rceil$ show the floor and ceiling of $y \in \mathbb{R}$. 

\section{System Model}
\label{section: system model}
\begin{figure}[ht]
    \centering
    \includegraphics[scale=0.3]{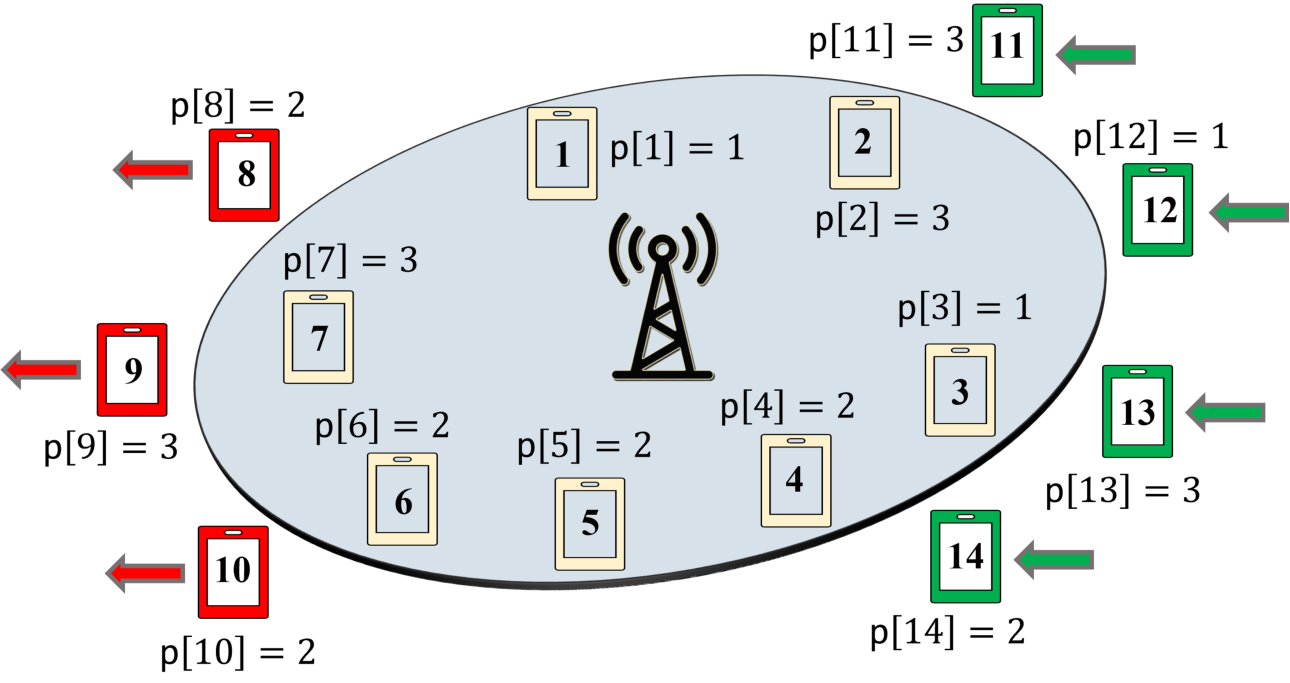}
    \vspace{-.5em}
    \caption{A dynamic cache-aided network with $P=3$  profiles, where users $\left\lbrace 1,2,\cdots, 7 \right\rbrace$ are present and do not depart the network, users $\left\lbrace 8,9,10 \right\rbrace$ and $\left\lbrace 11,12,13,14 \right\rbrace$ are departing and entering the network, respectively.}
    \label{fig: system}
\end{figure}

In this paper, we focus on a dynamic MISO network, where a base station (BS) equipped with $L$ transmit antennas and the spatial multiplexing gain of $\alpha \leq L$ serves several cache-enabled single-antenna users. The BS has access to a library $\CF$ with $N$ unit-sized files, and each user is equipped with a large enough memory to store a portion $0< \gamma= \frac{M}{N} < 1$ of the entire library, where $M$ is the number of files that can be stored in cache memory of each user. Accordingly, for a given value of cache ratio $\gamma$, we design the natural numbers $P$ and $\Brt$, such that $\Brt=P \gamma$. In this dynamic setup, as illustrated in Fig.~\ref{fig: system}, users can move, enter and depart the network at any time. Thus, the BS does not have any prior knowledge about the number of available users during the transmission. When a user $k$ enters the network, it is assigned to a profile represented by $\Sfp [k] \in \left[ P \right]$, and its cache content is updated based on a \textit{content placement algorithm}.

In the placement phase, by following the same way as in~\cite{maddah2014fundamental}, each file  $W^{n} \in \CF$, $n \in [N]$,  is split into $\binom{P}{\Brt}$ equal-sized mini-files $W^{n}_{\CP}$ such that $W^{n} \rightarrow \left\lbrace W^{n}_{\CP}: \CP \subseteq [P] ,  \left\vert \CP \right\vert=\Brt  \right\rbrace$. The cache content associated with profile $p \in [P]$, represented by $\CZ_{p}$, includes a portion $\gamma$  of each file $W^{n}$ as 
\begin{equation*}
    \CZ_{p}=\left\lbrace W^{n}_{\CP}: \CP \ni p, \CP \subseteq \left[ P \right], \left\vert \CP \right\vert = \Brt, \forall n \in \left[ N \right] \right\rbrace.
\end{equation*}
Then, defining $\CU_{p}$ as the set of users assigned to profile $p$, i.e., $\CU_{p}=\left\lbrace k: \Sfp \left[k \right]=p \right\rbrace$, each user $k \in \CU_{p}$ stores the cache content $\CZ_{p}$ during the  placement phase.

The dynamic nature of the network causes a fluctuating user population throughout time.  During regular intervals, the network's demanding users reveal their required files from the library $\CF$ to the BS. Then, using a \textit{content delivery algorithm}, the BS constructs and transmits a set of codewords, enabling $K$ present users to retrieve their requested files. Accordingly, during each transmission, the BS broadcasts a transmission vector $\Bx \in \mathbb{C}^{L \times 1}$ containing a superposition of the precoded codewords such that each target user $k$ receives the signal
\begin{equation}
\label{eq: Yk system}
y_{k}=\Bh_{k}^{\rm H} \Bx + n_{k},
\end{equation}
where $\Bh_{k} \in \mathbb{C}^{L\times 1}$ is the channel gain of user $k$ and $n_{k}$ is the zero-mean additive white Gaussian noise with variance $N_{0}$. 

\section{Resource Allocation and Data Delivery}
\label{section: data delivery}
In this section, we discuss the resource allocation and transmission strategies during the content delivery phase. This phase commences once the active users reveal their requested files and comprises two consecutive steps: \textit{1) Coded caching (CC) data delivery}; and \textit{2) Unicast (UC) data delivery}.

Let us define the number of users assigned to profile $p$ as the length of profile $p$ denoted by $\eta_{p}$, where without loss of generality, it is assumed that $\eta_{1} \geq \eta_{2} \geq  \cdots \geq \eta_{P}$. By choosing a \textit{delivery parameter}  $\hat{\eta}\leq \max_{p}\eta_{p}$,\footnote{Here, the delivery parameter plays a similar role as the unifying length parameter in \cite{Abolpour2022CodedNetworks}. Both delivery and unifying length parameters tune the DoF loss caused by the non-uniformness in the user-to-profile association.} the BS builds and transmits a set of codewords to serve at most $\hat\eta$ users assigned to each profile with a novel CC-based approach. In this regard, for every profile $p$
\\$\bullet$ if $\hat{\eta} < \eta_{p}$, we exclude $\eta_{p}-\hat{\eta}$ users, and exempt BS to serve these users during the CC delivery step. Accordingly, the excluded users are served in the UC delivery step.  
 \\$\bullet$ if $\hat{\eta} \geq \eta_{p}$, all $\eta_{p}$ users are served via the CC delivery step. 
 \begin{figure}[t]
      \centering
      \includegraphics[width=\columnwidth]{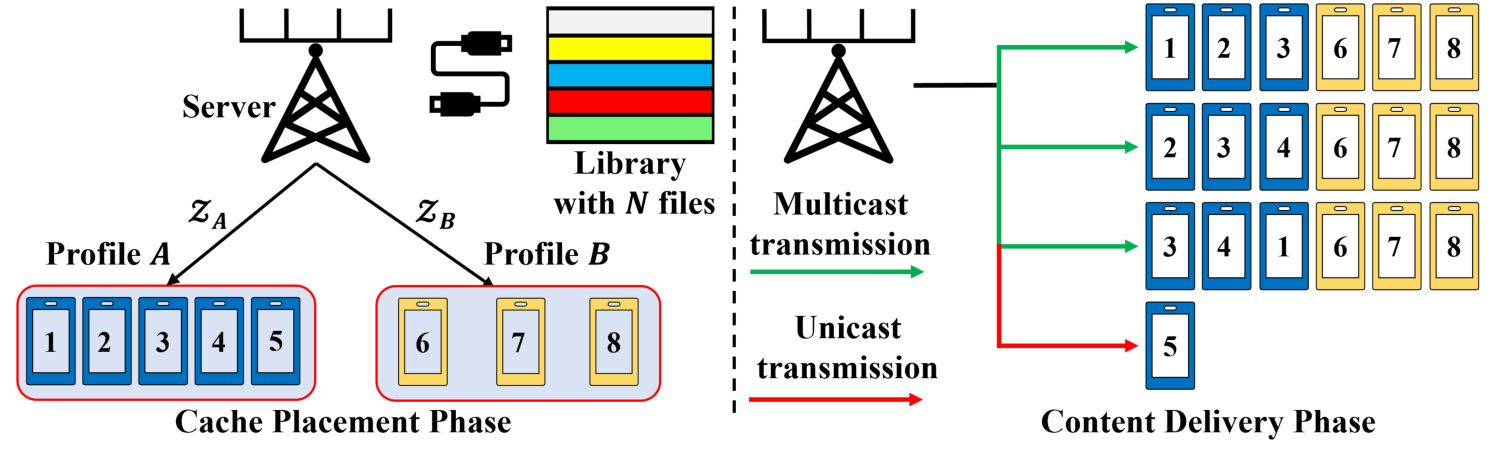}
      \vspace{-1em}
      \caption{System model for a dynamic coded caching setup, where $P=2$, $\gamma=\frac{1}{2}$, $\Brt=1$, $\alpha=4$, $\hat{\eta}=4$, $Q=2$ and $\beta=3$. During the placement phase, each user assigned to profiles $A$ and $B$ stores the cache content associated with those profiles. For the delivery phase, user $5$ is served via unicasting and other users are served via $3$ multicast transmissions.}
      \label{fig:system}
      \vspace{-1.5em}
  \end{figure}
 
 Now, let us suppose that the set of users assigned to profile $p \in [P]$ and served during the CC delivery step is denoted by $\CV_{p}$ such that $\left\vert \CV_{p} \right\vert=\delta_{p}$, $\CV_{p}=\left\lbrace v_{p,1},v_{p,2}\cdots, v_{p,\delta_{p}} \right\rbrace$, and $v_{p,i} \in \CU_{p}$ for $i \in [\delta_{p}]$. We note that $\delta_{p}=\min \left( \hat{\eta},\eta_{p} \right)$, and clearly, $\delta_{1}=\hat{\eta}$ and $\delta_{1} \geq \delta_{2} \geq \cdots \geq \delta_{P}$.
 
In order to build the transmission vectors, as depicted in Fig.~\ref{fig:system}, the BS selects a parameter $Q$, which represents the number of profiles served in each transmission, and a parameter $\beta$, which expresses the number of users chosen from each profile to serve in each transmission. The necessary conditions for choosing any arbitrary values for $Q$ and $\beta$ are defined in Remark~\ref{rem: conditions}.
\begin{rem}
\label{rem: conditions}
In order to serve $Q$ profiles each with a maximum of $\beta$ users, the network parameters should satisfy the constraints $\Brt+1 \leq Q \leq \Brt+ \left\lceil \nicefrac{\alpha}{\beta} \right\rceil$ and $\beta \leq \min \left( \alpha, \hat{\eta} \right)$.
\end{rem}
\begin{IEEEproof}
    The proof is relegated to Appendix~\ref{apx: proof conditions}.
\end{IEEEproof}
\begin{table}[h]
\Large
\centering
\caption{'System' and 'Design' Parameters}
\label{table: Parameters}
\resizebox{\columnwidth}{!}{
\begin{tabular}{|c|c|c|c|}
\hline
\huge \multirow{2}{*}{Parameter} & \huge \multirow{2}{*}{Description} &  \huge \multirow{2}{*}{Parameter}  & \huge \multirow{2}{*}{Description} \\  
 &&& \\
\hline
\huge \multirow{2}{*}{$K$ ("S")} & \huge  \multirow{2}{*}{Number of active users} &   \huge \multirow{2}{*}{$L$ ("S")} &  \huge \multirow{2}{*}{number of transmit antennas} \\
 &&& \\
 \hline
\huge \multirow{2}{*}{$M$ ("S")} & \huge \multirow{2}{*}{\begin{tabular}{@{}c@{}} Number of files that can be stored  \\ in the cache memory of users\end{tabular}} & \huge \multirow{2}{*}{$N$ ("S")} & \huge \multirow{2}{*}{\begin{tabular}{@{}c@{}} Total number of files available \\ in the data library\end{tabular}}  \\ 
 &&& \\
 &&& \\
  \hline
\huge \multirow{2}{*}{$\gamma$ ("S")} & \huge \multirow{2}{*}{Cache ratio of users}  & \huge \multirow{2}{*}{$P$ ("D")} & \huge \multirow{2}{*}{Number of caching profiles}   \\
 &&& \\
  \hline
 \huge \multirow{2}{*}{$\eta_{p}$ ("D")} &  \huge \multirow{2}{*}{\begin{tabular}{@{}c@{}} Number of users assigned to \\ profile $p$ \end{tabular}} &  \huge \multirow{2}{*}{$\delta_{p}$ ("D")} & \huge \multirow{2}{*}{\begin{tabular}{@{}c@{}} Number of users  assigned to  profile \\ $p$ and served in CC delivery step \end{tabular}}   \\
  &&&\\
  &&& \\
   \hline
   \huge \multirow{2}{*}{$\hat{\eta}$ ("D")} &  \huge \multirow{2}{*}{Delivery parameter} &  \huge \multirow{2}{*}{$\beta$ ("D")} & \huge \multirow{2}{*}{\begin{tabular}{@{}c@{}} Maximum number of users served \\ from each profile per transmission \end{tabular}}\\
   &&& \\
    &&& \\
     \hline
    \huge \multirow{2}{*}{$\alpha$ ("D")} &  \huge \multirow{2}{*}{Spatial multiplexing gain} & \huge \multirow{2}{*}{$Q$ ("D")}  & \huge \multirow{2}{*}{\begin{tabular}{@{}c@{}} Number of selected caching profiles in \\ each transmission\end{tabular}}\\
     &&& \\
      &&& \\
\hline
\end{tabular}}
\end{table}
It is worth mentioning that the key parameters of this work are classified into two categories: 'system' and 'design' parameters. The 'system' parameters rely on the physical characteristics and structure of the network. Moreover, the 'design' parameters are used to construct the transmission strategies and manage the resource allocation. Table~\ref{table: Parameters} illustrates the 'system' and 'design' parameters, while the letters "S" and "D" represent the categories 'system' and 'design', respectively. We note that the 'system' and 'design' parameters are not completely independent, and some correlation exists between them. For instance, $\alpha$ is limited by $L$. 

In the following, we present the system operation maximizing the DoF performance separately for two regimes: \textit{1) $\alpha \leq  \hat{\eta}$} and \textit{2) $\alpha > \hat{\eta}$}. For the CC delivery step, each of these cases operates either with \textit{Strategy A} (cf. Section~\ref{subsection: Strategy A}) or \textit{Strategy~B} (cf. Section~\ref{subsection: strategy B}). The chosen transmission strategy only depends on the parameters $\alpha$ and $\hat{\eta}$, and it is independent of the content placement phase. In other words, the server performs the placement phase only based on parameter $\gamma$ without considering which transmission strategy the system will use for the CC delivery step.

\subsubsection{System operation for $\alpha \leq \hat{\eta}$}
\label{subsection: alpha l eta}
For the case of $\alpha \leq \hat{\eta}$, we set $\beta=\alpha$, and the only option for $Q$ to maximize the DoF is $Q=\Brt+1$. Here, \textit{Strategy A} is utilized to build the transmission vectors. Replacing the constraint $\alpha \leq \hat{\eta}$ with $\alpha \leq  \min_{p} \delta_{p}$ reduces our system model to \cite{parrinello2019fundamental}, while our proposed scheme also works for the scenarios with $\min_{p} \delta_{p} < \alpha \le \hat{\eta}$. 
\subsubsection{System operation for $\alpha > \hat{\eta}$}
\label{subsection: alpha g eta}
For this case, we set $\beta=\hat{\eta}$, and define $\hat{\alpha}=\frac{\alpha}{\hat{\eta}}$. If $\hat{\alpha}$ is an integer, then we follow \textit{Strategy~A} to build the transmission vectors. For non-integer $\frac{\alpha}{\hat{\eta}}$, the server can serve users via \textit{Strategy~A} by setting $\Brt+1 \leq Q \leq \Brt + \left\lfloor  \nicefrac{\alpha}{\hat{\eta}} \right\rfloor$, and via \textit{Strategy B} by choosing $Q=\Brt +\left\lceil \nicefrac{\alpha}{\hat{\eta}} \right\rceil$. In this case, if we assume that users are uniformly distributed among caching profiles, and  $\frac{\alpha}{\hat{\eta}}$ is an integer, the system performance is simplified to \cite{parrinello2020extending}, while our proposed scheme also covers the uneven user-to-profile associations with non-integer $\frac{\alpha}{\hat{\eta}}$. 
\subsection{Transmission Strategy A}
\label{subsection: Strategy A}
In this strategy, each mini-file $W_{\CP}^{n}$  is split into $S_{\rm A}=\beta \binom{P-\Brt -1}{Q-\Brt-1}$ subpackets $ W_{\CP ,q }^{n}$,  where $q \in [S_{\rm A}]$ increases sequentially after each transmission to ensure none of the subpackets is transmitted twice. Next, we follow the so-called \textit{elevation process} to serve $Q$ caching profiles each with at most $\beta$ users. 

\begin{figure*}[htb]
\centering
\subfloat[Creating  $\CV_{p}$]{\includegraphics[scale=0.33]{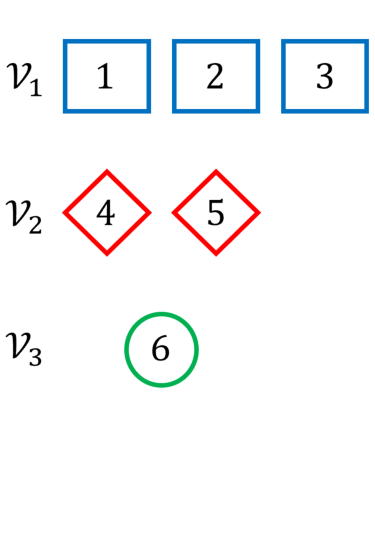} \label{fig: e1}} \hspace{1.2em}
\subfloat[Creating $\CR_{p}$]{\includegraphics[scale=0.33]{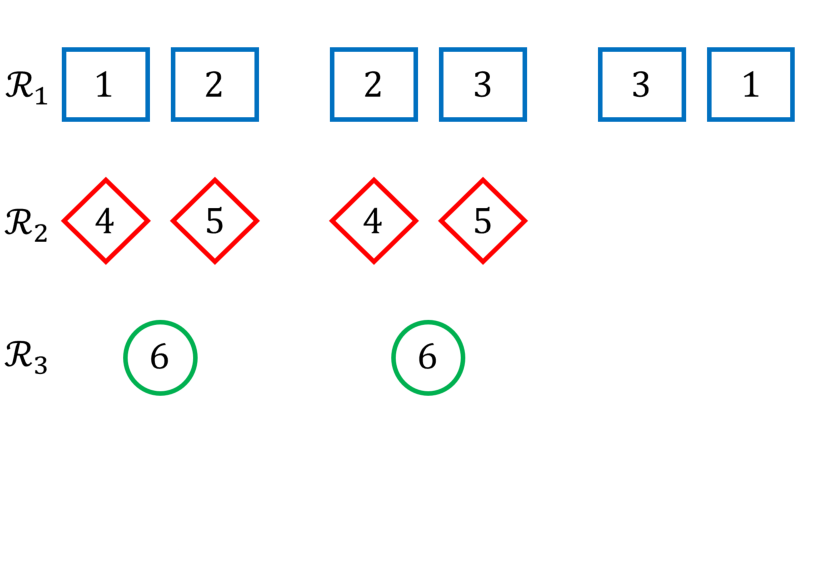} \label{fig: e2}} \hspace{1.2em}
\subfloat[Creating $\CS_{p}$]{\includegraphics[scale=0.33]{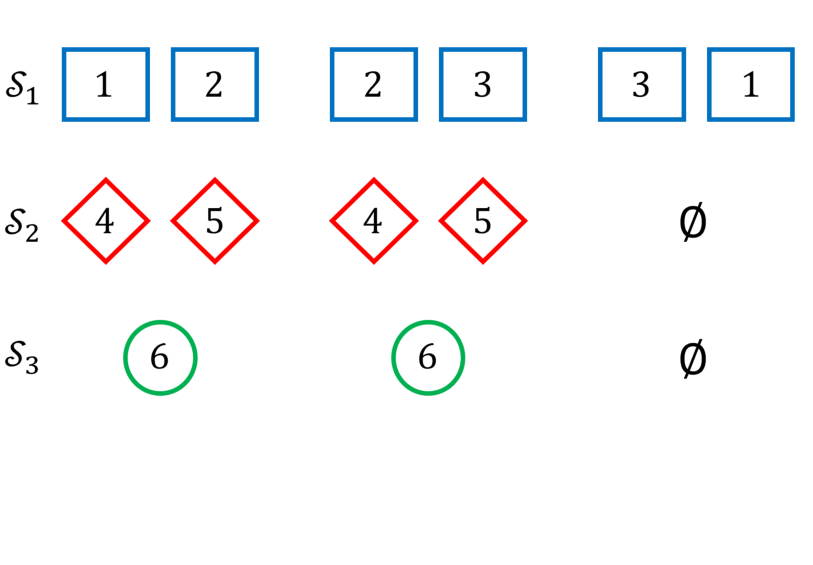}\label{fig: e3}} \hspace{1.2em}
\subfloat[Creating $\CT_{\SfA_{i}}$]{\includegraphics[scale=0.33]{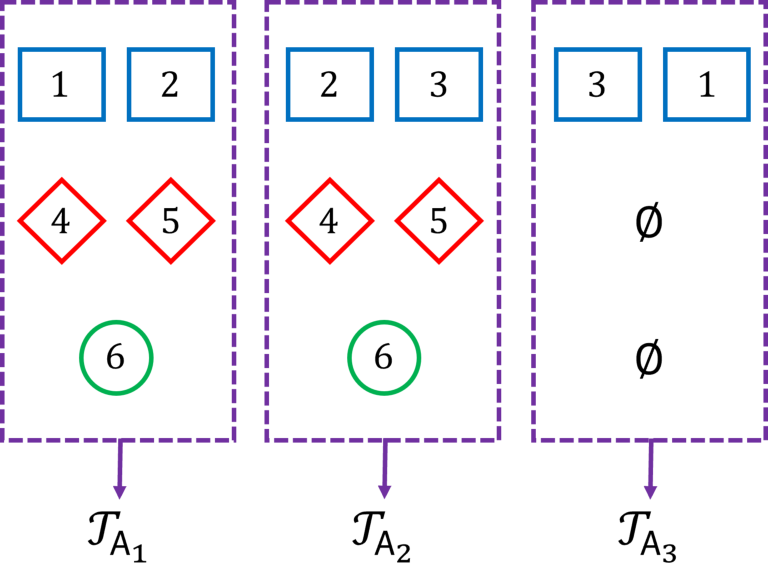}\label{fig: e4}}
\caption{The elevation process to create the set of served users during the transmission triple $\SfA_{i}$ for \emph{Strategy~A} in a network with $\gamma=\frac{2}{3}$, $\Brt =2$, $P=3$, $\alpha = 2$, $\hat{\eta} =3$, $\eta_{1} = \delta_{1} = 3$, $\eta_{2} = \delta_{2} = 2$, $\eta_{3} = \delta_{3} = 1$, $Q=3$, $\beta = 2$, $\SfA_{1} = (1,1,1)$, $\SfA_{2} = (1,2,1)$ and $\SfA_{3} = (1,3,1)$. }
 \label{fig: simple exmp}
 \vspace{-1em}
\end{figure*}
\subsubsection*{Elevation process}
    In this process, the aim is to characterize the set of users that are served in each transmission. To this end, we utilize the \emph{sliding-window approach}. Recall that the set of users assigned to profile $p$ and served during the CC delivery step is denoted by $\CV_{p}$ with $\left\vert \CV_{p} \right\vert = \delta_{p}$. This sliding-window approach extends the set $\CV_{p}$ to $\hat{\eta}$ windows, such that each of the first $\phi_{p} = \max \left( \beta, \delta_{p} \right)$ windows contains $\min (\delta_{p}, \beta)$ users from the set $\CV_{p}$ circularly shifted to the right, and the other $\hat{\eta}-\phi_{p}$ windows are the empty windows without any users. Accordingly, during each transmission, in order to serve the users assigned to profile $p$, the BS selects one of these $\hat{\eta}$ windows and serves users associated with that window. 
    Here, we use the so-called \textit{transmission triple} $\SfA = \left( r,c,l \right)$, where $r \in \left[ P-Q+1 \right]$, $c \in \left[ \phi_{r} \right]$ and $l \in [ \binom{P-r}{Q-1} ]$. In each transmission triple $\SfA =(r,c,l)$, $Q$ profiles are selected, such that the first selected profile is the caching profile $r$, and the remaining $Q-1$ profiles are chosen from the set $[r+1: P]$, and hence, there exist $\binom{P-r}{Q-1}$ possibilities to choose these $Q-1$ profiles. The $l$-th possibility (out of  $\binom{P-r}{Q-1}$ possibilities) is represented by index $l$ in the transmission triple $\SfA =(r,c,l)$. Generally speaking, the indices $r$ and $l$ determine the $Q$ caching profiles that are selected in the transmission triple $\SfA =(r,c,l)$.
    The elevation process creates $\CT_{\SfA}$, the set of users that are served during the transmission triple $\SfA$, which is characterized by concatenating the $c$-th window of each of the $Q$ selected profiles. As mentioned earlier, it is assumed that $\eta_{1} \geq \eta_{2} \geq \cdots \geq \eta_{P}$, which results in $\phi_{1}\geq \phi_{2}\geq \cdots \geq \phi_{P}$. In other words, if the $c$-th window of profile $p$ is empty, then the $c$-th window of profile $p+1$ is also empty. Hence, for fixed values of $r$ and $l$, the total number of transmission tiples serving $Q$ selected profiles is equal to $\phi_{r}$, as each of these $Q$ profiles has only $\phi_{r}$ non-empty windows. Following the elevation process, the BS constructs the transmission vector $\Bx_{\SfA}$ to serve users in the set $\CT_{\SfA}$ during the transmission triple $\SfA$. In the proceeding, we present the mathematical representation of the elevation process.  
In this regard, first, for every $p \in [P]$, we elevate the set $\CV_{p}$ to the set $\CR_{p}$ as follows.
\begin{equation}
\label{eq: Rp final}
    \CR_{p}=\CR_{p,1} \Vert \cdots \Vert \CR_{p,\phi_{p}},
\end{equation}
where, for $j \in \left[  \phi_{p} \right]$, $\CR_{p,j}$ is defined as:
\begin{equation*}
\begin{array}{c}
\label{eq: Rpj}
\CR_{p,j}=
    \begin{cases}
      \CV_{p} & \delta_{p} \leq  {\beta} \\
       \left\lbrace v_{p,l}: l=\left( i+j -1 \right) \% \delta_{p}, 1\leq i \leq \beta \right\rbrace & \delta_{p} > \beta
    \end{cases}.
    \end{array}
\end{equation*}
 In~\eqref{eq: Rp final}, $\CR_{p}$ is comprised of $\phi_{p}$ sets (windows) $R_{p,j}$ with $ j \in [\phi_{p}]$, such that if $\lvert \CV_{p} \rvert =\delta_{p} \leq \beta$, then each set $\CR_{p,j} = \CV_{p}$. Otherwise, $\CR_{p,j}$ is obtained by making a circular shift on $\CV_{p}$ and picking $\beta$ elements from that.    
Now, for $p \in [P]$,  we define  $\CS_{p}=\CS_{p,1} \Vert \cdots \Vert \CS_{p, \hat{\eta}}$, while
  \begin{equation}
  \label{eq: Spj}
  \CS_{p,j}=
      \begin{cases}
        \CR_{p,j} & 1\leq j \leq \phi_{p}\\
       \varnothing &\phi_{p}+1 \leq j \leq \hat{\eta}
      \end{cases}.
  \end{equation}
   As per~\eqref{eq: Spj}, for each $p \in [P]$, $\CS_{p}$ consists of $\hat{\eta}$ sets (windows) $\CS_{p,j}$, such that for the first $\phi_{p}$ windows, i.e., $j \leq [\phi_{p}]$, we have $\CS_{p,j} = \CR_{p,j}$, while the remaining $\hat{\eta} -\phi_{r}$ windows are the empty windows. Indeed, \eqref{eq: Spj} equalizes the number of available windows associated with each profile to $\hat{\eta}$. Then, for each $r \in [P-Q+1]$, we define the set
  $\CM_{r}$ as:
  \begin{equation}
  \label{eq: Mfinal}
      \begin{array}{c}
            \CM_{r}=\left\lbrace \CD : \CD \subseteq \left[  r+1: P \right], \,  \left\vert \CD \right\vert =Q-1  \right\rbrace.
      \end{array}
  \end{equation}
 In the proceeding, we use $\CM_{r} \left( l \right)$ to indicate the $l$-th $\left( Q-1 \right)$-tuple of $\CM_{r}$. Finally,  the set $\CT_{\SfA}$ for the transmission triple $\SfA = \left( r,c,l \right)$  is given by: 
  \begin{equation}
  \label{eq: Tfinal}
      \begin{aligned}
               & \CT_{\SfA} \!= \!
            \left\lbrace \CS_{r,c} \Vert \CS_{b_{1},c}  \Vert \! \cdots \! \Vert \CS_{b_{Q-1},c} \! : \! b_{i} \! \in \! \CM_{r} \left( l \right), \forall i  \!\in \! [Q \!- \! 1]   \right\rbrace. 
      \end{aligned}
  \end{equation}
  In Fig.~\ref{fig: simple exmp}, an illustrative example for the elevation process to create the set of served users for a network with $K=6$, $\gamma=\frac{2}{3}$, $\Brt = 2$, $P = 3$, $\CU_{1} = \CV_{1} = \lbrace 1,2,3 \rbrace $, $\CU_{2} = \CV_{2} = \lbrace 4,5 \rbrace $, and $\CU_{3} = \CV_{3} = \lbrace 6 \rbrace $ is presented. In this network, it is assumed that $\alpha =2$, $\hat{\eta} = 3$, and therefore, we have $\eta_{1} = \delta_{1} =3$, $\eta_{2} = \delta_{2} =2$, $\eta_{3} = \delta_{3} =1$, $\beta = \min (\hat{\eta}, \alpha) = 2$, $\phi_{1} = \max (\beta, \delta_{1}) = 3 $, $\phi_{2} = \max (\beta, \delta_{2}) = 2 $, $\phi_{3} = \max (\beta, \delta_{3}) = 2 $ and $Q = \Brt + 1 = 3$. In this setup, $r =1$, $c \in [3]$ and $l =1$, hence, there exist $3$ transmission triples $\SfA_{1} = (1,1,1)$, $\SfA_{2} = (1,2,1)$ and $\SfA_{3} = (1,3,1)$, for which the BS creates the set of served users $\CT_{\SfA_{1}}$, $\CT_{\SfA_{2}}$ and $\CT_{\SfA_{3}}$, respectively.

 Generally speaking, for the transmission triple $ \SfA = \left( r,c,l \right)$, if $\delta_{r}=0$, the BS does not transmit any signal; otherwise, the BS serves users assigned to  $\CT_{\SfA}=\CS_{r,c} \Vert \CS_{b_{1},c}\Vert \cdots \Vert \CS_{b_{Q-1},c}$, which represents the concatenation of the $c$-th window of profile $r$ and the profiles in the set $ \CM_{r} \left( l \right) =  \lbrace b_{1},\cdots , b_{Q-1} \rbrace$.  During the transmission triple $ \SfA  =  \left( r, c,l \right)$, by defining $\CN=\left\lbrace r \right\rbrace \cup \CM_{r} \left( l \right)$, the BS constructs the transmission vector as follows. 
  \begin{equation} 
  \label{eq: x A}
      \begin{array}{c}
           \Bx_{\SfA}  = \sum\limits_{\Lambda \subseteq \CN: \left\vert \Lambda \right\vert=\Brt} \, \,  \sum\limits_{k \in \CS_{p,c}: p \in \CN _{\backslash \Lambda}}  \Bw_{k}^{\Lambda} W_{\Lambda,q}^{k},
      \end{array}
  \end{equation}
  where $\Bw_{k}^{\Lambda} \in \mathbb{C}^{L\times 1}$ is the precoder that suppresses the interference of user $k$ at the set $\CG_{k}^{\Lambda}=\left\lbrace j \in \CS_{p,c}: \forall p \in \CN_{\backslash \Lambda}, j\neq k \right\rbrace$. In Appendix~\ref{apx: proof DoF}, it is proven that all users served with \textit{Strategy~A} can decode their requested files at the end of the CC delivery step. Accordingly, at the end of the transmission triple $ \SfA $, user $k$ receives a signal as described in~\eqref{eq: Yk system}. The following example shows the data delivery via \textit{Strategy~A}.

  \begin{exmp}
  \label{exmp: Strategy A}
For a cache-aided dynamic setup with $P=3$, $\gamma=\frac{1}{3}$, $\Brt=1$, $\alpha=6$, $\beta=3$ and $Q=\Brt+ \nicefrac{\alpha}{\beta} = 3$, let us assume that $\CU_{1}=\left\lbrace 1,2,3,4,5 \right\rbrace$, $\CU_{2}=\left\lbrace 6,7,8,9 \right\rbrace$ and $\CU_{3}=\left\lbrace 10,11,12 \right\rbrace$ are the set of users assigned to profiles $1,2$ and $3$, respectively. 
During the placement phase, each file $W^{n}$ is split into $\binom{P}{\Brt}=3$ mini-files $W_{\CP}^{n}$, where $\CP \in \left\lbrace 1,2,3 \right\rbrace$. For $n \in \left[ N \right]$, users assigned to profiles $1,2$ and $3$ store the mini-files $W_{1}^{n}$, $W_{2}^{n}$ and $W_{3}^{n}$, respectively.  For the delivery phase, user $k \in [K]$ requests the file $W^{k}$, and then, each mini-file $W_{\CP}^{n}$ is further split into $\beta \binom{P-\Brt-1}{Q-\Brt-1}=3$ subpackets $W_{\CP,q}^{n}$, where $q \in \left[ 3 \right]$. 

Assuming, $\hat{\eta}\!= \! 4$, we have $\CV_{1} \!= \! \left\lbrace 1,2,3,4 \right\rbrace$, $\CV_{2}=\left\lbrace 6,7,8,9 \right\rbrace$ and $\CV_{3}=\left\lbrace 10,11,12 \right\rbrace$, while user $5$ will be served during the UC delivery step.\footnote{Although  it is mentioned in Section~\ref{section: data delivery} to set $\beta=\hat{\eta}$ to maximize the achievable DoF for the case of $\alpha>\hat{\eta}$, here, we set $\beta<\hat{\eta}$ to give further insight on the system operation with \textit{Strategy A}.} Then, we obtain $\CR_{1}-\CR_{3}$ as follows. 
\begin{equation*}
\begin{array}{lll}
    &\CR_{1}=\left(  \left\lbrace 1,2,3 \right\rbrace  \Vert \left\lbrace 2,3,4 \right\rbrace  \Vert  \left\lbrace 3,4,1 \right\rbrace \Vert  \left\lbrace 4,1,2 \right\rbrace    \right),\\
     &\CR_{2}=\left(  \left\lbrace 6,7,8 \right\rbrace\Vert  \left\lbrace 7,8,9 \right\rbrace\Vert   \left\lbrace 8,9,6 \right\rbrace\Vert   \left\lbrace 9,6,7 \right\rbrace  \right), \\
     & \CR_{3}=\left( \left\lbrace 10,11,12 \right\rbrace \Vert  \left\lbrace 10,11,12 \right\rbrace\Vert  \left\lbrace 10,11,12 \right\rbrace \right),\\
    \end{array}
    \end{equation*}
where $\CR_{1,1}=\left\lbrace 1,2,3 \right\rbrace$ and so on. Accordingly, using $\CR_{1}-\CR_{3}$, $\CS_{1}-\CS_{3}$ are given by:
 \begin{equation*}
     \begin{array}{lll}
           &\CS_{1}=\left(  \left\lbrace 1,2,3 \right\rbrace  \Vert \left\lbrace 2,3,4 \right\rbrace  \Vert  \left\lbrace 3,4,1 \right\rbrace \Vert  \left\lbrace 4,1,2 \right\rbrace    \right),\\
           &\CS_{2}=\left(  \left\lbrace 6,7,8 \right\rbrace\Vert  \left\lbrace 7,8,9 \right\rbrace\Vert   \left\lbrace 8,9,6 \right\rbrace\Vert   \left\lbrace 9,6,7 \right\rbrace  \right), \\
           & \CS_{3}=\left( \left\lbrace 10,11,12 \right\rbrace \Vert  \left\lbrace 10,11,12 \right\rbrace\Vert  \left\lbrace 10,11,12 \right\rbrace \Vert \varnothing \right).
     \end{array}
 \end{equation*}
In order to build the transmission vector for the transmission triple $ \SfA_{\circ} = \left( 1,1,1 \right)$, first, we obtain $\CM_{1}=\CM_{1}(1)= \left\lbrace 2,3 \right\rbrace$. Therefore, the set of users that are served during the transmission triple $\SfA_{\circ} =  (1,1,1)$ is equal to:
\begin{equation*}
          \begin{aligned}
               \CT_{\SfA_{\circ}} &=
                \left( \CS_{1,1} \Vert \CS_{2,1} \Vert \CS_{3,1}\right)
               = \left\lbrace 1,2,3,6,7,8,10,11,12 \right\rbrace.
          \end{aligned}
      \end{equation*}
      Finally, the transmission vector corresponding to the transmission triple $ \SfA_{\circ} =  \left( 1,1,1 \right)$ is constructed as follows. 
  \begin{equation}
  \label{eq: x1 A}
      \begin{aligned}
        \Bx_{\SfA_{\circ}} = & \sum\limits_{\Lambda \subseteq \left\lbrace 1,2,3 \right\rbrace: \left\vert \Lambda \right\vert=1} \, \, \sum\limits_{k \in \CS_{p,c}: p \in \left\lbrace 1,2,3 \right\rbrace_{\backslash \Lambda}} \Bw_{k}^{\Lambda}  W_{\Lambda,1}^{k}  \\
           =  &\sum\nolimits_{k \in \CS_{2,1} \Vert \CS_{3,1}} \Bw_{k}^{1} W_{1,1}^{k}  +
            \sum\nolimits_{k \in \CS_{1,1} \Vert \CS_{3,1}} \Bw_{k}^{2} W_{2,1}^{k}  \\
          + & \sum\nolimits_{k \in \CS_{1,1} \Vert \CS_{2,1}} \Bw_{k}^{3} W_{3,1}^{k}.
      \end{aligned}
  \end{equation}
 Now, in order to show the decoding process, for any $k,j \in \CS_{p,c}$ with $p \in \CN_{\backslash \Lambda}$, if $j\neq k$, using ZF precoders, we have $\Bh_{k}^{\rm H} \Bw_{j}^{\Lambda}=0$; otherwise ($j=k$), $\Bh_{k}^{\rm H} \Bw_{k}^{\Lambda} \neq 0$. For example, $\Bh_{1}^{\rm H} \Bw_{1}^{3} \neq 0$ and $\Bh_{1}^{\rm H} \Bw_{2}^{3}=0$. Now, the received  signal at user~$1$ is given by:
  \begin{equation*}
  \label{eq: Y1 A exmp}
      \begin{array}{l}
            y_{1} \! = \! \sum\limits_{k \in \CS_{2,1} \Vert \CS_{3,1}} \! \Bh_{1}^{\rm H}  \Bw_{k}^{1}W_{1,1}^{k} \! + \! \Bh_{1}^{\rm H} \Bw_{1}^{2} W_{2,1}^{1} \! + \! \Bh_{1}^{\rm H} \Bw_{1}^{3}W_{3,1}^{1}  +n_{1}.
      \end{array}
  \end{equation*}
  User $1$ has all subpackets $W_{1,1}^{n}$, $\forall n \in [N]$ in its cache memory. So, it can regenerate $\sum\nolimits_{k \in \CS_{2,1} \Vert \CS_{3,1}} \Bh_{1}^{\rm H}  \Bw_{k}^{1}W_{1,1}^{k}$\footnote{Similarly to \cite{salehi2020lowcomplexity}, for each $k \in \CS_{2,1} \Vert \CS_{3,1}$, the term $\Bh_{1}^{\rm H} \Bw_{k}^{1}W_{1,1}^{k}$ is estimated by using downlink precoded pilots, leading user $1$ to regenerate $\sum_{k \in \CS_{2,1} \Vert \CS_{3,1}} \Bh_{1}^{\rm H}  \Bw_{k}^{1}W_{1,1}^{k}$.} and subtract it from $y_{1}$, which results in observing the superposition signal $\Tilde{y}_{1}=\Bh_{1}^{\rm H} \Bw_{1}^{2} W_{2,1}^{1}+\Bh_{1}^{\rm H} \Bw_{1}^{3}W_{3,1}^{1} +n_{1}$. Here, we note that user~$1$ needs to implement SIC to decode the subpackets~\cite{tolli2017multi},~\cite{parrinello2019fundamental}. 
  \end{exmp}

\subsection{Transmission Strategy B}
\label{subsection: strategy B}
When $\alpha > \hat{\eta}$ and $\frac{\alpha}{\hat{\eta}}$ is not an integer,  we can serve users by setting $\beta=\hat{\eta}$, and $Q=\Brt + \left\lceil \nicefrac{\alpha}{\hat{\eta}} \right\rceil$.\footnote{For $\alpha > \hat{\eta}$ and non-integer $\frac{\alpha}{\hat{\eta}}$, we can still serve users with \textit{Strategy A}. However, the achievable DoF is less than the one with \textit{Strategy B}.} Here, first, we split each mini-file $W_{\CP}^{n}$ into $S_{\rm B}=\left( \hat{\eta}\Brt+ \alpha \right) \binom{P-\Brt-1}{Q-\Brt-1} \binom{Q-2}{Q-\Brt-2}$ subpackets $W_{\CP,q}^{n}$, where $q \in \left[ S_{\rm B} \right]$. Then, we use the elevation process to serve $Q$ profiles each with at most $\beta$ users. 
\subsubsection*{Elevation process}
\label{subsection: alphageta nonint}
This process builds the set of users served in each transmission. Similarly to \emph{Strategy~A}, this process uses the sliding-window approach to create the set of users served in each transmission. In this regard, the set of users assigned to profile $r \in [P]$ and served in the CC delivery step, described as $\CV_{r}$, is extended to the set $\CY_{r}$, such that $\lvert \CY_{r} \rvert = \hat{\eta}$. To this end, if $\lvert  \CV_{r} \rvert = \delta_{r} < \hat{\eta} $, then, $\hat{\eta} - \delta_{r}$ phantom (non-existing) users are added to the set $\CV_{r}$; otherwise, $\CY_{r} = \CV_{r}$. In other words,    for $r \! \in \! \left[ P \right]$,  we define $\CY_{r}$~as: 
\begin{equation}
    \begin{aligned}
           \CY_{r}=
           \begin{cases}
           \CV_{r} & \delta_{r}=\hat{\eta}\\
           \CV_{r} \Vert \left( f^{*} \Vert f^{*} \right)_{ \hat{\eta}-\delta_{r}} & \mathrm{o.w.}
           \end{cases},
    \end{aligned}
\end{equation}
where $f^{*}$ denotes the phantom users. Generally speaking, in each transmission of \textit{Strategy B}, we serve the users assigned to $Q$  profiles such that we select at most $\hat{\eta}$ users from $Q-1 = \Brt + \lfloor \nicefrac{\alpha}{\hat{\eta}} \rfloor$ profiles, and pick $\theta=\alpha -\hat{\eta}\left\lfloor \nicefrac{\alpha}{\hat{\eta}} \right\rfloor$ users from another profile, which results in serving $\hat{\eta} \Brt + \alpha$ users per transmission. Here, we note that if $K$ users are uniformly assigned to $P$ caching profiles, i.e., $\hat{\eta}= \frac{K}{P}$, the number of served users per transmission is obtained as $\hat{\eta} \Brt + \alpha =  K\gamma+\alpha$, which coincides with the optimal DoF in MISO CC schemes as described in~\cite{shariatpanahi2016multi}.  In this process, first, we select a profile $r \in [P]$, and then, pick $\theta$ users from $\CY_{r}$. To this end,  for $r \in \left[ P \right]$ and $m \in \left[ \hat{\eta} \right]$,  we consider the set $\CE_{r}^{m}$ as:
\begin{equation}
\label{eq: erm}
    \begin{array}{c}
           \CE_{r}^{m}= \bigcup\nolimits_{i=0}^{\theta-1} \CY_{r} \left( \left( i+m \right) \% \hat{\eta} \right),
    \end{array}
\end{equation}
where $\CY_{r}(i)$ is the $i$-th element of $\CY_{r}$.  Indeed, $\CE_{r}^{m}$ shifts $\CY_{r}$ to the right for $m$ times, and picks $\theta$ elements from it. This circular shift on $\CY_{r}$ to create $\CE_{r}^{m}$ is used to make the user selection independent of the users' indices.

Moreover, we assume that $\overline{\CP}_{r}=\left[ P \right]_{\backslash r}$ for $r \in \left[ P \right]$, and $\Bar{\delta}_{c}=\overline{\CP}_{r}(c)$ is the $c$-th element of $\overline{\CP}_{r}$. The caching profiles in $\overline{\CP}_{r}$ are sorted in descending order such that if $i<j$, then $\delta_{\overline{\CP}_{r}(i)} \geq \delta_{\overline{\CP}_{r}(j)}$. Next, for a given $r$ and $c \in \left[ P-Q+1 \right]$, let $\CI_{c}^{r}$ as:
\begin{equation}
\small
\label{eq: Icr non-int}
    \begin{array}{c}
          \CI_{c}^{r} \!=
          \left\lbrace \CD: \! \CD \! \subseteq \! \left\lbrace \overline{\CP}_{r}(c\!+\!1),\cdots, \overline{\CP}_{r}(P\!-\!1) \right\rbrace, \left\lvert \CD \right\rvert\!=\! Q\!-\!2 \right\rbrace.
    \end{array}
\end{equation}
Furthermore, denote the $l$-th $\left( Q-2 \right)$-tuple of $\CI_{c}^{r}$ by  $\CI_{c}^{r}(l)$, where $l \in [ \binom{P-c-1}{Q-2} ]$.  Now, for a given $r \in [P]$, $c \in [P-Q+1]$ and $l \in [\binom{P-c+1}{Q-2}]$, the BS serves the users assigned to the set $\CE_{r}^{m} \cup \lbrace \CY_{p} : p \in \lbrace \Bar{\delta}_{c} \rbrace \cup \CI_{c}^{r} (l) \rbrace$. In order to create a transmission vector to serve these users, first,  all subsets of $\lbrace \Bar{\delta}_{c} \rbrace \cup \CI_{c}^{r} (l)  $ with the cardinality of $Q-\Brt-1 = \lfloor \nicefrac{\alpha}{\hat{\eta}} \rfloor$ is aggregated in the set $\CB$ as follows
\begin{equation*}
    \begin{array}{c}
        \CB = \lbrace  ( \CY_{b_{1}}, \cdots, \CY_{b_{\lfloor \nicefrac{\alpha}{\hat{\eta}} \rfloor}}   ) : i \in [\lfloor \nicefrac{\alpha}{\hat{\eta}} \rfloor], b_{i} \in \lbrace \Bar{\delta}_{c} \rbrace \cup \CI_{c}^{r} (l)    \rbrace,
    \end{array}
\end{equation*}
\noindent
where $\lvert \CB \rvert=\nu_{2}=\binom{Q-1}{Q-\Brt-1}$ and $\CB(n)$ represents the $n$-th $\lfloor \nicefrac{\alpha}{\hat{\eta}} \rfloor$-tuple of $\CB$. Here, we note that each $k \in \CY_{p}$ with $p \in \lbrace \Bar{\delta}_{c} \rbrace \cup \CI_{c}^{r} (l)  $ appears $\nu_{1}  = \binom{Q-2}{Q-\Brt-2} $ times in the set $\CB$. Now, each user $u \in \CE_{r}^{m}$ is assigned to $\nu_{1}$ tuples of $\CB$ to ensure that each user receives $\nu_{1}$ subpackets per transmission. To this end, for any $u \in \CE_{r}^{m}$, we define the set $\CK_{r}^{m,u}$ as follows 
\begin{equation}
    \begin{array}{c}
         \CK_{r}^{m,u}= \left( u  \Vert u \right)_{\nu_{1}} \Vert \left( f^{*}  \Vert f^{*} \right)_{\nu_{2}-\nu_{1}}.
    \end{array}
\end{equation}
As observed, $\nu_{1}$ elements of $\CK_{r}^{m,u}$ are equal to $u$, and the remaining $\nu_{2} - \nu_{1}$ elements are the phantom users $f^{*}$. Next, by considering $\CK_{r}^{m,u} (n)$ as the $n$-th element of $\CK_{r}^{m,u}$, in order to create the transmission vectors, the element $\CK_{r}^{m,u} (n)$ should be merged with $\CB(n)$ for $n \in [\nu_{2}]$ and $u \in \CE_{r}^{m}$. However, in order to make this transmission strategy independent of user indexing, we add another round of circular shift on $\CK_{r}^{m,u}$, such that for any $s \in [\nu_{2}]$, the set 
\begin{equation}
\label{eq: Krsmu}
    \begin{array}{c}
         \CK_{r,s}^{m,u}=\bigcup\nolimits_{i=1}^{\nu_{2}} \CK_{r}^{m,u} \left( \left( i+s \right) \% \nu_{2} \right),
    \end{array}
\end{equation}
represents $s$ circular shifts of $\CK_{r}^{m,u}$ to the right. Moreover, $\CK_{r}^{m,u}(n)$ shows the $n$-th element of $\CK_{r}^{m,u}$ with $n \in [\nu_{2}]$.

For the delivery process with \textit{Strategy~B}, we use the so-called \textit{transmission quintuple} $ \SfB = \left( r,c,l,m,s \right)$, where $r \in \left[ P \right]$, $c \in \left[ P-Q+1 \right]$, $l \in [ \binom{P-c-1}{Q-2} ]$, $m \in \left[ \hat{\eta} \right]$ and $s \in \left[ \nu_{2} \right]$. In each transmission quintuple $\SfB = \left( r,c,l,m,s \right)$, for any $n \in [\nu_{2}]$, we create the set $\CC(n) = \CK_{r,s}^{m,u}(n) \cup \CB(n)$, 

and define the function $I^{+} \left( \Bar{\delta}_{c}, \CE_{r}^{m} \right)$ such that $I^{+} \left( \Bar{\delta}_{c}, \CE_{r}^{m} \right)=0$, if $\CE_{r}^{m}=f^{*}$ and $\Bar{\delta}_{c}=0$; otherwise, $I^{+} \left( \Bar{\delta}_{c}, \CE_{r}^{m} \right)=1$. 
 If $I^{+} \left( \Bar{\delta}_{c}, \CE_{r}^{m} \right)=1$, after eliminating the impacts of the phantom  users $f^{*}$, the BS builds the transmission vector for the transmission quintuple $  \SfB =  \left( r,c,l,m,s \right)$ as follows. 
 \begin{equation}
\label{eq: x B}
      \begin{array}{c}
        \Bx_{\SfB} = \sum\nolimits_{n=1}^{\nu_{2}} \, \, \sum\nolimits_{k \in  \CC(n)_{\backslash \{ f^{*} \}} } \Bw_{k}^{\Lambda} W_{\Lambda,q}^{k},
      \end{array}
  \end{equation}
where $\Lambda= \lbrace p:  p \in \lbrace \Bar{\delta}_{c} \rbrace \cup \CI_{c}^{r} (l),  \CY_{p} \notin \CB(n) \rbrace$, $\left\lvert \Lambda \right\rvert = \Brt$,   
and $\Bw_{k}^{\Lambda} \in \mathbb{C}^{L\times 1}$ is the precoder that suppresses the interference of user $k$ at the set
$\CH_{k}^{\Lambda}=\lbrace  j \in \CE_{r}^{m} \cup \CV_{p}:  \CY_{p} \in \CB(n), j\neq k, f^{*} \rbrace$. 
In Appendix~\ref{apx: proof DoF}, we prove that all users can decode their requested files with \textit{Strategy~B}. Accordingly, user $k$ receives the signal shown in~\eqref{eq: Yk system} at the end of the transmission quintuple $\SfB$. The following example studies the data delivery via \textit{Strategy~B}.

\begin{exmp}
  \label{exmp: Strategy B}
With the given $\CU_{1}-\CU_{3}$ and $\CV_{1}-\CV_{3}$ in Example~\ref{exmp: Strategy A}, suppose that $P=3$, $\gamma=\frac{1}{3}$, $\Brt=1$, $\alpha=6$, $\beta=\hat{\eta}=4$, $\theta=\alpha -\hat{\eta}\left\lfloor \nicefrac{\alpha}{\hat{\eta}} \right\rfloor=2$ and $Q=\Brt+\left\lceil \nicefrac{\alpha}{\hat{\eta}} \right\rceil=3$. For the placement phase, similarly to Example~\ref{exmp: Strategy A}, each file $W^{n}$ is split into 3 mini-files, and users assigned to profiles $1,2$ and $3$, store the mini-files $W_{1}^{n}$, $W_{2}^{n}$ and $W_{3}^{n}$, respectively. Then, during the content delivery phase, user $k \in [K]$ requests the file $W^{k}$, and each mini-file is split into  $\left( \hat{\eta}\Brt+ \alpha \right) \binom{P-\Brt-1}{Q-\Brt-1} \binom{Q-2}{Q-\Brt-2}=10$ subpackets. Now, for the transmission quintuple $ 
\SfB_{\circ} =  (1,1,1,1,1)$, we have:
\begin{equation*}
    \begin{array}{l}
           \CY_{1}=\left\lbrace 1,2,3,4 \right\rbrace,   \, \, \CY_{2}=\left\lbrace 6,7,8,9 \right\rbrace, \, \,   \CY_{3}=\left\lbrace 10,11,12,f^{*} \right\rbrace.
    \end{array}
\end{equation*}
At the next step, for $r=c=1$, it is found that $\overline{\CP}_{1}=\left\lbrace 2,3 \right\rbrace$,  $\overline{\CP}_{1}(1)= 2$, and $\CI_{1}^{1}=\CI_{1}^{1}(1)=\left\lbrace 3 \right\rbrace$. During this transmission, the users of the set $\CE_{1}^{1}=\left\lbrace 1,2 \right\rbrace$, and the users assigned to the set $\CY_{2} \cup \CY_{3}$   
are served. Furthermore, $\CK_{1,1}^{1,1}=\left\lbrace 1,f^{*} \right\rbrace$ and $\CK_{1,1}^{1,2}=\left\lbrace 2,f^{*} \right\rbrace$, and, it is found that $\CB = \lbrace \CY_{2} ,\CY_{3} \rbrace$, $\CC  = \lbrace \lbrace 1,2, \CY_{2} \rbrace, \lbrace f^{*}, f^{*}, \CY_{3} \rbrace \rbrace $, such that $\CC(1) = \lbrace 1,2, \CY_{2} \rbrace$ and $\CC(2) = \lbrace f^{*}, f^{*}, \CY_{3} \rbrace$. Here, we note that we use the generalized multiset definition, where the same element can appear multiple times in the set.  
Hence, after removing the phantom users $f^{*}$, the transmission vector for the transmission quintuple $\SfB_{\circ}  =  (1,1,1,1,1)$ takes the form of:
\begin{equation*}
      \begin{array}{ll}
            \Bx_{\SfB_{\circ}} &= \sum\nolimits_{n=1}^{2} \, \sum\limits_{ k \in  \CC(n)_{\backslash \{ f^{*} \}}  } \Bw_{k}^{\Lambda} W_{\Lambda,1}^{k}\\
             &=\sum\nolimits_{k \in \left\lbrace 1,2,6,7,8,9 \right\rbrace} \Bw_{k}^{3} W_{3,1}^{k} +\sum\limits_{k \in \left\lbrace 10,11,12 \right\rbrace} \Bw_{k}^{2}  W_{2,1}^{k},
      \end{array}
  \end{equation*}
  where  $\CH_{6}^{3}=\left\lbrace 1,2,7,8,9 \right\rbrace$, $\CH_{10}^{2}=\left\lbrace 1,2,11,12 \right\rbrace$ and so on. Similarly to Example~\ref{exmp: Strategy A}, all users that are served via \textit{Strategy~B} are able to decode their requested files.
  \end{exmp}


\subsection{Unicast (UC) Data Delivery}
\label{section: UC}
In the UC delivery step, the BS transmits data to the users excluded from the CC delivery step. Here, unlike the CC delivery step that benefits from the global coded caching and spatial multiplexing gains, only local coded caching and spatial multiplexing gains are available. Suppose that the BS serves $K_{U}$ users during the UC delivery step such that $K_{U}=\sum_{p=1}^{P} \max \left( 0, \eta_{p}-\hat{\eta} \right)$. Each of the requested files by these users is split into the same number of subpackets as in the CC delivery step. Then, in order to transmit these missing subpackets, we follow a greedy algorithm similar to \cite{salehi2021low}, which comprises 3 processes:  1) sort users based on the number of their missing subpackets in descending order; 2)~create a transmission vector to deliver one missing subpacket to each of the first $\min \left( \alpha, K_{U} \right)$ users; 3) repeat processes 1 and 2 until all missing files are transmitted.

\section{SIC-Free Data Delivery}
\label{sec: sic-free}
As stated in Section~\ref{section: data delivery}, during every transmission of the CC delivery step, each user served via \textit{Strategy~A} and \textit{Strategy~B} receives $\nu_{2}=\binom{Q-1}{Q-\Brt-1}$ and $\nu_{1}=\binom{Q-2}{Q-\Brt-2}$ subpackets of its requested file, respectively. Consequently, the retrieval of subpackets may require the implementation of a SIC receiver at each user, which is complex and undesired \cite{10052083}.  
To address this issue, we propose a data delivery algorithm that eliminates the SIC requirement. 
In the following, we review this algorithm for both transmission strategies \emph{A} and \emph{B}.

\noindent \textbf{\textit{Strategy A}.} Let us denote the set of users served during the transmission triple $\SfA = \left( r,c,l \right)$, $r \in \left[ P-Q+1 \right]$, $c \in \left[ \phi_{r} \right]$, and $l \in [ \binom{P-r}{Q-1} ]$ with $\CT_{\SfA}$, and represent the set of subpackets received by user $k \in \CT_{\SfA}$ in transmission triple $\SfA$ with $\CO_{\SfA}^{k}$, i.e.,
\begin{equation*}
    \begin{array}{c}
        \CO_{\SfA}^{k}= \left\lbrace W^{k}_{\Lambda,q}: \Lambda \subseteq \CN, \left\vert \Lambda \right\vert =\Brt, \Sfp\left[ k \right] \in \CN_{\backslash \Lambda} \right\rbrace,
    \end{array}
\end{equation*}
and $\vert \CO_{\SfA}^{k} \vert=\nu_{2}$. Now, to eliminate the SIC requirement, we split each transmission triple $\SfA $ into $\nu_{2}$ transmissions, denoted by ${\SfA}^j$, $j \in [\nu_{2}]$, such that in transmission ${\SfA}^j$, the BS transmits the superposition signal 
\begin{equation}
\label{eq: x A sic-free}
    \begin{array}{c}
        \Bx_{\SfA}^{j}=\sum\nolimits_{k \in \CT_{\SfA}} \Bw_{k}^{\Lambda^{\prime}}  W^{k}_{\Lambda^{\prime},q},
    \end{array}
\end{equation}
where $ W^{k}_{\Lambda^{\prime},q}$ is the $j$-th entry of $\CO_{\SfA}^{k}$, and $\Bw_{k}^{\Lambda^{\prime}} \in \mathbb{C}^{L\times 1}$ is the beamforming vector suppressing the interference of subpacket $ W^{k}_{\Lambda^{\prime},q}$ at the set $\CG_{k}^{\Lambda^{\prime}}$, defined in~\eqref{eq: x A}. Note that, following~\eqref{eq: x A} and~\eqref{eq: x A sic-free}, it is guaranteed that $\Bx_{\SfA}=\sum_{j=1}^{\nu_{2}} \Bx_{\SfA}^{j}$. 

\noindent \textbf{\textit{Strategy B}.} Let us denote the set of users served during the transmission quintuple $ \SfB = \left( r,c,l,m,s \right)$, $r \in \left[ P \right]$, $c \in \left[ P-Q+1 \right]$, $l \in [ \binom{P-c-1}{Q-2} ]$, $m \in \left[ \hat{\eta} \right]$, $s \in \left[ \nu_{2} \right]$, with $\Tilde{\CT}_{\SfB}$, and represent the set of subpackets received by user $k \in \Tilde{\CT}_{\SfB}$ in transmission quintuple $\SfB= (r,c,l,m,s)$ with $\Tilde{\CO}_{\SfB}^{k}$, i.e., 
\begin{equation*}
    \begin{array}{l}
        \Tilde{\CO}_{\SfB}^{k} =   \left\lbrace W^{k}_{\Lambda,q}  :  \Sfp \left[ k \right]  \in   \left\lbrace r,\Bar{\delta}_{c} \right\rbrace \cup \CI^{r}_{c}(l)  \right\rbrace,
    \end{array} 
\end{equation*}
 where $\Lambda= \lbrace p:  p \in \lbrace \Bar{\delta}_{c} \rbrace \cup \CI_{c}^{r} (l), n   \in  \left[ \nu_{2} \right], \CY_{p} \notin \CB(n) \rbrace$, $\left\lvert \Lambda \right\rvert = \Brt$,  and $ \vert \Tilde{\CO}_{\SfB}^{k} \vert =\nu_{1}$. To eliminate the SIC requirement, we split
every transmission quintuple $\SfB$  into $\nu_{1}$ transmissions, denoted by $\SfB^j$, $j \in [\nu_{1}]$, such that in transmission $\SfB^j$, the BS transmits the superposition signal
\begin{equation}
\label{eq: x B sic-free}
    \begin{array}{c}
        \Bx_{\SfB}^{j}=\sum_{k \in \Tilde{\CT}_{\SfB}}  \Bw_{k}^{\tilde{\Lambda}} W^{k}_{\tilde{\Lambda},q},
    \end{array}
\end{equation}
where $W^{k}_{\tilde{\Lambda},q}$ denotes the $j$-th entry of $\Tilde{\CO}_{\SfB}^{k}$ and $\Bw_{k}^{\tilde{\Lambda}}\in \mathbb{C}^{L\times 1}$ is the precoder suppressing the interference of subpacket $W^{k}_{\tilde{\Lambda},q}$ at every user in $\CH_{k}^{\Tilde{\Lambda}}$, defined in~\eqref{eq: x B}. Clearly, according to \eqref{eq: x B} and \eqref{eq: x B sic-free}, we have $\Bx_{\SfB}=\sum_{j=1}^{\nu_{1}}\Bx_{\SfB}^{j}$.

\begin{exmp}
    Let us focus on the cache-aided network considered in Example~\ref{exmp: Strategy A} and the transmission vector $\Bx_{\SfA_{\circ}}$ in \eqref{eq: x1 A}. For this transmission vector, the sets of subpackets received by users 1 and 2, for example, are given as $\CO_{\SfA_{\circ}}^{1}=\lbrace W_{2,1}^{1},W_{3,1}^{1} \rbrace$ and $\CO_{\SfA_{\circ}}^{2}=\lbrace W_{2,1}^{2},W_{3,1}^{2} \rbrace$, respectively. So, for SIC-free data delivery, $\Bx_{\SfA_{\circ}}$ should be split into $\nu_{2}=2$ transmission vectors  
    \begin{equation*}
    \begin{array}{cc}
         & \Bx_{\SfA_{\circ}}^{1}=\sum\limits_{k \in \CS_{2,1} \Vert \CS_{3,1}}  \Bw_{k}^{1}  W_{1,1}^{k}+
            \sum\limits_{k \in \CS_{1,1}}  \Bw_{k}^{2} W_{2,1}^{k}, \\
         & \Bx_{\SfA_{\circ}}^{2}= \sum\limits_{k \in \CS_{3,1}} \Bw_{k}^{2}  W_{2,1}^{k} 
          +  \sum\limits_{k \in \CS_{1,1} \Vert \CS_{2,1}}  \Bw_{k}^{3} W_{3,1}^{k},
    \end{array}
    \end{equation*}
    where, for example, $\Bw_{1}^{2}$ and $\Bw_{1}^{3}$ suppresses the interference at the sets of users $ \CG_{1}^{2}=\lbrace 2,3,10,11,12 \rbrace$ and $\CG_{1}^{3}=\lbrace 2,3,6,7,8 \rbrace$, respectively.
    
\end{exmp}

\section{DoF Analysis}
In this section, we use DoF as the metric of interest to measure performance. Here, it is assumed that both strategies \emph{A} and \emph{B} employ the SIC-free data delivery algorithm, and the DoF is defined as the average number of users served concurrently during the delivery phase.\footnote{Note that utilizing the SIC-free data delivery algorithm does not alter the DoF~\cite{tolli2017multi}. The assumption here is only for the sake of simplicity.} In CC and UC delivery steps, we denote the total transmissions by $T_{M}$ and $T_{U}$, respectively, and the number of transmitted subpackets by $J_{M}$ and $J_{U}$. Therefore, DoF is computed as follows. 
\begin{equation}
\label{eq: DoF def}
    \mathrm{DoF}=\frac{J_{M}+J_{U}}{T_{M}+T_{U}}.
\end{equation}
Furthermore, we suppose that $K_{M}$ and $K_{U}$ users are served during the CC and UC delivery steps such that $K_{M}=\sum_{p} \min \left( \hat{\eta},\eta_{p} \right)$ and $K_{U}=\sum_{p} \max \left( 0, \eta_{p}-\hat{\eta} \right)$. The next theorem characterizes the DoF for the cache-aided networks operating with strategies \textit{A} and \textit{B} during the CC delivery step. 
\begin{thm}
\label{thm: DoF}
Consider a dynamic MISO network with the spatial multiplexing gain of $\alpha$, cache ratio $\gamma$ and the delivery parameter $\hat{\eta}$. If the system operates with \textit{Strategy A} in the CC delivery step, the DoF is given by:
    \begin{equation}
    \label{eq: DoF thm}
        \mathrm{DoF}=
        \begin{cases}
        \frac{K  \binom{P-1}{Q-1}\beta}{\sum\limits_{r=1}^{P-Q+1} D\left( \delta_{r}\right) \binom{P-r}{Q-1}} & K_{U}=0 \\
            \frac{K_{M} \binom{P-1}{Q-1}\beta \nu_{2} + K_{U} \left( 1-\gamma  \right) \binom{P}{\Brt} \beta^{\prime}}{\sum\limits_{r=1}^{P-Q+1} D\left( \delta_{r}\right) \binom{P-r}{Q-1}\nu_{2}+\left\lceil \frac{K_{U} \left( 1-\gamma \right) \binom{P}{\Brt}\beta^{\prime}}{\min \left( K_{U},\alpha \right)} \right\rceil} & K_{U} \neq 0
        \end{cases},
    \end{equation}
    where $\beta^{\prime}=\beta \binom{P-\Brt-1}{Q-\Brt-1}$ and $D(\delta_{r})= \phi_{r}$ if $\delta_{r} \neq 0$; otherwise, $D\left( \delta_{r} \right)=0$. If \textit{Strategy B} is applied during the CC delivery step, the DoF takes the form as follows.  
    \begin{equation}
    \label{eq: DoF thm non-int}
        \mathrm{DoF}=
        \begin{cases}
        \frac{K \binom{P-1}{Q-1}\left( \hat{\eta}\Brt+\alpha \right) \nu_{2}}{N_{M}} & K_{U}=0 \\
            \frac{K_{M} \binom{P-1}{Q-1}\left( \hat{\eta}\Brt+\alpha \right)\nu_{2} \nu_{1} + K_{U} \left( 1-\gamma  \right) \binom{P}{\Brt} \alpha^{\prime}}{N_{M}\nu_{1}+N_{U}} & K_{U} \neq 0
        \end{cases},
    \end{equation}
    where $\alpha^{\prime}=\left( \hat{\eta}\Brt+\alpha \right) \nu_{1} \binom{P-\Brt-1}{Q-\Brt-1}$, $N_{U}=\bigg\lceil \frac{K_{U} \left( 1-\gamma \right) \binom{P}{\Brt}\alpha^{\prime}}{\min \left( K_{U},\alpha \right)} \bigg\rceil$ and
    \begin{equation}
    \label{eq: NM def B}
    \begin{array}{c}
        N_{M}=\sum_{r=1}^{P} \sum_{c=1}^{P-Q+1} \sum_{m=1}^{\hat{\eta}} \sum_{s=1}^{\nu_{2}} \binom{P-c-1}{Q-2}  I^{+} \left( \Bar{\delta}_{c}, \CE_{r}^{m} \right).
        \end{array}
    \end{equation}
\end{thm}

\begin{IEEEproof}
The proof is relegated to Appendix~\ref{apx: proof DoF}.
\end{IEEEproof}

\begin{cor}
    To give further insight into Theorem~\ref{thm: DoF}, assume that the length of all caching profiles is non-zero, i.e., $\eta_{p} \neq 0$ for $p \in [P]$, all users are served during the CC delivery step, i.e., $\eta_{p} = \delta_{p}$ for $p \in [P]$, and $\hat{\eta} = \eta_{1} = \max_{p} \eta_{p}$. If the system operates with \emph{Strategy~A} and $\alpha \geq \hat{\eta}$, then  $\beta = \min (\alpha, \hat{\eta}) =\hat{\eta} $ and the DoF in~\eqref{eq: DoF thm} is simplified to $\mathrm{DoF} = \nicefrac{KQ}{P}$. 
    Similarly, when the system operates with \emph{Strategy~B} in the CC delivery step, the DoF in~\eqref{eq: DoF thm non-int} is rearranged as $\mathrm{DoF} = \nicefrac{K \left( \hat{\eta} \Brt + \alpha \right)}{P \hat{\eta}}$.
\end{cor}

\begin{rem}
\label{rem: optimal}
Suppose that $K$ users are uniformly assigned to the caching profiles, i.e., $K=P\hat{\eta}$, and all users are served in the CC delivery step, i.e., $K_{M}=K$ and $K_{U}=0$. If $\alpha \leq \hat{\eta}$, our schemes achieves the optimal DoF $\alpha \left( P\gamma+1 \right)$ obtained in~\cite{parrinello2019fundamental}. If $\alpha > \hat{\eta}$,  the achievable DoF of our scheme is simplified to the optimal DoF $K\gamma+\alpha$ obtained in~\cite{parrinello2020extending}. However, unlike the scheme proposed in~\cite{parrinello2020extending}, our scheme supports the networks with non-integer $\frac{\alpha}{\hat{\eta}}$ (cf. Appendix~\ref{apx: proof DoF alphaleta uniform}). 
\end{rem}

Indeed, increasing non-uniformness in user distribution prevents the system from achieving optimal DoF performance. For instance, for the case $\alpha > \hat{\eta}$, suppose that all users are served with \textit{Strategy A} during the CC delivery step, such that $\beta=\hat{\eta}$ and $Q \leq \Brt + \left\lfloor \nicefrac{\alpha}{\hat{\eta}} \right\rfloor$. In this setup,  we can set $Q= \Brt+ \left\lfloor \nicefrac{\alpha}{\hat{\eta}} \right\rfloor$ to boost the DoF performance. By defining $\eta_{\rm avg}=\frac{K}{P}$ and assuming $\alpha$ is divisible by $\hat{\eta}$ and $\eta_{\rm avg}$, the DoF loss (compared to uniform user distribution) is  $\alpha \left( 1-\nicefrac{\eta_{\rm avg}}{\hat{\eta}} \right)$.

\begin{rem}
\label{rem: DoF max alphaleta}
For the non-uniform user-to-profile association with  $Q=\Brt +1$ and $\eta_{1} \leq \alpha$, the best possible DoF is achievable by setting $\hat{\eta}=\max_{p}\eta_{p}=\eta_{1}$ (cf. Appendix~\ref{apx: Proof DoF max alphaleta}).
\end{rem}

Generally speaking, for any network parameters, we should compare the  achievable DoF of our proposed scheme with $\alpha$ such that: \textit{i)} if $\mathrm{DoF} \geq \alpha$, we use the proposed CC scheme to simultaneously benefit from global CC and spatial multiplexing gains; \textit{ii)} if $\mathrm{DoF}<\alpha$, we serve users via unicasting to not have any loss in  DoF. 

\begin{rem}
In scenarios with non-uniform user-to-profile associations, some transmission vectors may serve fewer than $\alpha$ users in the CC delivery step, preventing the system from fully utilizing the available multiplexing gain. To ensure that at least $\alpha$ users are served per transmission during the CC delivery step, we can adopt the so-called \textit{efficient multicast} method. In this method, the BS disregards transmission vectors serving fewer than $\alpha$ users, and consequently, subpackets associated with these transmission vectors are not transmitted during the CC delivery step. Instead, all these omitted subpackets, in addition to the subpackets for $K_{U}$ users excluded from the CC delivery step, are transmitted during the UC delivery step by following a similar way as described in Section~\ref{section: UC}. The derivation of a closed-form expression for the DoF of our proposed scheme via the efficient multicast method is beyond the scope of this paper, and we will explore it in our future work. It's worth noting that, without employing the efficient multicast method, \textit{Strategy~B} outperforms \textit{Strategy~A} in terms of achievable DoF, as outlined in Theorem~\ref{thm: DoF}. However, by using efficient multicast transmission, \textit{Strategy~A} is not necessarily inferior to \textit{Strategy~B} in terms of DoF.  
\end{rem}

\section{Beamforming Design}
In this section, we design optimized beamformers to deliver data during the CC and UC delivery steps.  In this regard, let us assume that the system operates with \textit{Strategy A} via the SIC-free data delivery algorithm during the CC delivery step, and without loss of generality, suppose that \emph{Strategy~A} is comprised of $N_{\rm A}$ transmissions. Obviously, if the length of all caching profiles is non-zero, i.e., $\eta_{p} \neq 0$ for all $p \in [P]$, we have $N_{\rm A}= \nu_{2} \sum_{r=1}^{P-Q+1} \phi_{r} \binom{P-r}{Q-1}$. Accordingly, we merge all transmissions $\SfA^{d} = (r,c,l)^{d}$ with $d \in [\nu_{2}]$, $r \in [P-Q+1]$, $c \in [\phi_{r} ]$ and $l \in [\binom{P-r}{Q-1}]$ to the set $\Pi$ as follows
\begin{equation*}
\small
    \begin{array}{l}
    \Pi = \left\lbrace (r,c,l)^{d}: d \in [\nu_{2}], r \in [P-Q+1], c \in [\phi_{r} ], l \in [\binom{P-r}{Q-1}] \right\rbrace,
    \end{array}
\end{equation*}
where $\left\lvert \Pi \right\rvert = N_{\rm A}$ and $\Pi(n)$ represents the $n$-th element of $\Pi$. Hereafter, for national simplicity, the aim is to design the beamformers for the transmission $n \in [N_{\rm A}]$, for which $\Pi(n) = \SfA^{d} = (r,c,l)^{d}$ and the set of users served in this transmission is given by $\CT_{\SfA}$ (cf. Section~\ref{sec: sic-free}).
  To this end, by using \eqref{eq: x A sic-free} and \eqref{eq: Yk system}, and  
enabling the cache-aided interference removal,
the achievable rate of user $k \in {\CT}_{\SfA}$ is given by $R_{k}=\log \left( 1+\Gamma_{k} \right)$,  where $\Gamma_{k}$ is  the  SINR at user $k$ in this transmission, and is equal to: 
\begin{equation}
\label{eq: gammaj def}
\begin{array}{c}
    \Gamma_{k}= \frac{\left\vert \Bh_{k}^{\rm H} \Bw_{k}^{\Lambda} \right\vert ^{2}}{\sum\nolimits_{j \in \CG_{k}^{\Lambda}} \left\vert \Bh_{k}^{\rm H}  \Bw_{j}^{\Lambda} \right\vert^{2} +N_{0} }.
    \end{array}
\end{equation}
We note that if the system operates with \textit{Strategy B}, by employing the SIC-free data delivery algorithm, each transmission $\SfB^{d}$ with $d \in [\nu_{1}]$ serves the set of users $\Tilde{\CT}_{\SfB}$. Here, if $\Sfp[k]=r$, $\CG_{k}^{\Lambda}$ in~\eqref{eq: gammaj def} is replaced by $\Tilde{\CT}_{\SfB}- \lbrace k \rbrace$; otherwise, we replace $\CG_{k}^{\Lambda}$ with $\CH_{k}^{\Lambda}$~(cf. Section~\ref{subsection: strategy B}). Similarly, for each transmission of the UC delivery step, by considering $\hat{\CT}$ as the set of served users, the interfering set $\CG_{k}^{\Lambda}$ in \eqref{eq: gammaj def} is substituted with $\hat{\CT}-\lbrace k \rbrace$.

As stated in Section~\ref{subsection: Strategy A}, by using \textit{Strategy A}, each mini-file $W_{\CP}^{n}$  is split into $\beta \binom{P-\Brt -1}{Q-\Brt-1}$ subpackets $ W_{\CP ,q }^{n}$,  where $q \in [\beta \binom{P-\Brt -1}{Q-\Brt-1}]$. On the other hand, the size of each mini-file is $\nicefrac{1}{\binom{P}{\Brt}}$ of the size of the original file. Hence, user $k \in {\CT}_{\SfA}$ receives a $\frac{1}{\mu_{k}}$ portion of its requested file at the end of the transmission $n \in [N_{\rm A}]$, where $\mu_{k}=\beta \binom{P-\Brt -1}{Q-\Brt-1} \binom{P}{\Brt}$. As a result, during this transmission, the required time to deliver a subpacket to user $k$, denoted by $T_{k}$, takes the form of $T_{k}=\frac{1}{\mu_{k} R_{k}}$.  Now, our aim is to design the beamformers to minimize the delivery time or, equivalently, maximize the achievable rate for the worst user served in the transmission $n \in [N_{\rm A}]$. Accordingly, given a fixed transmit power $P_{\rm T}$, we should solve the following optimization problem in order to obtain the optimized beamformers. 
\begin{equation}
\label{eq: opt prob rate def}
\begin{array}{ll}
  &\underset{ \Bw_{k}^{\Lambda}: k \in {\CT}_{\SfA} }{\max\min}     \quad  {\mu_{k}R_{k}} \\
 & \quad \quad  \mathrm{s.t.} \quad \sum\nolimits_{k \in {\CT}_{\SfA}} \left\Vert \Bw_{k}^{\Lambda} \right\Vert^{2} \leq P_{\rm T}.
\end{array} 
\end{equation}

Here, problem \eqref{eq: opt prob rate def} can be efficiently solved as a max-min problem using iterative methods based on Lagrangian duality, as demonstrated in various studies including \cite{salehi2020lowcomplexity,9838415}. In this sense, employing a Lagrangian duality-based technique described in \cite{salehi2020lowcomplexity}, we can iteratively find the optimum rate $R_{\rm e}$ via bisection search, which is given by $R_{\rm e}=\mu_{k}R_{k}$. For each fixed $R_{\rm e}$, the fixed point iteration method~\cite{wiesel2005linear} is used to obtain the dual variables $\lambda_{k}$ for $k \in {\CT}_{\SfA}$. In this regard, at the first step, for $k \in {\CT}_{\SfA}$, we initialize $\lambda_{k}  \leftarrow \lambda_{k}[1] $. Then, until achieving the desired level of convergence,  we iteratively update the dual variables such that for iteration $i+1$, the dual variable $\lambda_{k}$ is given by: 
\begin{equation}
\label{eq: lambda final}
    \begin{aligned}
    \lambda_{k}[i+1]=
     \frac{\omega_{k}}{\omega_{k}+1}   \left( \Bh_{k}^{\rm H} \boldsymbol{\Sigma}_{k,i}^{-1} \Bh_{k} \right)^{-1},
    \end{aligned}
\end{equation}
where $\omega_{k}=2^{\nicefrac{R_{\rm e}}{\mu_{k}}}-1$, $\lambda_{k} [i]$ denotes the value of $\lambda_{k}$ during the $i$-th iteration,  and $\boldsymbol{\Sigma}_{k,i}= \BI_{L} + \sum_{j \in \CG_{k}^{\Lambda} \cup \left\lbrace k \right\rbrace}  \lambda_{j} [i] \Bh_{j} \Bh_{j}^{\rm H}$.

After the convergence of \eqref{eq: lambda final}, we obtain the normalized beamformer $ \Tilde{\Bw}_{k}$. Accordingly, for $k \in  { \CT}_{\SfA}$, as per~\cite{salehi2020lowcomplexity}, $\Tilde{\Bw}_{k}$ is given by $\Tilde{\Bw}_{k}=\nicefrac{\Tilde{\boldsymbol{\Sigma}}_{k}^{-1} \Bh_{k}}{\Vert \Tilde{\boldsymbol{\Sigma}}_{k}^{-1} \Bh_{k} \Vert}$, where $\Tilde{\boldsymbol{\Sigma}}_{k}=\BI_{L} + \sum_{j \in \CG_{k}^{\Lambda} \cup\lbrace k \rbrace}  \lambda_{j} \Bh_{j} \Bh_{j}^{\rm H}$.  
In order to find the power vector of the beamformers, denoted by $\Tilde{\Bp}$, we simply follow the same steps as in~\cite[eq. (26)]{salehi2020lowcomplexity}. Finally,  during the transmission $n \in [N_{\rm A}]$, the beamforming vector $ \Bw_{k}^{\Lambda}$ takes the form of $\Bw_{k}^{\Lambda}= \sqrt{ \Tilde{\Bp}(k) } \Tilde{\Bw}_{k}$, where $\Tilde{\Bp}(k)$ is the $k$-th entry of $\Tilde{\Bp}$. Consequently, the transmission rate for the transmission $n \in [N_{\rm A}]$ during the CC delivery step, denoted by $R_{n}^{\rm A}$, is given by $R_{n}^{\rm A}=\min_{k \in \CT_{\SfA}} \log \left( 1+\Gamma_{k} \right)$. Similarly, we can follow the same steps to obtain the transmission rates for each transmission of \textit{Strategy B} and UC delivery step. Finally, we evaluate the system performance at finite-SNR regime based on the \textit{symmetric rate per user}, which is denoted by $R_{\rm sym}$. Assuming that the system operates with \textit{Strategy A} during the CC delivery step, $R_{\rm sym}$ is defined as follows. 
\begin{equation}
\label{eq: Rsym def}
    \begin{array}{l}
    R_{\rm sym}  = \big( \sum\nolimits_{n=1}^{N_{\rm A}} \frac{1}{R_{n}^{\rm A}}  +  \sum\nolimits_{j=1}^{\Tilde{N}_{\rm UC}} \frac{1}{R_{j}^{\rm UC}}  \big)^{-1},
    \end{array}
\end{equation}
where $\Tilde{N}_{\rm UC}$ is the total number of transmissions during the UC delivery step, and $R_{j}^{\rm UC}$ is the achievable rate associated with the $j$-th transmission of the UC delivery step. Here, we note that if we use the efficient multicast method during the CC delivery step, then $\Tilde{N}_{\rm UC}$ is not necessarily equal to $T_{U}$ in \eqref{eq: DoF def}. Additionally, as mentioned earlier, for the network setup operating with \textit{Strategy B} in the CC delivery step, we can follow the similar steps as in \eqref{eq: gammaj def}-\eqref{eq: lambda final} to obtain the achievable rate for each transmission in the CC delivery step. Then, we can simply compute $R_{\rm sym}$ in \eqref{eq: Rsym def} based on the total number of transmissions during the CC delivery step via \textit{Strategy B}. 
Here, we note that any beamforming design can be applied to the proposed data delivery scheme, and the finite-SNR performance can be measured via various metrics such as data delivery time~\cite{9723190}.

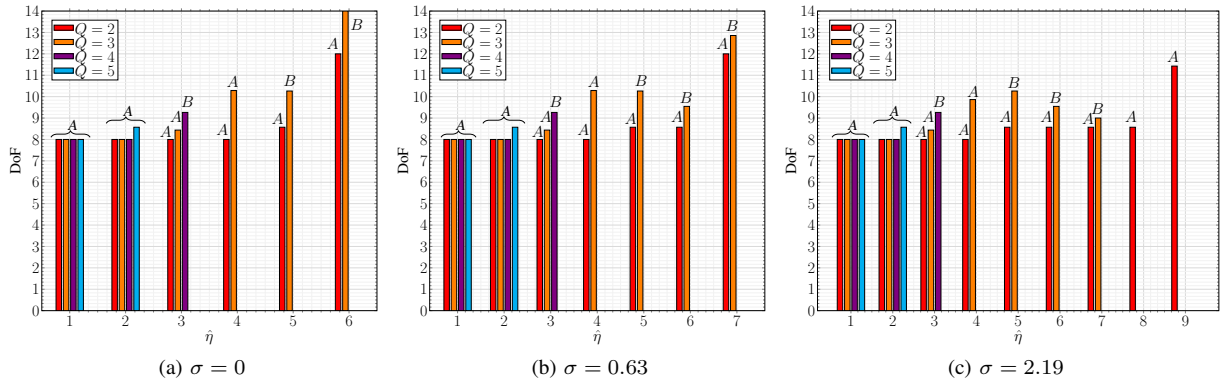
\begin{figure*}[ht]
\centering
\subfloat[$\sigma=0$]{\pgfplotstableread[row sep=\\,col sep=&]{
    interval & carT & carD & carR& carE \\
    1     & 8 & 8 & 8 & 8  \\
    2     & 8 & 8  & 8 & 8.5714  \\
    3    & 8 &  8.4375 & 9.2632 &   \\
    4   & 8 &  10.2857 & & \\
    5   & 8.5714 & 10.2632 &   &  \\
    6      & 12  & 14 & &  \\
    }\mydata

    \pgfplotstableread[row sep=\\,col sep=&]{
    interval & carT & carD & carR& carE \\
    1     & $A$ & $A$ & $A$ & $A$  \\
    2     & $A$ & $A$  & $A$ & $A$  \\
    3    & $A$ &  $A$ & $B$ &   \\
    4   & $A$ &  $A$ & & \\
    5   & $A$ & $B$ &   &  \\
    6      & $A$  & $B$ & &  \\
    7      & $A$  & $B$ & & \\
    8      & $A$ & & & \\
    9      & $A$ & & & \\
    }\mystrategy

\begin{tikzpicture}[scale=.24]
    \begin{axis}[
            ybar=1mm,
            bar width=3mm,
            width=1.1\textwidth,
            height=1\textwidth,
            legend pos = north west,
            legend style={legend columns=1},
            xtick=data,
             xtick=data,
            ymin=0,ymax=14,
            ylabel={\Huge $\text{DoF}$},
            xlabel={\Huge $\hat{\eta}$},
    grid=both,
    major grid style={line width=.2pt,draw=gray!30},
    grid style={line width=.1pt, draw=gray!10},
    minor tick num=5,
            axis line style={thick},
            tick label style={font=\Huge},
        ]   
        \addplot[draw=black, fill=red] table[x=interval,y=carT]{\mydata};
        \addplot[draw=black, fill=orange] table[x=interval,y=carD]{\mydata};
        \addplot[draw=black, fill=violet] table[x=interval,y=carR]{\mydata};
        \addplot[draw=black, fill=cyan] table[x=interval,y=carE]{\mydata};
        \node[above,font=\Huge] at (2.72,8) {$A$};
        \node[above,font=\Huge] at (2.9,8.7) {$A$};
        \node[above,font=\Huge] at (3.11,9.4) {$B$};
        \node[above,font=\Huge] at (3.73,8.1) {$A$};
        \node[above,font=\Huge] at (3.92,10.35) {$A$};
        \node[above,font=\Huge] at (4.72,8.6) {$A$};
        \node[above,font=\Huge] at (4.94,10.4) {$B$};
        \node[above,font=\Huge] at (5.71,12.1) {$A$};
        \node[above,font=\Huge] at (6.14,13) {$B$};
        \draw [decorate,decoration={brace,amplitude=10pt},xshift=-4pt,yshift=0pt]
    (.71,8.1) -- (1.41,8.1) node [black,midway,yshift=20pt] {\Huge \textit{ A}};
    \draw [decorate,decoration={brace,amplitude=10pt},xshift=-4pt,yshift=0pt]
    (1.7,8.75) -- (2.4,8.75) node [black,midway,yshift=20pt] {\Huge \textit{ A}};
        \legend{\Huge $Q=2$, \Huge $Q=3$, \Huge $Q=4$, \Huge $Q=5$}
    \end{axis}
\end{tikzpicture} \label{fig: sigma0}}
\subfloat[$\sigma=0.63$]{\pgfplotstableread[row sep=\\,col sep=&]{
    interval & carT & carD & carR& carE \\
    1     & 8 & 8 & 8 & 8  \\
    2     & 8 & 8  & 8 & 8.5714  \\
    3    & 8 &  8.4375 & 9.2632 &   \\
    4   & 8 &  10.2857 & & \\
    5   & 8.5714 & 10.2632 &   &  \\
    6      & 8.5714  & 9.5455 & &  \\
    7      & 12 &12.8571 & & \\
    }\mydata

    \pgfplotstableread[row sep=\\,col sep=&]{
    interval & carT & carD & carR& carE \\
    1     & $A$ & $A$ & $A$ & $A$  \\
    2     & $A$ & $A$  & $A$ & $A$  \\
    3    & $A$ &  $A$ & $B$ &   \\
    4   & $A$ &  $A$ & & \\
    5   & $A$ & $B$ &   &  \\
    6      & $A$  & $B$ & &  \\
    7      & $A$  & $B$ & & \\
    8      & $A$ & & & \\
    9      & $A$ & & & \\
    }\mystrategy

\begin{tikzpicture}[scale=.24]
    \begin{axis}[
            ybar=1mm,
            bar width=3mm,
            width=1.1\textwidth,
            height=1\textwidth,
            legend pos = north west,
            legend style={legend columns=1},
            xtick=data,
             xtick=data,
            ymin=0,ymax=14,
            ylabel={$\Huge \text{DoF}$},
            xlabel={\Huge $\hat{\eta}$},
    grid=both,
    major grid style={line width=.2pt,draw=gray!30},
    grid style={line width=.1pt, draw=gray!10},
    minor tick num=5,
            axis line style={thick},
            tick label style={font=\Huge},
        ]   
        \addplot[draw=black, fill=red] table[x=interval,y=carT]{\mydata};
        \addplot[draw=black, fill=orange] table[x=interval,y=carD]{\mydata};
        \addplot[draw=black, fill=violet] table[x=interval,y=carR]{\mydata};
        \addplot[draw=black, fill=cyan] table[x=interval,y=carE]{\mydata};
        \node[above,font=\Huge] at (2.72,8) {$A$};
        \node[above,font=\Huge] at (2.89,8.7) {$A$};
        \node[above,font=\Huge] at (3.11,9.4) {$B$};
        \node[above,font=\Huge] at (3.7,8.1) {$A$};
        \node[above,font=\Huge] at (3.92,10.35) {$A$};
        \node[above,font=\Huge] at (4.72,8.6) {$A$};
        \node[above,font=\Huge] at (4.94,10.4) {$B$};
        \node[above,font=\Huge] at (5.71,8.7) {$A$};
        \node[above,font=\Huge] at (5.91,9.6) {$B$};
        \node[above,font=\Huge] at (6.69,12.1) {$A$};
        \node[above,font=\Huge] at (6.9,12.95) {$B$};
        \draw [decorate,decoration={brace,amplitude=10pt},xshift=-4pt,yshift=0pt]
    (.7,8.1) -- (1.45,8.1) node [black,midway,yshift=20pt] {\Huge \textit{ A}};
    \draw [decorate,decoration={brace,amplitude=10pt},xshift=-4pt,yshift=0pt]
    (1.6,8.75) -- (2.5,8.75) node [black,midway,yshift=20pt] {\Huge \textit{ A}};
        \legend{\Huge $Q=2$, \Huge $Q=3$, \Huge $Q=4$, \Huge $Q=5$}
    \end{axis}
\end{tikzpicture} \label{fig: sigma03}} 
\subfloat[$\sigma=2.19$]{\pgfplotstableread[row sep=\\,col sep=&]{
    interval & carT & carD & carR& carE \\
    1     & 8 & 8 & 8 & 8  \\
    2     & 8 & 8  & 8 & 8.5714  \\
    3    & 8 &  8.4375 & 9.2632 &   \\
    4   & 8 &  9.8630 & & \\
    5   & 8.5714 & 10.2632 &   &  \\
    6      & 8.5714  & 9.5455 & &  \\
    7      & 8.5714  & 9 & & \\
    8      & 8.5714 & & & \\
    9      & 11.4286 & & & \\
    }\mydata

    \pgfplotstableread[row sep=\\,col sep=&]{
    interval & carT & carD & carR& carE \\
    1     & $A$ & $A$ & $A$ & $A$  \\
    2     & $A$ & $A$  & $A$ & $A$  \\
    3    & $A$ &  $A$ & $B$ &   \\
    4   & $A$ &  $A$ & & \\
    5   & $A$ & $B$ &   &  \\
    6      & $A$  & $B$ & &  \\
    7      & $A$  & $B$ & & \\
    8      & $A$ & & & \\
    9      & $A$ & & & \\
    }\mystrategy

\begin{tikzpicture}[scale=.24]
    \begin{axis}[
            ybar=1mm,
            bar width=3mm,
            width=1.3\textwidth,
            height=1\textwidth,
            legend pos = north west,
            legend style={legend columns=1},
            xtick=data,
             xtick=data,
            ymin=0,ymax=14,
            ylabel={$\Huge \text{DoF}$},
            xlabel={\Huge $\hat{\eta}$},
    grid=both,
    major grid style={line width=.2pt,draw=gray!30},
    grid style={line width=.1pt, draw=gray!10},
    minor tick num=5,
            axis line style={thick},
            tick label style={font=\Huge},
        ]   
        \addplot[draw=black, fill=red] table[x=interval,y=carT]{\mydata};
        \addplot[draw=black, fill=orange] table[x=interval,y=carD]{\mydata};
        \addplot[draw=black, fill=violet] table[x=interval,y=carR]{\mydata};
        \addplot[draw=black, fill=cyan] table[x=interval,y=carE]{\mydata};
        \node[above,font=\Huge] at (2.68,8) {$A$};
        \node[above,font=\Huge] at (2.85,8.7) {$A$};
        \node[above,font=\Huge] at (3.11,9.4) {$B$};
        \node[above,font=\Huge] at (3.7,8.1) {$A$};
        \node[above,font=\Huge] at (3.92,9.9) {$A$};
        \node[above,font=\Huge] at (4.69,8.7) {$A$};
        \node[above,font=\Huge] at (4.9,10.3) {$B$};
        \node[above,font=\Huge] at (5.67,8.7) {$A$};
        \node[above,font=\Huge] at (5.89,9.6) {$B$};
        \node[above,font=\Huge] at (6.67,8.6) {$A$};
        \node[above,font=\Huge] at (6.9,9.1) {$B$};
        \node[above,font=\Huge] at (7.7,8.7) {$A$};
        \node[above,font=\Huge] at (8.7,11.5) {$A$};
        \draw [decorate,decoration={brace,amplitude=10pt},xshift=-4pt,yshift=0pt]
    (.61,8.1) -- (1.49,8.1) node [black,midway,yshift=20pt] {\Huge \textit{ A}};
    \draw [decorate,decoration={brace,amplitude=10pt},xshift=-4pt,yshift=0pt]
        (1.6,8.75) -- (2.5,8.75) node [black,midway,yshift=20pt] {\Huge \textit{ A}};
        \legend{\Huge $Q=2$, \Huge $Q=3$, \Huge $Q=4$, \Huge $Q=5$}
    \end{axis}
\end{tikzpicture}\label{fig: sigma3}}
\caption{The DoF versus $\hat{\eta}$ with different values of $Q$, $K=30$, $\alpha=8$, $P=5$, $\gamma=0.2$ and $\Brt=1$ for: (a) $\sigma=0$, (b) $\sigma=0.63$ and (c) $\sigma=2.19$. Here, \emph{A} and \emph{B} show that the system operates with \emph{Strategy~A} and \emph{Strategy~B}, respectively. The number of user profiles is fixed in all subfigures (a), (b), and (c). }
 \label{fig: DoF}
 \vspace{-1.5em}
\end{figure*}
\section{Numerical Results}
We use numerical simulations to analyze the performance of the proposed coded caching scheme for dynamic networks in terms of both the achievable DoF and the symmetric rate.

\subsection{DoF Analysis}
\label{sec: DoF num}
We consider a cache-aided MISO setup with $\Brt=1$, $P=5$, and $\gamma=0.2$, in which  $K=30$ users are present during the delivery phase. Here, for each user-to-profile association, we compute the standard deviation $\sigma$ as $\sigma^{2}=\frac{1}{P} \sum_{p=1}^{P} \left( \eta_{p} -\eta_{\rm avg} \right)^{2}$, where $\eta_{\rm avg}= \nicefrac{K}{P} = 6$. 

Fig.~\ref{fig: DoF} illustrates the achievable DoF for different $Q$ and $\sigma$ values when $\alpha=8$. In the figure, we have also shown which transmission strategy (i.e., \emph{A} or \emph{B}) should be used for the selected parameters. From Fig.~\ref{fig: DoF}, we see that even though choosing $\hat{\eta} = \max_{p} \eta_p$ is proven to maximize the DoF only when $\alpha$ is large (i.e. $\alpha\geq \max_{p} \eta_{p}$, cf.~Remark~\ref{rem: DoF max alphaleta}), it is also a safe choice for smaller $\alpha$ values. Moreover, for a given $\hat{\eta}$ value, selecting the maximum possible $Q$, i.e., $Q= \Brt +\lceil \nicefrac{\alpha}{\hat{\eta}} \rceil$, always maximizes the achievable DoF. Moreover, \emph{Strategy B} enables our proposed scheme to be truly universal and apply to any user-to-profile association. For instance, although no existing scheme in the literature can achieve the optimal DoF of $K\gamma+\alpha=14$ for the considered network with the uniform user-to-profile association (as $\alpha> \hat{\eta}$ and $\nicefrac{\alpha}{\hat{\eta}}$ is a non-integer, cf. Section~\ref{subsection: strategy B}), our proposed scheme can achieve this DoF with \emph{Strategy B} by setting $\hat{\eta}=\eta_{\rm avg}=6$ and $Q = 3$.

\begin{figure}[t]
    \centering 
    \input{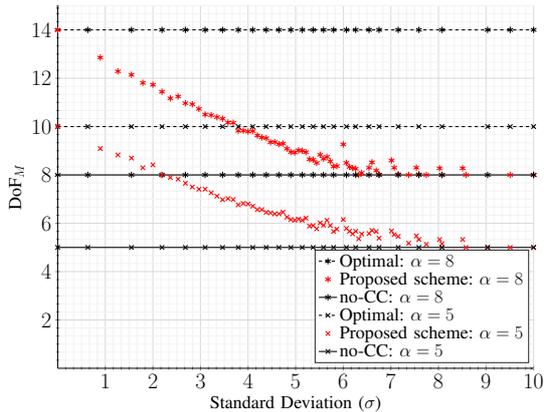}
    \vspace{-.5em}
    \caption{The average of the maximum achievable DoF $\left( {\rm DoF}_{M} \right)$ versus the standard deviation $(\sigma)$ with $K=30$, $\gamma=0.2$, $P=5$, $\Brt=1$ and $\alpha \in \left\lbrace 5,8 \right\rbrace$.}
    \label{fig: DoFM}
    \vspace{-1em}
\end{figure}

In Fig.~\ref{fig: DoFM}, we analyze the effect of the non-uniformness in user-to-profile associations (measured by $\sigma$) on the achievable DoF and compare the results with two extreme cases: \textit{i) Optimal}, where the users are uniformly assigned to the caching profiles, and \textit{ii) no-CC}, where the users only benefit from the local CC gain, and $\min \left( \alpha, K \right)$ users are served per transmission in a TDMA framework. The parameter $\mathrm{DoF}_{M}$ denotes the average of the maximum achievable DoF over the user-to-profile associations, and is calculated as follows:
for any possible user-to-profile association for the considered network, we find $\mathrm{DoF}_{\max}$ which is the maximum achievable DoF obtained by a line search over $\hat{\eta}$ and $Q$ values, and then, for the associations with the same $\sigma$ value, we set $\mathrm{DoF}_{M}$ as the average of all the corresponding $\mathrm{DoF}_{\max}$ values. As can be seen, although the placement phase of our scheme was designed solely based on the knowledge of $\gamma$ (and not the user count $K$), it can still boost the maximum achievable DoF by $10\%-70\%$ over the no-CC solution for moderate $\sigma$ values~(e.g., $1 \le \sigma \le 4.5$). This shows the strength of our algorithm in enabling a coded caching gain for any given non-uniform user distribution.

\begin{table}[htb]
    \centering
    \caption{Comparison between  the Existing Works in the Literature}
    \label{tab: DoF benchmarks}
    \Large
    \resizebox{\columnwidth}{!}{
    \begin{tabular}{|c|c|c|c|c|c|}
        \hline
          User distribution: $\alpha$ &   \cite{parrinello2019fundamental} &   \cite{parrinello2020extending} &   \cite{Abolpour2022CodedNetworks} &   no-CC &   Proposed scheme \\
        \hline
        $(6,6,6,6,6): \alpha =8$ & - & $\mathrm{DoF} = 14$ & $\mathrm{DoF} = 14$ & $\mathrm{DoF} = 8$ & $\mathrm{DoF} = 14$ \\
        \hline
       $(9,8,6,5,2): \alpha =8$ & - & - & $\mathrm{DoF} = 11.43$ & $\mathrm{DoF} = 8$ & $\mathrm{DoF} = 11.43$ \\
        \hline
         $(6,6,6,6,6): \alpha =4$ & $\mathrm{DoF} = 8$ & - & - & $\mathrm{DoF} = 4$ & $\mathrm{DoF} = 8$ \\
        \hline
         $(9,8,6,5,2): \alpha =4$ & - & - & - & $\mathrm{DoF} = 4$ & $\mathrm{DoF} = 6.23$ \\
        \hline
        \end{tabular}}
        \vspace{-1em}
\end{table}
In Table~\ref{tab: DoF benchmarks}, for different user-to-profile associations and $\alpha$ values,  the achievable DoF of our proposed scheme is compared with the schemes \cite{parrinello2019fundamental, parrinello2020extending, Abolpour2022CodedNetworks} and no-CC. Here, "-" shows that the scheme is not applicable to the region. Moreover, it is assumed that $K=30$, $\gamma=0.2$, $\Brt =1 $, $P =5$ and the length of caching profiles is represented as $(\eta_{1}, \eta_{2},\eta_{3}, \eta_{4}, \eta_{5})$. 
For the association $(6,6,6,6,6)$ and $\alpha =8$, our scheme achieves the optimal DoF of $14$ as same as the methods introduced in \cite{parrinello2020extending, Abolpour2022CodedNetworks}. However, the scheme \cite{parrinello2019fundamental} is not applicable to these network parameters, as the constraint $\alpha \leq \min_{p} \eta_{p}$ is not satisfied. Since the scheme \cite{parrinello2020extending} only supports the networks with the uniform distribution of users, only the introduced method in \cite{Abolpour2022CodedNetworks} and our proposed scheme can manage the data delivery in a setup with the association $(9,8,6,5,2)$ and $\alpha =8$ and achieve the DoF of $11.43$. For  $\alpha =4$ and uniform user distribution $(6,6,6,6,6)$, our proposed scheme achieves the DoF of $8$ as same as the method described in \cite{parrinello2019fundamental}. For the association $(9,8,6,5,2)$ and $\alpha =4$, none of the works \cite{parrinello2019fundamental, parrinello2020extending, Abolpour2022CodedNetworks} can handle the data delivery, while our proposed scheme supports this regime and achieves the DoF of $6.23$. 
\vspace{-1em}

\subsection{Symmetric Rate}
\label{sec: rate num}
\begin{figure}[ht]
    \centering 
    \pgfplotstableread[row sep=\\,col sep=&]{
sigmax & MC8\\
0	&	0.333345088	\\
5	&	0.659283455	\\
10	&	1.123898944	\\
15	&	1.723127636	\\
20	&	2.428337065	\\
25	&	3.197235638	\\
30	&	4.000166609	\\
35	&	4.821639133	\\
40	&	5.652467963	\\
45	&	6.486286863	\\
50	&	7.319390458	\\
}\AL

\pgfplotstableread[row sep=\\,col sep=&]{
sigmax & MC8\\
0	&	0.3320132	\\
5	&	0.646901408	\\
10	&	1.071038395	\\
15	&	1.587018893	\\
20	&	2.194174729	\\
25	&	2.903884906	\\
30	&	3.717288283	\\
35	&	4.615601783	\\
40	&	5.571591812	\\
45	&	6.561217011	\\
50	&	7.566725663	\\
}\BL

\pgfplotstableread[row sep=\\,col sep=&]{
sigmax & MC8\\
0	&	0.276715364	\\
5	&	0.495018863	\\
10	&	0.761578966	\\
15	&	1.060473044	\\
20	&	1.399061055	\\
25	&	1.804490699	\\
30	&	2.2965376	\\
35	&	2.872472171	\\
40	&	3.510999967	\\
45	&	4.186180642	\\
50	&	4.87890216	\\
}\UCL

\pgfplotstableread[row sep=\\,col sep=&]{
sigmax & MC8\\
0	&	0.336017016	\\
5	&	0.671705509	\\
10	&	1.154776214	\\
15	&	1.781733747	\\
20	&	2.52061122	\\
25	&	3.318872389	\\
30	&	4.140112474	\\
35	&	4.968747725	\\
40	&	5.799485835	\\
45	&	6.630316158	\\
50	&	7.460064808	\\
}\AS

\pgfplotstableread[row sep=\\,col sep=&]{
sigmax & MC8\\
0	&	0.334117289	\\
5	&	0.663404012	\\
10	&	1.130590572	\\
15	&	1.73770415	\\
20	&	2.475512739	\\
25	&	3.302440587	\\
30	&	4.171965995	\\
35	&	5.057101594	\\
40	&	5.947021059	\\
45	&	6.837748981	\\
50	&	7.727218341	\\
}\BS

\pgfplotstableread[row sep=\\,col sep=&]{
sigmax & MC8\\
0	&	0.271500956	\\
5	&	0.494191032	\\
10	&	0.792094229	\\
15	&	1.173393837	\\
20	&	1.635823738	\\
25	&	2.152750942	\\
30	&	2.695094149	\\
35	&	3.246472731	\\
40	&	3.800453465	\\
45	&	4.354814553	\\
50	&	4.908641158	\\
}\UCS

\begin{tikzpicture}[scale=0.45]

\begin{axis}[%
width=2.1\columnwidth,
height=1.45\columnwidth,
axis lines = center,
xmin=0,
xlabel near ticks,
xlabel={\LARGE SNR [dB]},
ymin=0,
ymax=8,
ylabel={\LARGE Symmetric Rate [bits/s/Hz]},
ylabel near ticks,
    grid=both,
    major grid style={line width=.2pt,draw=gray!30},
    grid style={line width=.1pt, draw=gray!10},
    minor tick num=5,
    legend pos = north west,
legend style={at={(0.01,0.98)},legend cell align=left, align=left, draw=white!15!black},
ticklabel style={font=\huge},
]
\addplot[mark=triangle, mark size=3pt,mark options={solid,blue},dashed,draw=blue,line width=1.2pt] table[x=sigmax,y=MC8]{\AL};
\addlegendentry{ \LARGE \emph{Strategy A}: $\alpha=10$}

\addplot[mark=triangle,  mark size=3pt,mark options={solid,red},dashed,draw=red,line width=1.2pt] table[x=sigmax,y=MC8]{\BL};
\addlegendentry{\LARGE \emph{Strategy B}: $\alpha=10$}

\addplot[mark=triangle,  mark size=3pt,mark options={solid,black},dashed,draw=black,line width=1.2pt] table[x=sigmax,y=MC8]{\UCL};
\addlegendentry{\LARGE no-CC: $\alpha=10$}

\addplot[solid,draw=blue,line width=1.2pt] table[x=sigmax,y=MC8]{\AS};
\addlegendentry{\LARGE \emph{Strategy A}: $\alpha=8$}

\addplot[solid,draw=red,line width=1.2pt] table[x=sigmax,y=MC8]{\BS};
\addlegendentry{\LARGE \emph{Strategy B}: $\alpha=8$}

\addplot[solid,draw=black,line width=1.2pt] table[x=sigmax,y=MC8]{\UCS};
\addlegendentry{\LARGE no-CC: $\alpha=8$}
\end{axis}
\end{tikzpicture}%
    \vspace{-.9em}
    \caption{Average symmetric rate versus SNR with random user-to-profile association, $K=30$, $L=10$, $\Brt=1$, $P=5$, $\gamma=0.2$, $\hat{\eta}=\eta_{1}$ 
    and $ \alpha \in \lbrace 8,10 \rbrace$. The average symmetric rate is taken over all possible $\eta_{p}$ values.   }
    \label{fig: random}
    \vspace{-1.7em}
\end{figure}
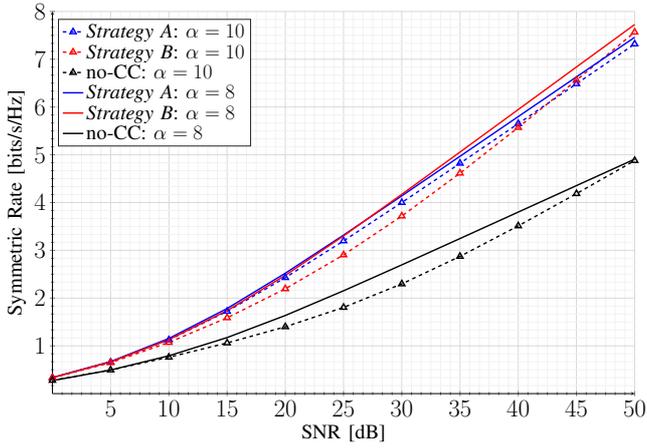

Here, we study the symmetric rate performance of cache-aided setups at finite-SNR regimes for the same network parameters considered in Section~\ref{sec: DoF num}, assuming the number of antennas at the transmitter is $L=10$. 
Moreover, the block-fading (quasi-static) i.i.d. Rayleigh fading channels $\Bh_{k}  \sim \CC \CN (0, \BI)$ with $k \in [K]$ are considered in simulations.  Fig.~\ref{fig: random} demonstrates the average symmetric rate versus SNR for both transmission strategies \emph{A} and \emph{B}, for $\alpha \in \{8,10\}$. The results are calculated for 60 random user-to-profile associations. Upon observation, it is evident that our proposed scheme yields a significant enhancement under both \textit{Strategy A} with $Q=\Brt+\left\lfloor \nicefrac{\alpha}{\hat{\eta}} \right\rfloor$ and \textit{Strategy B} with $Q=\Brt+\left\lceil \nicefrac{\alpha}{\hat{\eta}} \right\rceil$ in comparison to the no-CC scheme. This enhanced performance is attributed to our scheme's utilization of both the global caching and the spatial multiplexing gains. 

As observed, when $\alpha=L=10$,  $\textit{Strategy B}$ tends to surpass $\textit{Strategy A}$ at high-SNR (e.g., 50dB). This is because $\textit{Strategy A}$ attains lower DoF compared to $\textit{Strategy B}$, i.e., it serves a smaller number of users per transmission in comparison to $\textit{Strategy B}$. However, maximizing the number of users served in each transmission also causes $\textit{Strategy B}$ to keep the system fully loaded in many transmission intervals, which is not an appropriate choice at finite SNR as discussed in~\cite{tolli2017multi}. Therefore, as can be seen in Fig.~\ref{fig: random}, by setting $\alpha=L$, the performance of \textit{Strategy B} is inferior to \textit{Strategy A} for smaller SNR values. 
On the other hand, when $\alpha = L-2 = 8$, the system will never be fully loaded even with \emph{Strategy B}, and hence, it outperforms \emph{Strategy A} throughout the entire SNR range. Nevertheless, in both comparisons, the difference between the two strategies is small, while \textit{Strategy B} poses a higher subpacketization complexity on the network. Therefore, for real-world applications where the goal is to implement a cache-aided dynamic setup with low complexity, $\textit{Strategy A}$ could be safely chosen to serve users, albeit with a negligible loss in symmetric rate at finite-SNR regimes.

\begin{figure}[t] 
   \centering
       \pgfplotstableread[row sep=\\,col sep=&]{
sigmax & MC8\\
0	&	0.333345088	\\
5	&	0.659283455	\\
10	&	1.123898944	\\
15	&	1.723127636	\\
20	&	2.428337065	\\
25	&	3.197235638	\\
30	&	4.000166609	\\
35	&	4.821639133	\\
40	&	5.652467963	\\
45	&	6.486286863	\\
50	&	7.319390458	\\
}\AS

\pgfplotstableread[row sep=\\,col sep=&]{
sigmax & MC8\\
0	&	0.334843038	\\
5	&	0.667316904	\\
10	&	1.144426041	\\
15	&	1.768013147	\\
20	&	2.520460946	\\
25	&	3.354855667	\\
30	&	4.226690785	\\
35	&	5.112044484	\\
40	&	6.001591301	\\
45	&	6.891835771	\\
50	&	7.780941912	\\
}\BS

\pgfplotstableread[row sep=\\,col sep=&]{
sigmax & MC8\\
0	&	0.322586098	\\
5	&	0.628894206	\\
10	&	1.054419994	\\
15	&	1.580350142	\\
20	&	2.165826208	\\
25	&	2.772886903	\\
30	&	3.38358864	\\
35	&	3.992566464	\\
40	&	4.598896177	\\
45	&	5.202678935	\\
50	&	5.803976043	\\
}\AX

\pgfplotstableread[row sep=\\,col sep=&]{
sigmax & MC8\\
0	&	0.318395291	\\
5	&	0.611822457	\\
10	&	1.008394001	\\
15	&	1.503924813	\\
20	&	2.094154702	\\
25	&	2.749190062	\\
30	&	3.431238949	\\
35	&	4.118587578	\\
40	&	4.803729543	\\
45	&	5.484955399	\\
50	&	6.161829446	\\
}\BX

\pgfplotstableread[row sep=\\,col sep=&]{
sigmax & MC8\\
0	&	0.28978559	\\
5	&	0.542428407	\\
10	&	0.888830966	\\
15	&	1.327563418	\\
20	&	1.841506498	\\
25	&	2.399621731	\\
30	&	2.976666037	\\
35	&	3.559997042	\\
40	&	4.144939345	\\
45	&	4.729855059	\\
50	&	5.313829454	\\
}\AT

\pgfplotstableread[row sep=\\,col sep=&]{
sigmax & MC8\\
0	&	0.289616383	\\
5	&	0.538804316	\\
10	&	0.876085789	\\
15	&	1.30821452	\\
20	&	1.834053252	\\
25	&	2.42622451	\\
30	&	3.051276933	\\
35	&	3.688703518	\\
40	&	4.329948305	\\
45	&	4.971861571	\\
50	&	5.612846134	\\
}\BT

\pgfplotstableread[row sep=\\,col sep=&]{
sigmax & MC8\\
0	&	0.272523575	\\
5	&	0.495434854	\\
10	&	0.793413576	\\
15	&	1.174558357	\\
20	&	1.63685652	\\
25	&	2.153852609	\\
30	&	2.696314514	\\
35	&	3.247767423	\\
40	&	3.801786823	\\
45	&	4.356168295	\\
50	&	4.910001676	\\
}\UC

\begin{tikzpicture}[scale=.45]

\begin{axis}[%
width=2.1\columnwidth,
height=1.45\columnwidth,
axis lines = center,
xmin=0,
xlabel near ticks,
xlabel={\LARGE SNR [dB]},
ymin=0,
ymax=8,
ylabel={\LARGE Symmetric Rate [bits/s/Hz]},
ylabel near ticks,
    grid=both,
    major grid style={line width=.2pt,draw=gray!30},
    grid style={line width=.1pt, draw=gray!10},
    minor tick num=5,
    legend pos = north west,
legend style={at={(0.01,.98)},legend cell align=left, align=left, draw=white!15!black},
ticklabel style={font=\huge},
]
\addplot[solid,draw=blue,line width=1.2pt] table[x=sigmax,y=MC8]{\AS};
\addlegendentry{\LARGE \emph{Strategy A}: $\hat{\eta}=7$, $\mathrm{DoF}=12$}

\addplot[solid,draw=red,line width=1.2pt] table[x=sigmax,y=MC8]{\BS};
\addlegendentry{\LARGE \emph{Strategy B}: $\hat{\eta}=7$, $\mathrm{DoF}=12.86$}

\addplot[mark=o, mark size=2pt,mark options={solid,blue},dashdotted,draw=blue,line width=1.2pt] table[x=sigmax,y=MC8]{\AX};
\addlegendentry{\LARGE \emph{Strategy A}: $\hat{\eta}=6$, $\mathrm{DoF}=8.57$}

\addplot[mark=o, mark size=2pt,mark options={solid,red},dashdotted,draw=red,line width=1.2pt] table[x=sigmax,y=MC8]{\BX};
\addlegendentry{\LARGE \emph{Strategy B}: $\hat{\eta}=6$, $\mathrm{DoF}=9.55$}

\addplot[mark=square, mark size=2pt,mark options={solid,blue},dashed,draw=blue,line width=1.2pt] table[x=sigmax,y=MC8]{\AT};
    \addlegendentry{\LARGE \emph{Strategy A}: $\hat{\eta}=3$, $\mathrm{DoF}=8.44$}

\addplot[mark=square, mark size=2pt,mark options={solid,red},dashed,draw=red,line width=1.2pt] table[x=sigmax,y=MC8]{\BT};
    \addlegendentry{\LARGE \emph{Strategy B}: $\hat{\eta}=3$, $\mathrm{DoF}=9.26$}

\addplot[solid,draw=black,line width=1.2pt] table[x=sigmax,y=MC8]{\UC};
    \addlegendentry{\LARGE no-CC: $\mathrm{DoF}=8$}

\end{axis}
\end{tikzpicture}%
       \vspace{-.9em}
    \caption{Symmetric rate versus SNR in a setup with $K=30$, $\sigma=0.63$, $\eta_{1}=7$, $\alpha=L-2=8$, $\Brt=1$, $P=5$, $\gamma=0.2$ for different $\hat{\eta}$ values.}
    \label{fig: sigma2}
   \vspace{-1.5em}
   \end{figure}
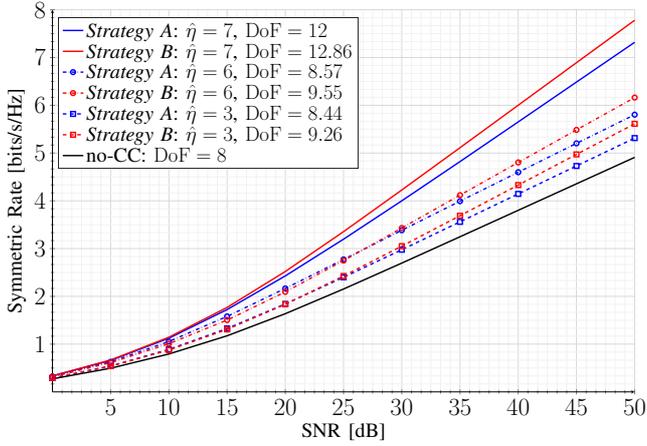
   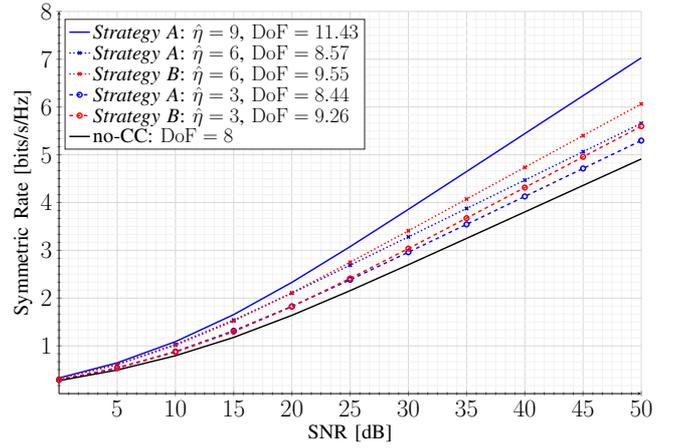
\begin{figure}[t]
   \centering
      \pgfplotstableread[row sep=\\,col sep=&]{
sigmax & MC8\\
0	&	0.327687688	\\
5	&	0.644563224	\\
10	&	1.088158865	\\
15	&	1.653981925	\\
20	&	2.329367232	\\
25	&	3.07704033	\\
30	&	3.857507669	\\
35	&	4.648879337	\\
40	&	5.442889386	\\
45	&	6.236792826	\\
50	&	7.029246314	\\
}\AM

\pgfplotstableread[row sep=\\,col sep=&]{
sigmax & MC8\\
0	&	0.315528221	\\
5	&	0.613704254	\\
10	&	1.029833903	\\
15	&	1.539707962	\\
20	&	2.103198012	\\
25	&	2.688635807	\\
30	&	3.281103793	\\
35	&	3.87539434	\\
40	&	4.469884697	\\
45	&	5.064017261	\\
50	&	5.65726528	\\
}\AS

\pgfplotstableread[row sep=\\,col sep=&]{
sigmax & MC8\\
0	&	0.313580764	\\
5	&	0.605357161	\\
10	&	1.010252912	\\
15	&	1.519490268	\\
20	&	2.111064091	\\
25	&	2.750107953	\\
30	&	3.408501299	\\
35	&	4.072560702	\\
40	&	4.737469881	\\
45	&	5.401659864	\\
50	&	6.064233599	\\
}\BS

\pgfplotstableread[row sep=\\,col sep=&]{
sigmax & MC8\\
0	&	0.287759802	\\
5	&	0.537784418	\\
10	&	0.880515021	\\
15	&	1.315576204	\\
20	&	1.827031832	\\
25	&	2.383843922	\\
30	&	2.960158371	\\
35	&	3.543189383	\\
40	&	4.128126998	\\
45	&	4.713142032	\\
50	&	5.29720553	\\
}\AT

\pgfplotstableread[row sep=\\,col sep=&]{
sigmax & MC8\\
0	&	0.287451479	\\
5	&	0.533965538	\\
10	&	0.867950814	\\
15	&	1.297135372	\\
20	&	1.82098506	\\
25	&	2.411920194	\\
30	&	3.036080823	\\
35	&	3.673047511	\\
40	&	4.314201185	\\
45	&	4.956180407	\\
50	&	5.59724801	\\
}\BT

\pgfplotstableread[row sep=\\,col sep=&]{
sigmax & MC8\\
0	&	0.271269728	\\
5	&	0.493928475	\\
10	&	0.793062924	\\
15	&	1.175889047	\\
20	&	1.638671246	\\
25	&	2.155443371	\\
30	&	2.697803873	\\
35	&	3.249332749	\\
40	&	3.803469126	\\
45	&	4.357957912	\\
50	&	4.911864198	\\
}\UC

\begin{tikzpicture}[scale=.45]

\begin{axis}[%
width=2.1\columnwidth,
height=1.45\columnwidth,
axis lines = center,
xmin=0,
xlabel near ticks,
xlabel={\LARGE SNR [dB]},
ymin=0,
ymax=8,
ylabel={\LARGE Symmetric Rate [bits/s/Hz]},
ylabel near ticks,
    grid=both,
    major grid style={line width=.2pt,draw=gray!30},
    grid style={line width=.1pt, draw=gray!10},
    minor tick num=5,
    legend pos = north west,
legend style={at={(0.01,.98)}, legend cell align=left, align=left, draw=white!15!black},
ticklabel style={font=\huge},
]
\addplot[solid,draw=blue,line width=1.2pt] table[x=sigmax,y=MC8]{\AM};
\addlegendentry{\LARGE \emph{Strategy A}: $\hat{\eta}=9$, $\mathrm{DoF}=11.43$}

\addplot[mark=x, mark size=2pt,mark options={solid,blue},dotted,draw=blue,line width=1.2pt] table[x=sigmax,y=MC8]{\AS};
\addlegendentry{\LARGE \emph{Strategy A}: $\hat{\eta}=6$, $\mathrm{DoF}=8.57$}

\addplot[mark=x, mark size=2pt,mark options={solid,red},dotted,draw=red,line width=1.2pt] table[x=sigmax,y=MC8]{\BS};
\addlegendentry{\LARGE \emph{Strategy B}: $\hat{\eta}=6$, $\mathrm{DoF}=9.55$}

\addplot[mark=o, mark size=2pt,mark options={solid,blue},dashed,draw=blue,line width=1.2pt] table[x=sigmax,y=MC8]{\AT};
\addlegendentry{\LARGE \emph{Strategy A}: $\hat{\eta}=3$, $\mathrm{DoF}=8.44$}

\addplot[mark=o, mark size=2pt,mark options={solid,red},dashed,draw=red,line width=1.2pt] table[x=sigmax,y=MC8]{\BT};
\addlegendentry{\LARGE \emph{Strategy B}: $\hat{\eta}=3$, $\mathrm{DoF}=9.26$}

\addplot[solid,draw=black,line width=1.2pt] table[x=sigmax,y=MC8]{\UC};
\addlegendentry{\LARGE no-CC: $\mathrm{DoF}=8$}

\end{axis}
\end{tikzpicture}%
      \vspace{-.9em}
    \caption{Symmetric rate versus SNR in a network with $K=30$, $\sigma=2.19$, $\eta_{1}=9$, $\alpha=L-2=8$, $\Brt=1$, $P=5$, $\gamma=0.2$ for different $\hat{\eta}$ values.}
    \label{fig: sigma1}
   \vspace{-1.5em}
\end{figure}

Figs.~\ref{fig: sigma2} and \ref{fig: sigma1} illustrate the symmetric rate of the considered network for different $\hat{\eta}$ values, when $\alpha = 8$. In Fig.~\ref{fig: sigma2}, it is assumed that $\sigma=0.63$ and $\eta_{1}=7$, while in Fig.~\ref{fig: sigma1}, we have $\sigma=2.19$ and $\eta_{1}=9$. 
As demonstrated, for both values of $\sigma$, the maximum symmetric rate is attained with $\hat{\eta}=\eta_{1}=\max_{p}\eta_{p}$. This is because setting $\hat{\eta}=\max_{p} \eta_{p}$ increases the number of users served during the CC delivery step and maximizes the benefit from the achievable global caching gain. Moreover, even though there is a notable discrepancy in non-uniformness of the two cases, it is demonstrated that when $\hat{\eta}=\max_p \eta_p$, the rate loss due to increased non-uniformness in Fig.~\ref{fig: sigma1} becomes negligible in comparison to Fig.~\ref{fig: sigma2}, particularly in the finite-SNR regime. This underscores the effectiveness of our approach to address the non-uniformness in the user-to-profile association by optimizing the number of users served per transmission.

From Figs.~\ref{fig: sigma2} and \ref{fig: sigma1}, it can also be seen that when $\hat{\eta}$ is significantly smaller than $\max_{p} \eta_{p}$, both the achievable DoF and the symmetric rate become almost independent of the $\sigma$ value.  For instance, considering $\hat{\eta}=3$, \textit{Strategy A} achieves exactly the same DoF and rate values in both Fig.~\ref{fig: sigma2} and Fig.~\ref{fig: sigma1}. This is because if $\hat{\eta}$ is small enough, the user-to-profile association of the users served in the CC delivery step becomes uniform, and the transmission vectors will be the same regardless of the $\sigma$ value. Moreover, a small increase in $\hat{\eta}$ results in a significant boost in the symmetric rate even though the DoF changes slightly.
For instance, when $\hat{\eta}=6$, \textit{Strategy A} still delivers a modest boost in DoF performance in comparison with the no-CC scheme ($7.13\%$). However, it significantly elevates the symmetric rate over the no-CC scheme for both high- and low-SNR scenarios ($18\%$ for 40dB and $27\%$ for 5dB). The reason for this behavior is that increasing $\hat{\eta}$ improves the achievable global caching gain, which means that fewer interference terms are suppressed by beamformers, and hence, we can design more directed beams that are beneficial in power-limited finite-SNR communications.

Table~\ref{tab: P} and Table~\ref{tab: P alpha small} show how choosing the parameter $P$ affects the performance and complexity of the proposed scheme for large $(\alpha > \hat{\eta})$ and small antenna arrays $(\alpha \leq \hat{\eta})$, respectively. In Table~\ref{tab: P}, we consider the same network of $K=30$ users with $\gamma = 0.2$ and $L=10$, assume uniform user-to-profile association and $\alpha = L-1=9$, and compare the system performance when $P\in \{5,10\}$ and either \emph{Strategy A} or \emph{Strategy B} is selected for delivery phase (note that when $P=10$, \emph{Strategy~B} is not applicable because $\nicefrac{\alpha}{\hat{\eta}}$ is an integer). Here, we note that setting $P \in \lbrace 5,10 \rbrace$ satisfies the large-antenna constraint $\alpha > \hat{\eta}$, as for the uniform user distributions $\hat{\eta} = \nicefrac{K}{P}$. As observed, the maximum achievable DoF and symmetric rate are almost independent of $P$. However, setting $P=10$ significantly increases the subpacketization complexity and the total number of transmissions compared to $P=5$. On the other hand, for $P=5$, it is demonstrated that employing \textit{Strategy A} achieves a higher symmetric rate at finite-SNR scenarios, e.g., 20dB, compared to \textit{Strategy B} with less subpacketization complexity. Hence, in real-world scenarios with $\alpha> \eta_{1}$ and a fixed $\gamma$ value, in order to lower the complexity of the cache-aided setups and achieve a superior finite-SNR rate, it is better to choose the smallest possible $P$, i.e., to find $P$ and $\Brt$ such that $\mathrm{gcd}(P,\Brt)=1$ and $\Brt=P\gamma$,  then, use \textit{Strategy A} for the data delivery during the CC delivery step.

\begin{table}[t]
\vspace{-1.2em}
    \centering
    \caption{Impacts of $P$ on the Performance for Large-Antenna Arrays}
    \begin{tabular}{|c|c|c|c|}
        \cline{2-4}
        \multicolumn{1}{c}{} & \multicolumn{2}{|c|}{$P=5$} & \multicolumn{1}{|c|}{$P=10$} \\
        \cline{2-4}
       \multicolumn{1}{c}{} & \multicolumn{1}{|c|}{\textit{Strategy A}} & \multicolumn{1}{|c|}{\textit{Strategy B}} & \multicolumn{1}{|c|}{\textit{Strategy A}} \\
       \cline{2-4}
       \multicolumn{1}{c}{} & \multicolumn{1}{|c|}{$Q=2$} & \multicolumn{1}{|c|}{$Q=3$} & \multicolumn{1}{|c|}{$Q=5$} \\
         \hline
         Maximum Achievable DoF & $12$ & $15$ & $15$ \\
         \hline
         Subpacketization & $30$ & $225$ & $2835$ \\
         \hline
         Number of Transmissions & $60$ & $360$ & $4536$ \\
         \hline 
         Rate at $20$dB (bits/s/Hz) & $2.5936$ & $ 2.2735$ & $ 2.2510$ \\
         \hline
    \end{tabular}
    \label{tab: P}
    \vspace{-1.5em}
\end{table}

\begin{table}[htb]
    \centering
    \caption{{Impacts of $P$ on the Performance for Small-Antenna Arrays}}
     \label{tab: P alpha small}
    \begin{tabular}{|c|c|c|c|}
        \cline{2-4}
       \multicolumn{1}{c}{} & \multicolumn{1}{|c|}{$P=5$} & \multicolumn{1}{|c|}{$P=10$} & \multicolumn{1}{|c|}{$P=15$} \\
       \cline{2-4}
       \multicolumn{1}{c}{} & \multicolumn{1}{|c|}{$Q=2$} & \multicolumn{1}{|c|}{$Q=3$} & \multicolumn{1}{|c|}{$Q=4$} \\
         \hline
         Maximum Achievable DoF & $4$ & $6$ & $8$ \\
         \hline
         Subpacketization & $10$ & $90$ & $910$ \\
         \hline
         Number of Transmissions & $60$ & $360$ & $2730$ \\
         \hline 
         Rate at $20$dB (bits/s/Hz) & $1.2757$ & $ 1.7623$ & $ 2.2061$ \\
         \hline
    \end{tabular} 
\end{table}

In Table~\ref{tab: P alpha small}, a network of $K =30$ users, $\gamma=0.2$, $L=10$, $\alpha = 2$, and a uniform association of users to profiles is considered for $P \in \lbrace 5,10,15 \rbrace$. We note that for $P \in \lbrace 5,10,15 \rbrace$, the small-antenna array constraint $\alpha \leq \hat{\eta}$ is satisfied, the system can only operate with \emph{Strategy~A}, and it functions similarly to the proposed schemes in~\cite{parrinello2019fundamental} and \cite{9923619} during the CC delivery step. As per Theorem~\ref{thm: DoF} and \cite{parrinello2019fundamental}, the optimal achievable DoF for the region $\alpha \leq \hat{\eta}$ is $\alpha (P\gamma+1)$. On the other hand, the optimal DoF for a network of $K$ users with the spatial multiplexing gain of $\alpha$ is given by $K\gamma + \alpha$. In other words, the achievable DoF $\alpha (P\gamma+1)$ is upper bounded by $K\gamma + \alpha$ for the region $\alpha \leq \hat{\eta}$. Therefore, as observed from Table~\ref{tab: P alpha small}, as long as the constraint $P \leq \lceil \nicefrac{K}{\alpha} \rceil$ hold, an increase in $P$ enhances the achievable DoF. Assuming $\nicefrac{K}{\alpha} \in \mathbb{N}$, setting $P=\nicefrac{K}{\alpha}$ maximizes the achievable rate, and the DoF takes the form of $\alpha (\nicefrac{K}{\alpha} \gamma + 1) = K \gamma + \alpha$ with the cost of increased subpacketization complexity. 

 \begin{figure}[htb]
 \centering
       \pgfplotstableread[row sep=\\,col sep=&]{
sigmax & MC8\\
2	&	1.26001484	\\
4	&	1.594641868	\\
8	&	2.208524381	\\
12	&	2.752590005	\\
16	&	3.226222711	\\
20	&	3.639815014	\\
24	&	4.003171723	\\
28	&	4.285901637	\\
32	&	4.48172828	\\
36	&	4.561409097	\\
40	&	4.557579991	\\
44	&	4.382780514	\\
48	&	4.254538707	\\
50	&	4.251500042	\\
}\AM

\pgfplotstableread[row sep=\\,col sep=&]{
sigmax & MC8\\
2	&	0.488655913	\\
4	&	0.883986469	\\
8	&	1.573056242	\\
12	&	2.167426132	\\
16	&	2.692782711	\\
20	&	3.14741375	\\
24	&	3.543917042	\\
28	&	3.860881381	\\
32	&	4.096420115	\\
36	&	4.217691449	\\
40	&	4.241684813	\\
44	&	4.104873828	\\
48	&	3.845333292	\\
50	&	3.657445055	\\
}\AT

\pgfplotstableread[row sep=\\,col sep=&]{
sigmax & MC8\\
2	&	0.617029672	\\
4	&	0.74004108	\\
8	&	0.940866529	\\
12	&	1.085744033	\\
16	&	1.199451947	\\
20	&	1.278175863	\\
24	&	1.326498719	\\
28	&	1.357522724	\\
32	&	1.375286088	\\
36	&	1.387678772	\\
40	&	1.391355767	\\
44	&	1.396510925	\\
48	&	1.391783957	\\
50	&	1.3882787	\\
}\AS

\pgfplotstableread[row sep=\\,col sep=&]{
sigmax & MC8\\
2	&	0.272834596	\\
4	&	0.453940297	\\
8	&	0.723949227	\\
12	&	0.90995491	\\
16	&	1.049951807	\\
20	&	1.148955432	\\
24	&	1.221879529	\\
28	&	1.261526351	\\
32	&	1.292236175	\\
36	&	1.30808877	\\
40	&	1.319826643	\\
44	&	1.330038424	\\
48	&	1.329201587	\\
50	&	1.325209963	\\
}\US


\pgfplotstableread[row sep=\\,col sep=&]{
sigmax & MC8\\
2	&	157.7818924	\\
4	&	80.40278843	\\
8	&	40.41444	\\
12	&	26.97295484	\\
16	&	19.77884206	\\
20	&	15.64965551	\\
24	&	12.96486099	\\
28	&	11.01044587	\\
32	&	9.357825873	\\
36	&	8.205500159	\\
40	&	7.420851292	\\
44	&	6.759956785	\\
48	&	10.75603186	\\
50	&	15.83241755	\\
}\AU

\begin{tikzpicture}[scale=.33]

\begin{axis}[%
width=2.6\columnwidth,
height=1.7\columnwidth,
axis lines = left,
xmin=2,
xmax=50,
xtick={2,5,10,15,20,25,30,35,40,45,50},
xlabel={\Huge $\alpha$},
ymin=0,
ymax=5,
ylabel={\Huge Symmetric Rate [bits/s/Hz]},
ylabel near ticks,
    grid=both,
    major grid style={line width=.2pt,draw=gray!30},
    grid style={line width=.1pt, draw=gray!10},
    minor tick num=5,
ticklabel style={font=\Huge},
]


\addplot[mark=triangle, line width=1pt, mark size=5pt,mark options={solid},dashed,draw=blue,line width=2pt] table[x=sigmax,y=MC8]{\AM};
\label{plot1}

\addplot[mark=triangle, line width=1pt, mark size=5pt,mark options={solid},dashed,draw=black,line width=2pt] table[x=sigmax,y=MC8]{\AT};
\label{plot2}




\end{axis}

\begin{axis}[
    width=2.6\columnwidth,
height=1.7\columnwidth,
axis lines=right,
 ymin=0,
ytick={0,20,40,60,80,100,120,140},
axis y line=right,
    axis x line=none,
ylabel={\Huge Performance Improvement [\%]},
    legend style={at={(1,.15)},
                anchor=south east,legend columns=1},
legend style={ legend cell align=left, align=left, draw=white!15!black},
ticklabel style={font=\Huge},
]
\addlegendimage{/pgfplots/refstyle=plot1}\addlegendentry{\Huge Rate of Proposed Scheme}
\addlegendimage{/pgfplots/refstyle=plot2}\addlegendentry{\Huge Rate of no-CC}
\addplot[mark=triangle, line width=1pt, mark size=5pt,mark options={solid},dashed,draw=red,line width=2pt] table[x=sigmax,y=MC8]{\AU};
\addlegendentry{\Huge Performance Improvement}
\end{axis}
\end{tikzpicture}%
       \caption{The rate performance of our proposed scheme versus no-CC for $K=50$, $\gamma=0.08$, $\Brt=2$, $L=50$, $\text{SNR} = 20$dB, and different $\alpha$ values.}
       \label{fig: LargeAlpha}
       \vspace{-.8em}
       \end{figure}
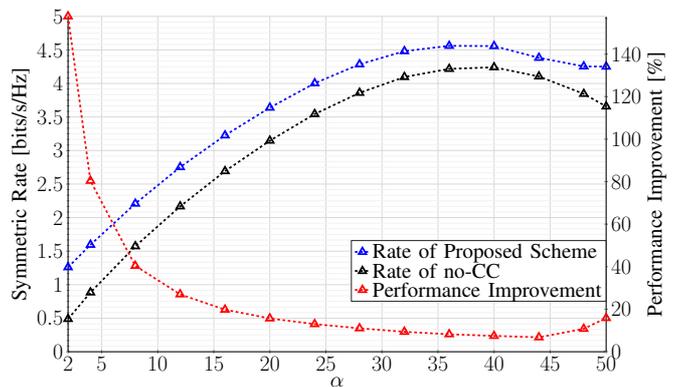

 Fig.~\ref{fig: LargeAlpha} compares the achievable symmetric of our proposed scheme with no-CC in a network with $K=50$, $\gamma=0.08$, $P=25$, $\Brt=2$,  $L=50$, $\text{SNR} = 20$dB, and different $\alpha$ values,  
where blue and black curves demonstrate the symmetric of our proposed scheme and no-CC, respectively, and the red curve illustrates the performance improvement of our proposed scheme over no-CC. As observed,   
increasing $\alpha$ does not monotonically increase the symmetric rate. For instance,  
increasing $\alpha$ from $2$ to $40$ improves the symmetric rate, while an increment in $\alpha$ from $40$ to $50$ reduces the achievable rate, due to that for mid-range SNR values (e.g., 20dB), we should serve fewer users at a time to achieve a higher symmetric rate~\cite{tolli2017multi}, revealing that for large-antenna arrays, setting $\alpha=L$ does not maximize the achievable rate, and we have to optimize $\alpha$ with respect to SNR and the other network parameters. Moreover, 
when $\alpha \leq 45$, the gap between our proposed scheme and the no-CC scheme decreases with $\alpha$, such that for $\alpha=2$ the performance improvement over the no-CC scheme is more than 150\%, while for $\alpha = 45$, the improvement is about $7\%$. On the other hand, increasing $\alpha$ from $46$ to $50$ increases the performance improvement to $18\%$, due to that for these values of $\alpha$, the achievable DoF is equal to $\min (K\gamma+\alpha,K)=K$, and consequently, we get some beamforming gain over the no-CC scheme.


\section{Conclusion} 
The dynamic nature of user behavior in practical applications poses a challenge for implementing coded caching schemes, which can be partially alleviated using a shared caching model. The conventional schemes in the literature, however, lack true adaptability as they are limited by specific network parameters. In this paper, we introduced a universally applicable shared-caching scheme for dynamic setups with no restrictions on network parameters. To this end, the placement phase was designed solely based on the users' cache ratio rather than the user count, and two transmission strategies were proposed to facilitate the data delivery during the delivery phase. Moreover, closed-form expressions were derived for the achievable DoF, demonstrating that our scheme achieved the optimal bounds of the shared-cache model. Additionally, we proposed a SIC-free data delivery algorithm and designed fast iterative beamformers to optimize the finite-SNR performance. Notably, it was shown that for uneven user distributions, although the achievable DoF may fall from the optimal DoF of shared caching, our scheme provided a significant performance improvement over unicasting at finite SNR regimes.

\appendices

\vspace{-.5em}
\section{Proof of Remark~\ref{rem: conditions}}
\label{apx: proof conditions}
As stated earlier, each transmission vector is constructed to serve at most $Q\beta$ users. Hence, each of these $Q\beta$ users is interfered by $Q\beta-1$ interfering users. Since each user has stored a $\gamma$ portion of the entire library, it can remove at most $\beta P \gamma=\beta \Brt$ interference terms using the stored cache content. Therefore, $\alpha$ must be large enough to suppress the remaining $Q\beta-1-\Brt \beta$ interference terms, i.e., 
\begin{equation}
    \alpha-1 \geq Q\beta-1-\Brt \beta,
\end{equation}
which results in $Q \leq \Brt + \left\lceil \nicefrac{\alpha}{\beta} \right\rceil$. Furthermore, $Q\geq \Brt+1$ should hold to benefit from the maximum coded caching gain.  

Let us assume that profile $p$ is one of the $Q$ selected profiles for the transmission. Therefore, $\min \left( \beta, \delta_{p} \right)$ users are served from profile $p$ during this transmission. On the one hand, it is supposed that the maximum number of users of each profile served in the CC delivery step is $\hat{\eta}$, which results in having $\beta \leq \hat{\eta}$. On the other hand, for each profile $p \in \left[ P \right]$, we should have $\min \left( \beta, \delta_{p} \right) \leq \alpha$ to be able to cancel out the inter-stream interference between users assigned to the same profile. Hence, according to the fact that $\delta_{1}=\hat{\eta}$, one can say $\min \left( \beta, \delta_{1} \right)=\min \left(\beta, \hat{\eta} \right)=\beta \leq \alpha$, which results in having the necessary condition $\beta \leq \min \left( \alpha,\hat{\eta} \right)$ to select the parameter $\beta$.

\section{Proof of Theorem~\ref{thm: DoF}}
\label{apx: proof DoF}
Here, first, we present a lemma to simplify the calculations, then, prove the theorem for strategies \textit{A} and \textit{B} in Sections~\ref{subsection: proof A} and \ref{subsection: proof B}, respectively.
\begin{lem}
    \label{lemma: comb}
    Given $P$ and $Q$, we have $\sum\limits_{r=1}^{P-Q+1} \binom{P-r}{Q-1}=\binom{P}{Q}$.
    \end{lem}
    
    \begin{IEEEproof}
        Assume a scenario at which we have a set $\CX=\left\lbrace b_{1}, \cdots, b_{P}  \right\rbrace$ with $\left\vert \CX \right\vert =P$, and we want to select the subsets $\Bar{\CX} \subseteq \CX$ such that $\vert \Bar{\CX} \vert=Q$. Here, our aim is to count the possibilities for the subsets $\Bar{\CX}$. One way to count these possibilities is to simply pick the subsets $\Bar{\CX}$ from $\CX$, and then count them. Here,  clearly, the number of possibilities is equal to $\binom{P}{Q}$. Another way to choose $\Bar{\CX}$ is to define $\left(P-Q+1 \right)$ subsets $\CL_{r}$ first, such that $r \in \left[ P-Q+1 \right]$, $\CL_{r}=\lbrace  b_{i}: r+1 \leq i \leq P \rbrace$ and $\vert \CL_{r} \vert=P-r$. Then, the selection of $\Bar{\CX}$ consists of $\left(P-Q+1 \right)$ rounds. In round $r \in \left[ P-Q+1 \right]$, we pick $b_{r}$ and choose $( Q-1 )$ remaining elements from $\CL_{r}$. In this regard, the number of possibilities to choose $\Bar{\CX}$ in round $r$ is equal to $\binom{P-r}{Q-1}$. By following this approach, we have $\sum_{r=1}^{P-Q+1} \binom{P-r}{Q-1}$ possibilities to choose $\Bar{\CX}$. Hence, it is sure that $\sum_{r=1}^{P-Q+1} \binom{P-r}{Q-1}=\binom{P}{Q}$.
    \end{IEEEproof}

\subsection{Proof of Theorem~\ref{thm: DoF} for \textit{Strategy A}}
\label{subsection: proof A}
In order to prove the theorem, we compute the total number of transmitted subpackets during the content delivery phase and divide it by the total number of transmissions in the CC and UC delivery steps.
First, we compute the number of transmission triples in \textit{Strategy A} that serve each user during the CC delivery step. Suppose that user $k$ with $p\left[k \right]=p$ is present in the CC delivery step. 
As per \eqref{eq: Mfinal} and \eqref{eq: Tfinal}, user $k$ is served in the transmission triple $\SfA =\left( r,c,l \right)$ such that $r \in \left[ p \right]$. On the other hand, according to \eqref{eq: Rp final} and \eqref{eq: Spj}, for each $r \in [p]$, there are $\beta = \min (\alpha, \hat{\eta})$ values for $c$ that user $k$ is served in the transmission triple $\SfA =\left( r,c,l \right)$. Now, for any $r \in [p]$, let $\CA_{k}^{(r)}$ be the set of values for $c$ at which user $k$ is served in the transmission triple $\SfA = \left( r,c,l \right)$ such that $\vert \CA_{k}^{(r)} \vert=\beta$. As mentioned in Section~\ref{subsection: Strategy A}, for $r=p$ and $c \in  \CA_{k}^{(p)} $, there are $ \binom{P-p}{Q-1}$ values for $l$ at which user $k$ is served during the transmission triple $\SfA $. Furthermore, for $r \in \left[ p-1 \right]$ and $c \in  \CA_{k}^{(r)}$, there exist $\binom{P-r-1}{Q-2}$  transmission triples  in which user $k$ receives the subpackets. Hence, the number of transmission triples serving user $k$ is given~by:
  \begin{equation}
  \label{eq: gpk def}
      \begin{array}{c}
        \beta \left( \binom{P-p}{Q-1}+ \sum\nolimits_{r=1}^{p-1}\binom{P-r-1}{Q-2} \right)=\beta \binom{P-1}{Q-1}.
      \end{array}
  \end{equation}
  \noindent
  As stated in Section~\ref{sec: sic-free}, by using the SIC-free data delivery, each transmission triple $\SfA$ is split into $\nu_{2} = \binom{Q-1}{Q-\Brt -1}$ transmissions $\SfA^{j}$ with $j \in [\nu_{2}]$, such that in each transmission $\SfA^{j}$, user $k \in \CT_{\SfA}$ receives only one subpacket of its requested file. Therefore, using \eqref{eq: gpk def}, the total number of subpackets received by each user via \emph{Strategy A} is $\beta \binom{P-1}{Q-1}\nu_{2}$. In order to give further insight into \eqref{eq: gpk def}, assume that user $k$ is assigned to profile $p$. To count the transmissions serving user $k$, the number of possibilities to choose $Q$ profiles with profile $p$ being selected is equal to $\binom{P-1}{Q-1}$. Now, we serve $\hat{\eta}$ times (assume that the length of each profile is greater than $\beta$) each of these $Q$ selected profiles (including profile $p$), such that in each transmission triple, the indices of users assigned to profiles are circularly shifted. Therefore, for each set of $Q$ selected profiles, user $k$ is served in $\beta$ transmission triples due to the fact that in each transmission triple, $\beta$ users from each profile are selected, and the indices of selected users are shifted in each transmission triple. However, each of these transmission triples contains $\nu_{2}$ subpackets for user $k$. Therefore, we split each transmission triple to $\nu_{2}$ transmissions to ensure that user $k$ receives only one of its desired subpackets per transmission. As a result, the total number of transmissions serving user $k$ and, equivalently, the total number of received subpackets by user $k$ is computed as $\beta \binom{P-1}{Q-1}\nu_{2}$.
  Now, according to that $K_{M}$ users are served via \textit{Strategy~A}, the total number of transmitted subpackets in the CC delivery step is equal to:
 \begin{equation}
 \begin{array}{c}
 \label{eq: JM A apx}
     J_{M}=K_{M} \beta \binom{P-1}{Q-1}\nu_{2}.
     \end{array}
 \end{equation}

Next, we show that all users can decode their requested files with \textit{Strategy A}. User $k$ stores a $\gamma$ portion of the entire library in the placement phase and needs to receive $\left( 1-\gamma \right) \beta \binom{P}{\Brt} \binom{P-\Brt -1}{Q-\Brt-1}$ subpackets in the CC delivery step. As mentioned,  user $k$ receives $\beta \binom{P-1}{Q-1}\nu_{2}=\left( 1-\gamma \right) \beta \binom{P}{\Brt} \binom{P-\Brt -1}{Q-\Brt-1}$ subpackets via \emph{Strategy~A}, demonstrating that user $k$ receives all its missing subpackets in the CC delivery step.
 
 Now, we calculate the total number of transmissions in the CC delivery step.
 As per \eqref{eq: Tfinal}, if $\delta_{r} = \min (\eta_{r}, \hat{\eta}) > 0$,  for any  $r \in \left[ P-Q+1 \right]$ and $c \in \left[ \phi_{r} \right]$, users assigned to the set $\CT_{\SfA}=\CS_{r,c} \Vert \CS_{b_{1},c} \Vert \cdots \Vert \CS_{b_{Q-1},c}$ are served, where $\left\lbrace b_{1},\cdots , b_{Q-1} \right\rbrace=\CM_{r} \left( l \right)$ and $l \in [ \binom{P-r}{Q-1} ]$. So, assuming $\delta_{r}>0$, the total number of transmission triples for any $r \in [ P-Q+1 ]$ is equal to $\phi_{r} \binom{P-r}{Q-1}$. We recall that $\phi_{r} = \max (\beta, \delta_{r})$.
 Furthermore, as mentioned in Section~\ref{sec: sic-free}, by employing the SIC-free data delivery, each transmission triple $\SfA$ is divided into $\nu_{2}$ transmissions $\SfA^{j}$ with $j \in [\nu_{2}]$. As a result, the total number of transmissions of \textit{Strategy~A}  is obtained as:
 \begin{equation}
 \label{eq: TM}
     \begin{array}{c}
         T_{M}=\sum\nolimits_{r=1}^{P-Q+1} D\left( \delta_{r}\right) \binom{P-r}{Q-1} \nu_{2}.
     \end{array}
 \end{equation}
 In order to give further observation into \eqref{eq: TM}, assume that the length of all profiles is greater than zero.  To create a transmission vector, first we choose a profile $p \in [P]$. Then, we select $Q-1$ profiles from $[p+1:P]$. Next, from these $Q$ selected profiles, we create $\max (\beta, \delta_{p})$ transmission triples. Now, since each transmission triple contains $\nu_{2}$ subpackets for each user, we divide each transmission triple into $\nu_{2}$ transmissions. As a result,  \eqref{eq: TM} is simplified to $ T_{M}=\sum\nolimits_{r=1}^{P-Q+1} \max (\beta, \delta_{p}) \binom{P-r}{Q-1} \nu_{2}$.  
\noindent
  In the next step, we compute the number of transmitted subpackets during the UC delivery step. By following a similar way as in~\cite{Abolpour2022CodedNetworks},   each user served with  UC delivery step must receive $\left( 1-\gamma \right) \beta \binom{P}{\Brt} \binom{P-\Brt -1}{Q-\Brt-1}$ subpackets. Since there are $K_{U}$ users in the UC delivery step, the number of transmitted subpackets during the UC delivery step is
 \begin{equation}
 \begin{array}{c}
 \label{eq: JU A apx}
     J_{U}=K_{U}\left( 1-\gamma \right) \binom{P}{\Brt}\beta^{\prime}.
     \end{array}
 \end{equation}
 
Then, we obtain the total number of transmissions during the UC delivery step. As stated earlier, during this step, we have to deliver $K_{U} \left( 1-\gamma \right)  \binom{P}{\Brt} \beta^{\prime} $ subpackets via $T_{U}$ transmissions, such that in each transmission, $\min \left( K_{U},\alpha \right)$ subpackets are transmitted. So, given $K_{U} \neq 0$, the total number of transmissions to deliver these subpackets is equal to:
 \begin{equation}
 \label{eq: TU}
     \begin{array}{c}
         T_{U}= \lceil \nicefrac{K_{U} \left( 1-\gamma \right) \binom{P}{\Brt}\beta^{\prime}}{\min \left( K_{U},\alpha \right)} \rceil.
     \end{array}
 \end{equation}
 
 Now, for $K_{U} \neq 0$, by substituting \eqref{eq: JM A apx}-\eqref{eq: TU} into \eqref{eq: DoF def}, the DoF is obtained as follows. 
 \begin{equation}
 \label{eq: DoF KUn0}
     \begin{array}{c}
         \mathrm{DoF}= \frac{K_{M} \binom{P-1}{Q-1}\beta \nu_{2}+ K_{U} \left( 1-\gamma  \right) \binom{P}{\Brt} \beta^{\prime}}{\sum\limits_{r=1}^{P-Q+1} D\left( \delta_{r}\right) \binom{P-r}{Q-1} \nu_{2}+\left\lceil \frac{K_{U} \left( 1-\gamma \right) \binom{P}{\Brt}\beta^{\prime}}{\min \left( K_{U},\alpha \right)} \right\rceil}.
     \end{array}
 \end{equation}
 Moreover, for $K_{U}=0$, we have $K_{M}=K$, and by applying \eqref{eq: JM A apx} and \eqref{eq: TM}  into \eqref{eq: DoF def}, the DoF takes the form as:
 \begin{equation}
 \label{eq: DoF KU0}
     \begin{array}{c}
         \mathrm{DoF}=\frac{K \binom{P-1}{Q-1}\beta}{\sum\nolimits_{r=1}^{P-Q+1} D\left( \delta_{r}\right) \binom{P-r}{Q-1}}.
     \end{array}
 \end{equation}
\subsection{Proof of Theorem~\ref{thm: DoF} for \textit{Strategy B}}
\label{subsection: proof B}
 In order to obtain the DoF for the dynamic MISO network with \textit{Strategy B} assisted by the SIC-free data delivery algorithm, first, we count the number of transmitted subpackets during the CC delivery step. In this regard, assume that user $k$ with $p \left[k \right]=p$ is present in the CC delivery step and receives its subpackets with \textit{Strategy~B}. According to \eqref{eq: erm}, \eqref{eq: Krsmu} and \eqref{eq: Icr non-int}, when $c \in \left[ P-Q+1 \right]$, $l \in [ \binom{P-c-1}{Q-2} ]$, $m \in [ \theta ]$ and $s \in \left[ \nu_{2} \right]$ , user $k$ is served during the transmission quintuples $\SfB = \left(p,c,l,m,s \right)$. Hence, for $r=p$, the number of transmission quintuples $\SfB =  \left( r,c,l,m,s \right)$ serving user $k$ is given by:
\begin{equation}
    \begin{array}{c}
           \sum\nolimits_{c=1}^{P-Q+1} \binom{P-c-1}{Q-2} \theta\nu_{2},
    \end{array}
\end{equation}
which by using Lemma~\ref{lemma: comb} is simplified to:
\begin{equation}
\label{eq: ser1}
    \begin{array}{c}
           \sum\nolimits_{c=1}^{P-Q+1} \binom{P-c-1}{Q-2} \theta\nu_{2}=\theta\nu_{2} \binom{P-1}{Q-1}.
    \end{array}
\end{equation}
Furthermore, as per \eqref{eq: erm}, \eqref{eq: Krsmu} and \eqref{eq: Icr non-int}, when $r \! \neq \! p$, $c \! \in \! \left[ P\!-\! Q \! + \! 1 \right]$, $l \! \in \! [ \binom{P-c-2}{Q-3}  ]$, $m \! \in \! \left[ \hat{\eta} \right]$ and $s \! \in \! \left[ \nu_{2} \right]$, the transmission quintuples $\SfB = \left( r,c,l,m,s \right)$ serve user $k$. Therefore, the number of transmission quintuples $\SfB = \left(r,c,l,m,s \right)$ serving user $k$ with $r\neq p$ is obtained as:
\begin{equation}
    \begin{array}{c}
          \sum\nolimits_{r=1}^{P-1} \sum\nolimits_{c=1}^{P-Q+1} \binom{P-c-2}{Q-3} \hat{\eta} \nu_{2},
    \end{array}
\end{equation}
which by using Lemma~\ref{lemma: comb} takes the form as follows. 
\begin{equation}
\label{eq: ser2}
    \begin{array}{c}
          \sum\nolimits_{r=1}^{P-1} \sum\nolimits_{c=1}^{P-Q+1} \binom{P-c-2}{Q-3} \hat{\eta} \nu_{2}=
         \hat{\eta} \nu_{2} \left( P-1 \right) \binom{P-2}{Q-2}.
    \end{array}
\end{equation}
As a result, by using \eqref{eq: ser1} and \eqref{eq: ser2}, and according to that $\theta=\alpha-\hat{\eta} \lfloor \nicefrac{\alpha}{\hat{\eta}} \rfloor$, the number of transmission quintuples serving user $k$ is equal to:
\begin{equation}
\label{eq: total service k}
    \begin{array}{c}
           \theta \nu_{2} \binom{P-1}{Q-1}+ \hat{\eta} \nu_{2} \left( P-1 \right) \binom{P-2}{Q-2}=
          \binom{P-1}{Q-1} \nu_{2} \left( \hat{\eta} \Brt +\alpha \right).
    \end{array}
\end{equation}
Moreover, as discussed in Section~\ref{sec: sic-free}, each transmission quintuple $\SfB $ is split into $\nu_{1} = \binom{Q-2}{Q-\Brt-2}$ transmissions $\SfB ^{j}$ with $j \in [\nu_{1}]$, where each transmission $\SfB ^{j}$ serves the user $k \in \Tilde{\CT}_{\SfB }$. Therefore, by using \eqref{eq: total service k}, the total number of subpackets that each user $k \in \Tilde{\CT}_{\SfB }$ receives during the CC delivery step is $\binom{P-1}{Q-1}\left( \hat{\eta} \Brt +\alpha \right)\nu_{2} \nu_{1}$. 
In order to give further observation, as mentioned in Section~\ref{subsection: strategy B},
during each transmission, $Q-1$ profiles are selected, and $\hat{\eta}$ users assigned to these profiles are served. Moreover, in each transmission, one profile is selected, and $\theta$ users assigned to this profile are served.
Assume user $k$ is assigned to profile $p$. If profile $p$ is one of the $Q-1$ profiles to serve $\hat{\eta}$ users, there are $P-1$ possibilities to choose a profile for serving $\theta$ users of it. Moreover, the number of possibilities to choose the remaining $Q-2$ profiles is $\binom{P-2}{Q-2}$.  
Now, from these $Q$ selected profiles, we make $\hat{\eta}\nu_{2}$ transmission quintuples by shifting the users' indices to ensure that each user receives $\nu_{1}$ subpackets per transmission quintuple. Here, we note that user $k$ assigned to profile $p$ receives $\nu_{1}$ subpackets in each transmission quintuple, as all users assigned to profile $p$ are served in each transmission quintuple. Hence, the total number of transmission quintuples serving user $k$ such that $\hat{\eta}$ users are selected from profile $p$ is obtained as $\hat{\eta}\nu_{2}(P-1) \binom{P-2}{Q-2}$. 
Now, if $\theta$ users are selected from profile $p$, the number of possibilities to select $Q-1$ profiles for serving $\hat{\eta}$ users assigned to them is equal to $\binom{P-1}{Q-1}$. Then, from these $Q$ selected profiles, we create $\hat{\eta} \nu_{2}$ transmission quintuples by shifting the users' indices to ensure that each user receives $\nu_{1}$ subpackets in each transmission quintuple. However, for each of these selected profiles,  user $k$ assigned to profile $p$ is served in $\theta \nu_{2}$ transmission quintuples, due to the fact that in each transmission quintuple $\theta$ users assigned to profile $p$ are served, and the indices of served users are circularly shifted in each transmission quintuple. 
Hence, the total number of transmission quintuples in which user $k$ is served and $\theta$ users from profile $p$ are selected takes the form of  $\theta \nu_{2} \binom{P-1}{Q-1}$.
Here, since each transmission quintuple contains $\nu_{1}$ subpackets for each user, each transmission quintuple is split into $\nu_{1}$ transmissions to ensure each user receives a single subpacket of its desired file per transmission. As a result, the total number of transmissions serving user $k$, and consequently, the total number of received subpackets by user $k$, is equal to $\binom{P-1}{Q-1}\left( \hat{\eta} \Brt +\alpha \right)\nu_{2} \nu_{1}$.  
Now, since $K_{M}$ users are served via \textit{Strategy~B}, the total number of transmitted subpackets in the CC delivery step is given by:
\begin{equation}
\label{eq: JM strategy B}
    \begin{array}{c}
         J_{M}=K_{M}\binom{P-1}{Q-1}  \left( \hat{\eta} \Brt +\alpha \right) \nu_{2} \nu_{1}.
    \end{array}
\end{equation}

Next, we show that user $k$ receives all its subpackets during the CC delivery step with \textit{Strategy B}. Since a $\gamma$ portion of the entire library is stored in the cache memory of user $k$, the total number of subpackets that must be transmitted to user $k$ during the CC delivery step is $\left( 1-\gamma\right) \binom{P}{\Brt}\binom{P-\Brt-1}{Q-\Brt-1} \left( \hat{\eta} \Brt +\alpha \right)\nu_{1}$. As stated earlier, the total number of subpackets that user $k$ receives with \textit{Strategy~B} is $\binom{P-1}{Q-1}  \left( \hat{\eta} \Brt +\alpha \right)\nu_{2} \nu_{1}=\left( 1-\gamma\right) \binom{P}{\Brt}\binom{P-\Brt-1}{Q-\Brt-1} \left( \hat{\eta} \Brt +\alpha \right)\nu_{1}$, showing that each user receives all its missing subpackets via \textit{Strategy~B}.

In the proceeding, we count the total number of transmissions with \textit{Strategy B}. As stated in Section~\ref{subsection: strategy B}, if $I^{+} \left( \Bar{\delta}_{c}, \CE_{r}^{m} \right)=1$, the server delivers the subpackets to the users during the transmission quintuple $\SfB =\left( r,c,l,m,s \right)$; otherwise, it does not transmit any signal. In addition, as per Section~\ref{sec: sic-free}, each transmission quintuple $\SfB $ is divided into $\nu_{1}$ transmissions $\SfB ^{j}$ with $j \in [\nu_{1}]$. Now, by following a similar approach to \eqref{eq: TM},  since $j \in [\nu_{1}]$, $r \in [P]$, $c \in \left[ P-Q+1 \right]$, $l \in [ \binom{P-c-1}{Q-2} ]$, $ m \in \left[ \hat{\eta} \right]$ and $s \in \left[ \nu_{2} \right]$, the total number of transmissions in \textit{Strategy B} is given by:
\begin{equation}
  \label{eq: TM B}
      \begin{array}{c}
       T_{M}= \sum\limits_{j=1}^{\nu_{1}} \sum\limits_{r=1}^{P} \sum\limits_{c=1}^{P-Q+1} \sum\limits_{m=1}^{\hat{\eta}} \sum\limits_{s=1}^{\nu_{2}} \binom{P-c-1}{Q-2}  I^{+} \left( \Bar{\delta}_{c}, \CE_{r}^{m} \right).
      \end{array}
  \end{equation}
Here, assuming the length of all caching profiles is non-zero, \eqref{eq: TM B} is simplified to $T_{M} = P \nu_{1} \nu_{2} \hat{\eta} \binom{P-1}{Q-1}$. In fact, we choose one profile to serve $\theta$ users and choose $Q-1$ profiles to serve $\hat{\eta}$ users. The total number of possibilities for selecting these $Q$ profiles is $P\binom{P-1}{Q-1}$. Finally, from each of these selected profiles, we create $\nu_{1} \nu_{2} \hat{\eta}$ transmissions.

Now, in order to obtain $J_{U}$ and $T_{U}$, we follow the same way as in \eqref{eq: JU A apx} and \eqref{eq: TU}. Accordingly, as stated in Section~\ref{subsection: strategy B}, each file is split into $\binom{P}{\Brt}\binom{P-\Brt-1}{Q-\Brt-1} \nu_{1} \left( \hat{\eta} \Brt +\alpha \right)$ subpackets. Therefore, according to the fact that the cache ratio at each user is $\gamma$ and $K_{U}$ users are served in the UC delivery step, the total number of transmitted subpackets during the UC delivery step is given by:
\begin{equation}
\label{eq: JU B}
    \begin{array}{cl}
         J_{U}&=K_{U} \left( 1-\gamma \right) \binom{P}{\Brt}\binom{P-\Brt-1}{Q-\Brt-1} \nu_{1} \left( \hat{\eta} \Brt +\alpha \right)\\
         &=K_{U} \left( 1-\gamma  \right) \binom{P}{\Brt} \alpha^{\prime}.
    \end{array}
\end{equation}
Furthermore, as mentioned in Section~\ref{section: UC}, assuming $K_{U}\neq 0$, each transmission of the UC delivery step serves $\min \left(K_{U}, \alpha \right)$ users. Hence, the total number of transmissions in the UC delivery step is obtained as follows. 
\begin{equation}
\label{eq: TU B}
    \begin{array}{l}
         T_{U}=\left\lceil \frac{K_{U} \left( 1-\gamma \right) \binom{P}{\Brt}\alpha^{\prime}}{\min \left( K_{U},\alpha \right)} \right\rceil.
    \end{array}
\end{equation}
As a result, for $K_{U}\neq 0$, by applying \eqref{eq: JM strategy B}-\eqref{eq: TU B} into \eqref{eq: DoF def}, the DoF is obtained as follows. 
\begin{equation}
    \begin{array}{c}
          \mathrm{DoF}= \frac{K_{M} \binom{P-1}{Q-1}\left( \hat{\eta}\Brt+\alpha \right)\nu_{2} \nu_{1}+ K_{U} \left( 1-\gamma  \right) \binom{P}{\Brt} \alpha^{\prime}}{N_{M} \nu_{1}+N_{U}}.
    \end{array}
\end{equation}
In addition, when $K_{U}=0$ and $K_{M}=K$, by substituting \eqref{eq: JM strategy B} and \eqref{eq: TM B} into \eqref{eq: DoF def}, the DoF is simplified to:
\begin{equation}
\begin{array}{c}
\label{eq: DoF Ku0 B}
    \mathrm{DoF}=  \frac{K \binom{P-1}{Q-1}\left( \hat{\eta}\Brt+\alpha \right) \nu_{2}}{N_{M}}.
    \end{array}
\end{equation}

\section{Proof of Remark~\ref{rem: optimal}}
\label{apx: proof DoF alphaleta uniform}
When $K$ users are uniformly assigned to $P$ caching profiles, i.e., $K=P\hat{\eta}$, and all users are served via CC delivery step, i.e., $K_{M}=K$ and $K_{U}=0$, we prove Remark~\ref{rem: optimal} for three regimes: \textit{1)} $\alpha \leq \hat{\eta}$, \textit{2)} $\alpha> \hat{\eta}$ with integer $\frac{\alpha}{\hat{\eta}}$ and \textit{3)} $\alpha> \hat{\eta}$ with non-integer $\nicefrac{\alpha}{\hat{\eta}}$.

\textit{1)}  $\alpha\leq \hat{\eta}$: In this regime, the system operates with \textit{Strategy~A} during the CC delivery step, where  $\beta=\alpha$ and $Q=\Brt+1$. By considering $\hat{\eta}=\eta_{r}$ for $\forall r \in [P]$, we know that $D\left( \delta_{r} \right)=\hat{\eta}$. Next. we reformulate \eqref{eq: DoF thm} as follows. 
    \begin{equation}
\label{eq: DoF alpha l eta simp}
    \begin{array}{c}
    \mathrm{DoF}=\frac{K \binom{P-1}{Q-1}\alpha}{ \hat{\eta} \sum\nolimits_{r=1}^{P-Q+1} \binom{P-r}{Q-1}}.
    \end{array}
\end{equation}
   Now, using Lemma~\ref{lemma: comb} and setting $Q=\Brt +1$, the DoF in \eqref{eq: DoF alpha l eta simp} is simplified to:
    \begin{equation}
\label{eq: DoF alpha l eta final}
    \begin{array}{c}
    \mathrm{DoF}=\frac{K \binom{P-1}{Q-1}\alpha}{ \hat{\eta} \binom{P}{Q}}= \frac{K Q \alpha}{P\hat{\eta}}=\alpha \left( P\gamma +1 \right).
    \end{array}
\end{equation}

\textit{2)}  $\alpha > \hat{\eta}$ and integer $\nicefrac{\alpha}{\hat{\eta}}$: In this regime, we know that the system uses \textit{Strategy A} to deliver data during the CC delivery step, where $\beta=\hat{\eta}$ and $\Brt +1 \leq Q \leq \Brt + \nicefrac{\alpha}{\hat{\eta}}$. Here, in order to maximize the achievable DoF, we set $Q=\Brt + \nicefrac{\alpha}{\hat{\eta}}$. Then, similarly to the previous case, we consider $\hat{\eta}=\eta_{r}$ and $D \left( \delta_{r} \right)=\hat{\eta}$ for all $r \in [P]$. So, the DoF in \eqref{eq: DoF thm} is given by:
 \begin{equation}
  \label{eq: DoF alpha g eta uni}
    \begin{array}{c}
    \mathrm{DoF}=\frac{K \binom{P-1}{Q-1}\hat{\eta}}{ \hat{\eta} \sum\nolimits_{r=1}^{P-Q+1} \binom{P-r}{Q-1}}.
    \end{array}
\end{equation}
Finally, by using Lemma~\ref{lemma: comb} and \eqref{eq: DoF alpha g eta uni}, the maximum achievable DoF corresponding to this regime is as follows. 
\begin{equation}
    \begin{array}{c}
    \mathrm{DoF}=\frac{K \binom{P-1}{Q-1}}{\binom{P}{Q}}=
    \frac{KQ}{P}=
    \hat{\eta}\Brt+\alpha=
    K\gamma+\alpha.
    \end{array}
\end{equation}

\textit{3)} $\alpha> \hat{\eta}$ with non-integer $\nicefrac{\alpha}{\hat{\eta}}$: In this regime, in order to maximize the achievable DoF, the server can transmit data via \textit{Strategy B} during the CC delivery step, such that $\hat{\eta}=\eta_{r}$ for all $r \in [P]$, $\beta=\hat{\eta}$ and $Q=\Brt+\left\lceil \nicefrac{\alpha}{\hat{\eta}} \right\rceil$. Accordingly, we have $I^{+} \left( \Bar{\delta_{c}}, \CE_{r}^{m} \right)=1$ for all $r \in [P]$, $c \in \left[P-Q+1 \right]$, $l \in [ \binom{P-c-1}{Q-2} ]$, $m \in \left[ \hat{\eta} \right]$ and $s \in \left[ \nu_{2} \right]$. Hence, by using Lemma~\ref{lemma: comb} and \eqref{eq: NM def B}, $N_{M}$ is simplified to:
\begin{equation}
\begin{array}{c}
\label{eq: NM uniform B}
    N_{M}=P\hat{\eta} \binom{P-1}{Q-1}\nu_{2}.
    \end{array}
\end{equation}
Finally, by substituting \eqref{eq: NM uniform B} into \eqref{eq: DoF Ku0 B} and using $\Brt=P\gamma$, the DoF takes the form of:
\begin{equation*}
    \begin{array}{c}
    \label{eq: DoF final uniform B}
         \mathrm{DoF}=\frac{K\binom{P-1}{Q-1}\left( \hat{\eta} \Brt +\alpha \right) \nu_{2}}{P\hat{\eta} \binom{P-1}{Q-1}\nu_{2}}= \hat{\eta} \Brt+\alpha=K\gamma+\alpha.
   \end{array}
\end{equation*}


\section{Proof of Remark~\ref{rem: DoF max alphaleta}}
\label{apx: Proof DoF max alphaleta}
   For $Q=\Brt+1$ and $\eta_{1} \leq \alpha$, the system operates with $\textit{Strategy~A}$ during the CC delivery step, $\nu_{2}=1$ and $\beta^{\prime}=\beta$, where $\beta=\alpha$, if $\alpha \leq \hat{\eta}$; otherwise, $\beta=\hat{\eta}$. Assuming $\hat{\eta} < \eta_{1}$ (i.e. $K_{U}>0$) and using \eqref{eq: TU}, clearly, it is sure that $ \binom{P}{Q} + \frac{1}{\beta} T_{U} >\binom{P}{Q} $. So, by using Lemma~\ref{lemma: comb}, we have:
    \begin{equation}
    \label{eq: ineq tx alpha g eta def2}
        \begin{array}{c}
      \frac{1}{\beta} \sum_{r=1}^{P-Q+1} \beta \binom{P-r}{Q-1} +\frac{1}{\beta}T_{U} >  
      \frac{1}{\eta_{1}} \sum_{r=1}^{P-Q+1} \eta_{1} \binom{P-r}{Q-1}.
        \end{array}
    \end{equation} 
    Now, according to the fact that for any arbitrary parameters $w$, $x$, $y$ and $z$, if $w+x \geq y$, then $\max ( w,z )+x \geq y$, we rearrange \eqref{eq: ineq tx alpha g eta def2} as follows. 
    \begin{equation}
    \label{eq: ineq tx alpha g eta def}
        \begin{array}{l}
      \frac{1}{\beta} \! \sum_{r=1}^{P-Q+1} \! \max \left( \beta, \delta_{r} \right) \binom{P-r}{Q-1} \! + \! \frac{T_{U}}{\beta} \! > \!
      \frac{1}{\eta_{1}} \! \sum_{r=1}^{P-Q+1} \! \eta_{1} \binom{P-r}{Q-1}.
        \end{array}
    \end{equation} 
    We know that $D \left( \delta_{r} \right)=D \left( \eta_{r} \right)=0$ if $\eta_{r}=0$; otherwise, $D \left( \delta_{r} \right)=\max \left( \beta , \delta_{r} \right)$ and $D \left( \eta_{r} \right)=\max \left( \eta_{r}, \eta_{1} \right)=\eta_{1}$. Hence, we reformulate \eqref{eq: ineq tx alpha g eta def} as follows. 
    \begin{equation}
    \label{eq: ineq tx alpha g eta simp}
        \begin{array}{c}
      \frac{ K \binom{P-1}{Q-1} \beta}{\sum\nolimits_{r=1}^{P-Q+1} D \left( \delta_{r} \right) \binom{P-r}{Q-1} +T_{U} }  < 
       \frac{K \binom{P-1}{Q-1} \eta_{1}}{\sum\nolimits_{r=1}^{P-Q+1} D \left( \eta_{r} \right) \binom{P-r}{Q-1}}.
        \end{array}
    \end{equation}
    Furthermore, using $\beta^{\prime}=\beta$, $Q=\Brt+1$, $\nu_{2}=1$, \eqref{eq: JM A apx} and \eqref{eq: JU A apx}, we have:
    \begin{equation}
    \label{eq: subpacket rem}
        \begin{array}{c}
        J_{M}+J_{U}=K\binom{P-1}{Q-1}\beta.
        \end{array}
    \end{equation}
    Using \eqref{eq: subpacket rem} and \eqref{eq: DoF thm}, and according to that $\nu_{2}=1$,  the left side of \eqref{eq: ineq tx alpha g eta simp} demonstrates the DoF of a setup with $Q=\Brt+1$ and $\hat{\eta}<\eta_{1}$ (i.e., $K_{U} \neq 0$), while the right side corresponds to the DoF with $Q=\Brt+1$ and $\hat{\eta}=\eta_{1}$ (i.e., $K_{U} = 0$). Hence, for $Q=\Brt+1$ and $\eta_{1} \leq \alpha$, we should set $\hat{\eta}=\eta_{1}$ to maximize the achievable DoF.
\vspace{-1em}

\end{document}